\documentclass[draftcls,onecolumn]{IEEEtran}

\usepackage{amssymb, amsthm, amscd, MnSymbol, mathrsfs}
\usepackage{graphicx}
\usepackage{nicefrac}
\usepackage{eufrak}
\usepackage{xcolor}
\usepackage[makeroom]{cancel}

\usepackage{url}

\usepackage{cite}

\usepackage[para,online,flushleft]{threeparttable}
\usepackage{multirow}
\renewcommand{\arraystretch}{1}

\usepackage[caption=false,font=normalsize,labelfont=sf,textfont=sf]{subfig}

\usepackage[linesnumberedhidden,ruled,vlined]{algorithm2e}
\usepackage{algorithmicx}
\usepackage{color}
\usepackage[none]{hyphenat}

\newtheorem{theorem}{Theorem}[section]
\newtheorem{definition}[theorem]{Definition}
\newtheorem{lemma}[theorem]{Lemma}

\numberwithin{equation}{section}

\newcommand{\bmu}{{\boldsymbol \mu}}
\newcommand{\bSigma}{{\boldsymbol \Sigma}}

\newcommand{\bepsilon}{{\boldsymbol \epsilon}}
\newcommand{\bA}{{\boldsymbol A}}

\newcommand{\bX}{{\boldsymbol X}}
\newcommand{\bY}{{\boldsymbol Y}}

\newcommand{\ba}{{\boldsymbol a}}

\newcommand{\bx}{{\boldsymbol x}}

\newcommand{\by}{{\boldsymbol y}}

\newcommand{\cS}{{\mathcal S}}

\newcommand{\cK}{{\mathcal K}}

\newcommand{\E}{{\mathbb E}}
\newcommand{\Pb}{{\mathbb P}}
\newcommand{\R}{{\mathbb R}}

\newcommand{\cN}{{\mathcal{N}}}

\newcommand{\cQ}{{\mathcal{Q}}}

\DeclareMathOperator{\diag}{diag}

\DeclareMathOperator{\argmin}{argmin}
\DeclareMathOperator{\argmax}{argmax}

\begin{document}
\title{Compressed Hypothesis Testing:\\ To Mix or Not to Mix?}
\author{Myung~Cho,
        Weiyu Xu and~Lifeng Lai
\thanks{Myung~Cho is with the Department of ECE, California State University, Northridge, CA, 91330 USA. Email: michael.cho@csun.edu.}
\thanks{Weiyu~Xu is with the Department of ECE, University of Iowa, Iowa City, IA, 52242 USA. Email: weiyu-xu@uiowa.edu.}
\thanks{Lifeng Lai is with the Department of ECE, University of California, Davis, CA 95616, USA. Email: lflai@ucdavis.edu.}
\thanks{The work of Lifeng Lai was supported by National Science Foundation (NSF) under grant ECCS-2000415.}
\thanks{The work of Weiyu Xu was supported by NSF under grants ECCS-2000425 and ECCS-2133205.}
\thanks{This paper in part was presented at the 51st Annual Allerton Conference on Communication, Control, and Computing (Allerton) in 2013 \cite{xu2013compressed}.}
}

\maketitle
\begin{abstract}
In this paper, we study the problem of determining $k$ anomalous random variables that have different probability distributions from the rest $(n-k)$ random variables. Instead of sampling each individual random variable separately as in the conventional hypothesis testing, we propose to perform hypothesis testing using mixed observations that are functions of multiple random variables. We characterize the error exponents for correctly identifying the $k$ anomalous random variables under fixed time-invariant mixed observations, random time-varying mixed observations, and deterministic time-varying mixed observations. For our error exponent characterization, we introduce the notions of \emph{inner conditional Chernoff information} and \emph{outer conditional Chernoff information}. We demonstrated that mixed observations can strictly improve the error exponents of hypothesis testing, over separate observations of individual random variables. We further characterize the optimal sensing vector maximizing the error exponents, which leads to explicit constructions of the optimal mixed observations in special cases of hypothesis testing for Gaussian random variables. These results show that mixed observations of random variables can reduce the number of required samples in hypothesis testing applications. In order to solve large-scale hypothesis testing problems, we also propose efficient algorithms - LASSO based and message passing based hypothesis testing algorithms.
\end{abstract}

\section{Introduction}

In many areas of science and engineering such as network tomography, cognitive radio, radar, and Internet of Things (IoTs), one needs to infer statistical information of signals of interest  \cite{romero2018blind,hassanien2012moving,poor2008quickest, DuffieldProbabilistic, ENV:ENV265, Pisarenko19874,PrestiProbabilistic, TseLink, LaiCognitive, cook1993radar}. Statistical information of interest can be the means, the variances or even the distributions of certain random variables. Obtaining such statistical information is essential in detecting anomalous behaviors of random signals. Espeically, inferring distributions of random variables has many important applications including quickest detections of potential hazards, detecting changes in statistical behaviors of random variables \cite{thottan2003anomaly,banerjee2015data,sun2022quickest,poor2008quickest, basseville, LaiCognitive}, and detecting congested links with abnormal delay statistics in network tomography \cite{TseLink,tsang2003network,shih2003unicast}.

In this paper, we consider a multiple hypothesis testing problem with few compressed measurements, which has applications in anomaly detection. In particular, we consider $n$ random variables, denoted by $X_i$, $i \in \cS = \{1,2,...,n\}$, out of which $k$ ($k \ll n$) random variables follow a probability distribution $f_2(\cdot)$ while the much larger set of remaining  $(n-k)$ random variables follow another probability distribution $f_1(\cdot)$. However, it is unknown which $k$ random variables follow the distribution $f_2(\cdot)$. Our goal in this paper is to infer the subset of random variables that follow $f_2(\cdot)$. In our problem setup, this is equivalent to determining whether $X_i$ follows the probability distribution $f_1(\cdot)$ or $f_2(\cdot)$ for each $i$. The system model of anomaly detection considered in this paper has appeared in various applications such as cognitive radio \cite{LaiCognitive,duan2010cooperative,quan2008optimal}, quickest detection and search \cite{LifengMultiple,poor2008quickest,MalloyRare,MalloyAsilomar,MalloyISIT,heydari2018quickest,lau2019quickest}, and communication systems \cite{shamir2014fundamental,steinhardt2016memory,tsitsiklis1988decentralized}.

In order to infer the probability distribution of the $n$ random variables, one conventional method is to obtain $l$ separate samples for each random variable $X_i$ and then use hypothesis testing techniques to determine whether $X_i$ follows the probability distribution $f_1(\cdot)$ or $f_2(\cdot)$ for each $i$. To ensure correctly identifying the $k$ anomalous random variables with high probability, at least $\Theta(n)$ samples are required for hypothesis testing with these samples involving only individual random variables. However, when the number of random variables $n$ grows large, the requirement on sampling rates and sensing resources can easily become a burden in the anomaly detection. For example,  in a sensor network, if the fusion center aims to track the anomalies in data generated by $n$ chemical sensors,  sending all the data samples of individual sensors to the fusion center will be energy-consuming and inefficient in the energy-limited sensor network. In this scenario, reducing the number of samples for inferring the probability distributions of the $n$ random variables is desired in order to lessen the communication burden in the energy-limited sensor network. Additionally, in some applications introduced in  \cite{DuffieldProbabilistic, PrestiProbabilistic, TseLink} for the inference of link delay in networks, due to physical constraints, we are sometimes unable to directly obtain separate samples of individual random variables. Those difficulties raise the question of whether we can perform hypothesis testing from a much smaller number of samples involving mixed observations.

Inspired by the compressed sensing technique \cite{candes2005decoding,Neighborlypolytope,DonohoTanner}, which recovers a \emph{deterministic} sparse vector from its linear projections, we introduce a new approach, so-called the \textit{compressed hypothesis testing}, to find $k$ anomalous random variables out of $n$ random variables using mixed measurements instead of separate measurements. Mathematically, for each measurement, we can express the measurement result as $\by=\ba^T \bx$, where $\ba \in \R^{n \times 1}$ is a sensing vector, $\bx$ is a vector of $n$ random variables among which $k$ random variables have different probability distributions than the other $n-k$ random variables, and $\by$ is the measurement result.  In our problem setup, one can take $m$ such measurements, where each random variable takes an independent realization for each measurement. By comparison, in the standard compressed sensing \cite{candes2005decoding,Neighborlypolytope,DonohoTanner}, we have $\by = \ba^T \bmu + \bepsilon$ with some unknown \emph{deterministic} spare vector $\bmu$ and additive noise $\bepsilon$. So in standard compressed sensing, the unknown vector $\bx$ takes the same value in each measurement. Unlike the standard compressed sensing, in compressed hypothesis testing, $\bx$ is a vector of $n$ random variables taking independent realizations across different measurements. As discussed below, our problem is relevant to multiple hypothesis testing such as in communication systems \cite{shamir2014fundamental,steinhardt2016memory,tsitsiklis1988decentralized}.

In addition to the multiple hypothesis testing with communication constraints, related works on identifying anomalous random variables include the detection of an anomalous cluster in a network \cite{arias2011detection}, Gaussian clustering \cite{hardt2015tight}, group testing \cite{du2000combinatorial,aldridge2014group,atia2012boolean,chan2014non,coja2020information,Cheraghchi2011Group}, and quickest detection \cite{MalloyAsilomar, MalloyISIT, MalloyRare,Caromi:TCOM:13,cao2019sketching}. Especially, in \cite{MalloyAsilomar, MalloyISIT, MalloyRare,Caromi:TCOM:13}, the authors optimized adaptive \textit{separate} samplings of individual random variables and reduced the number of needed samples for individual random variables by utilizing the sparsity of anomalous random variables. However, the total number of observations is still at least $\Theta(n)$ for these methods \cite{MalloyAsilomar, MalloyISIT, MalloyRare,Caromi:TCOM:13}, since one is restricted to individually sample the $n$ random variables. The major difference between the previous research \cite{MalloyAsilomar, MalloyISIT, MalloyRare,Caromi:TCOM:13,arias2011detection,hardt2015tight} and ours is that we consider compressed measurements instead of separate measurements of individual random variables. Additionally, group testing is different from our problem setting, since our sensing matrices and variables are general sensing matrices and vectors taking real-numbered values, while in group testing \cite{du2000combinatorial, aldridge2014group,atia2012boolean,chan2014non,coja2020information,Cheraghchi2011Group}, Bernoulli matrices are normally used. Moreover, in group testing, the unknown vector is often assumed to be deterministic across different measurements rather than assumed to be taking independent realizations as in this paper.

The contribution of our research is four-fold. Firstly, we introduce a new framework, called the \textit{compressed hypothesis testing}, to find $k$ anomalous random variables out of $n$ random variables from mixed measurements. In this framework, we make each observation a general function of the $n$ random variables instead of using separate observations of each individual random variable. When the number of anomalous random variables $k$ is much smaller than the total number of random variables $n$, we analytically demonstrate that the compressed hypothesis testing framework can reduce the number of measurements to figure out the $k$ anomalous random variables out of $n$ random variables, by considering the sparsity of anomalous random variables.  In particular, we show that the number of samples required to correctly identify the $k$ anomalous random variables can be reduced to $O\left({\displaystyle \frac{k \log(n)}{\min_{p_v, p_w} C(p_v, p_w)}}\right)$ observations, where $C(p_v, p_w)$ is the Chernoff information between two possible distributions $p_v$ and $p_w$ for the proposed mixed observations. Secondly, we also show that mixed observations can strictly increase error exponents of the hypothesis testing, compared to separate sampling of individual random variables. For special cases of Gaussian random variables, we derive optimal mixed measurements to maximize the error exponent of the hypothesis testing. Various examples provided in this paper clearly demonstrate the advantage of the compressed hypothesis testing framework in reducing the sample complexity. Thirdly, from the compressed hypothesis testing framework, we introduce novel statistical concepts - \emph{the inner conditional Chernoff information} and \emph{the outer conditional Chernoff information}, to characterize the error exponent of compressed hypothesis testing.  Finally, we propose numerical algorithms - the Least Absolute Shrinkage and Selection Operator (LASSO) and the Message Passing (MP) based hypothesis testing algorithms - to solve large-scale compressed hypothesis testing problems with reduced computational complexity.

The rest of the paper is organized as follows. Section \ref{sec:model} describes the mathematical model of the considered anomaly detection problem. In Section \ref{sec:fixed_time_invariant}, we investigate the hypothesis testing error performance using time-invariant mixed observations in hypothesis testing, and propose corresponding hypothesis testing algorithms. And then, we provide their performance analysis. Section \ref{sec:random_timevarying} describes the case using random time-varying mixed observations to identify the anomalous random variables, and we derive the error exponent of wrongly identifying the anomalous random variables. In Section \ref{sec:deterministic_timevarying}, we consider the case using deterministic time-varying mixed observations for hypothesis testing, and derive a bound on the error probability. In Section \ref{sec:undersampling}, we take into account undersampling measurement case, where the number of measurement is smaller than the number of random variables, to show the advantage of compressed hypothesis testing with mixed measurements against separate measurements. In Section \ref{sec:example}, we demonstrate, through examples of Gaussian random variables, that linear mixed observations can strictly improve the error exponent over separate sampling of each individual random variable. Section \ref{sec:optimalmixing} describes the optimal mixed measurements for Gaussian random variables maximizing the error exponent in hypothesis testing. Section \ref{sec:efficient_algo} introduces efficient algorithms to find abnormal random variables using mixed observations, for large values of $n$ and $k$. In Section \ref{sec:simulation}, we demonstrate the effectiveness of our hypothesis testing methods with mixed measurements in various numerical experiments. Section \ref{sec:conclusion} provides the conclusion of this paper.

\textbf{Notations}: We denote a random variable and its realization by an uppercase letter and the corresponding lowercase letter respectively. We use $X_i$ to refer to the $i$-th element of the random variable vector $\bX$. We reserve calligraphic uppercase letters $\cS$ and $\cK$ for index sets, where $\cS = \{1,2,...,n\}$, and $\cK \subseteq \cS$. We use superscripts to represent time indices. Hence, $\bx^j$ represents the realization of a random variable vector $\bX$ at time $j$. We reserve the lowercase letters $f$ and $p$ for Probability Density Functions (PDFs). We also denote the probability density function $p_X(x)$ as $p(x)$ or $p_{X}$ for notation convenience. In this paper, $\log$ represents the logarithm with the natural number $e$ as its base.

\section{Mathematical Models}
\label{sec:model}
We consider $n$ independent random variables $X_i$, $i=1,2,\cdots,n$. Out of these $n$ random variables, $(n-k)$ of them follow a known probability distribution $f_1(\cdot)$; while the other $k$ random variables follow another known probability distribution $f_2(\cdot)$:
\begin{align}
    X_i \sim \begin{cases}
                f_1(\cdot), \; & i \notin \cK \\
                f_2(\cdot), \; & i \in \cK,
           \end{cases}
\end{align}
where $\cK \subset \cS =\{1,2,\cdots,n \}$ is an unknown ``support'' index set, and $|\cK| = k \ll n$. We take $m$ mixed observations of the $n$ random variables at $m$ numbers of time indices. The measurement at time $j$ is stated as
\begin{align*}
	Y^j=g^j(X_1^j, X_2^j,..., X_n^j),
\end{align*}
which is a function of $n$ random variables, where $1\leq j \leq m$. Note that the random variable $X_i^j$ follows the probability distribution $f_1(\cdot)$ or $f_2(\cdot)$ depending on whether $i \in \cK$ or not, which is the same distribution as the random variable $X_i$. Our goal in this paper is to determine $\cK$ by identifying those $k$ anomalous random variables with as few measurements as possible. We assume that the realizations at different time slots are mutually independent. Additionally, although our results can be extended to nonlinear observations, in this paper, we specifically consider the case when the functions $g^j(\cdot)$'s are \emph{linear} due to its simplicity and its wide range of applications including network tomography \cite{TseLink} and cognitive radio \cite{LaiCognitive}. Especially, the network tomography problem is a good example of the considered linear measurement model, where the goal of the problem is figuring out congested links in a communication network by sending packets through probing paths that are composed of connected links.  The communication delay through a probing path is naturally a linear combination of the random variables representing the delays of that packet traveling through corresponding links.

Throughout this paper, when the measurements through functions $g^{j}$'s are taken, the decoder knows the functions $g^{j}$'s. In particular, when the functions are linear functions, the decoder knows the coefficients of these linear functions or the matrices $A$ as discussed later in this paper.

When the functions $g^j$'s are linear, the $j$-th measurement is stated as follows:
\begin{align}
\label{eq:observation}
    Y^j=g^j(X_1^j, X_2^j,..., X_n^j)=\sum_{i=1}^{n}a_i^j X_i^j = \langle \ba^j, \bX^j \rangle,
\end{align}
where a sensing vector $\ba^j=[a_1^j, a_2^j,~...~, a_n^j]^T \in \R^{n\times 1}$, and $\bX^j=[X_1^j, X_2^j,..., X_n^j]^T$. We obtain an estimate of the index set $\cK$ by using a decision function $\phi(\cdot)$ from $Y^j$, $j=1,\cdots,m$, as follows:
\begin{eqnarray}
\hat{\cK}=\phi(Y^1,\cdots,Y^m).
\end{eqnarray}
We would like to design the sampling functions $g^j(\cdot)$'s and the decision function $\phi(\cdot)$ such that the probability
\begin{align}
    \Pb(\hat{\cK}\neq \cK)\leq \epsilon,
\end{align}
for an arbitrarily small $\epsilon>0$.

Our approach is inspired by compressed sensing technique \cite{candes2005decoding,Neighborlypolytope,DonohoTanner,duarte2011structured}. However, our problem has a major difference from the conventional compressed sensing problem introduced in \cite{candes2005decoding,Neighborlypolytope,DonohoTanner,duarte2011structured}. In our problem setup, each random variable $X_i$ takes an \emph{independent} realization in each measurement, while in the conventional compressed sensing problem  $\by = \bA\bx$,  where $\by \in \R^{m \times 1}$ is the observation vector and $\bA  \in \R^{m \times n}$ is a sensing matrix, the vector $\bx = [x_1,x_2,x_3,...,x_n]^T$ takes the same \emph{deterministic} values across all the measurements. In some sense, our problem is a compressed sensing problem dealing with random variables taking different values across measurements. Bayesian compressed sensing stands out as one model where prior probability distributions of the vector $\bx$ is considered \cite{BayesianCS1, StatisticalCS2}. However, in \cite{BayesianCS1, StatisticalCS2}, the vector $\bx = [x_1,x_2,x_3,...,x_n]^T$ is \emph{fixed} once the random variables are realized from the prior probability distribution, and then remains \emph{unchanged} across different measurements. That is fundamentally different from our setting where random variables dramatically change across different measurements.  There also exists a collection of research works \cite{timevaryingGiannakis, Vaswani, DavidTimeVarying,ziniel2013dynamic} discussing compressed sensing for \emph{smoothly} time-varying signals. In contrast to \cite{timevaryingGiannakis, Vaswani, DavidTimeVarying,ziniel2013dynamic}, the objects of interest in this research are random variables taking \emph{completely independent} realizations at different time indices, and we are interested in recovering statistical information of random variables, rather than recovering the deterministic values.

\section{Compressed Hypothesis Testing}
\label{sec:compressed_hypothesis_testing}
In compressed hypothesis testing, we consider three different types of mixed observations, namely \emph{fixed time-invariant} mixed measurements, \emph{random time-varying} measurements, and \emph{deterministic time-varying} measurements. Table \ref{tbl:type} summarizes the definition of these types of measurements. For these different types of mixed observations, we characterize the number of measurements required to achieve a specified hypothesis testing error probability.

\begin{table}[t]
\caption{Three different types of mixed observations}
\label{tbl:type}
\begin{center}
\renewcommand{\arraystretch}{1.4}
\begin{tabular}{|c| p{12 cm}|}
\hline
    \textbf{Measurement type} &  \textbf{Definition} \\
\hline\hline
    Fixed time-invariant &  The measurement function is the same at every time index.\\
\hline
    Random time-varying & The measurement function is randomly generated from a distribution at each time index. \\
\hline
    Deterministic time-varying &  The measurement function is time-varying at each time index but predetermined.\\
\hline
\end{tabular}
\end{center}
\end{table}

\subsection{Fixed Time-Invariant Measurements}
\label{sec:fixed_time_invariant}
In this subsection, we focus on a simple case in which sensing vectors are time-invariant across different time indices, i.e., $\ba^1=\cdots=\ba^m := \ba$, where $\ba \in \R^{n\times 1}$. This simple case helps us to illustrate the main idea that will be generalized to more sophisticated schemes in later sections.

We first introduce the maximum likelihood estimate method with fixed time-invariant sensing vectors in Algorithm \ref{alg:likelihoodratiotest_timeinvariant}, over $l = \binom{n}{k}$ possible hypotheses. Since there are $l=\binom{n}{k}$ hypotheses, the complexity to calculate the probability (or likelihood)  is proportional to the number of hypotheses, i.e., $l$. Therefore, Algorithm \ref{alg:likelihoodratiotest_timeinvariant} is exponentially complex with respect to $k$ when $n$ is fixed. To analyze the number of required samples for achieving a certain hypothesis testing error probability, we consider another related hypothesis testing algorithm based on pairwise hypothesis testing in Algorithm \ref{alg:pairwise_timeinvariant}. Since there are $l = \binom{n}{k}$ possible probability distributions for the output of the function $g(\bX)= \langle \ba, \bX \rangle$, where $\bX \in \R^{n}$, depending on which $k$ random variable $X_i$'s are anomalous, we denote these possible distributions as $p_1$, $p_2$, ..., and $p_l$. Our likelihood ratio test algorithm is to find the true distribution by doing pairwise Neyman-Pearson hypothesis testing \cite{coverbook} over these $l$ distributions. We provide the sample complexity for finding the $k$ anomalous random variables by using the fixed time-invariant mixed measurements in Theorem \ref{thm:numberofsamples}.

\begin{algorithm}[t]
\LinesNumbered
 \KwData{Observation data $\by = [y^1, y^2, ..., y^m]^T$, $l = \binom{n}{k}$}
 \KwResult{$k$ anomalous random variables}


    \nlset{1} For each hypothesis $H_v$ ($1\leq v\leq l$), calculate the likelihood $p_{\bY|H_v}(\by|H_v)$.\\
    \nlset{2} Choose the hypothesis with the maximum likelihood.\\
    \nlset{3} Decide the corresponding $k$ random variables as the anomalous random variables.
 \caption{Maximum likelihood estimation with fixed time-invariant measurements}
 \label{alg:likelihoodratiotest_timeinvariant}
\end{algorithm}

\begin{algorithm}[t]
\LinesNumbered
 \KwData{Observation data $\by = [y^1, y^2, ..., y^m]^T$}
 \KwResult{$k$ anomalous random variables}

\nlset{1} For all pairs of distinct probability distributions $p_v$ and $p_w$ ($1\leq v, w \leq l$, $l= \binom{n}{k}$, and $v\neq w$), perform Neyman-Pearson testing for two hypotheses:
\begin{itemize}
\item $Y^1$, $Y^2$, ..., $Y^m$ follow probability distribution $p_v$
\item $Y^1$, $Y^2$, ..., $Y^m$ follow probability distribution $p_w$
\end{itemize}

 \nlset{2} Find a certain $w^*$ such that $p_{w^*}$ is the winning probability distribution whenever it is involved in a pairwise hypothesis testing.\\
 \nlset{3} Declare the $k$ random variables producing $p_{w^*}$ as anomalous random variables.\\
 \nlset{4} If there is no such $w^*$ in Step 2, then declare a failure in finding the $k$ anomalous random variables.

\caption{Likelihood ratio test with fixed time-invariant mixed measurements}
\label{alg:pairwise_timeinvariant}
\end{algorithm}

\begin{theorem} \label{thm:fixed_time_invariant_measurements}
Consider fixed time-invariant measurements $Y^j={\ba^j}^T \bX^j = \ba^T \bX^j$, $1 \leq j \leq m$, for $n$ random variables $X_1, X_2, ..., X_n$. Algorithms \ref{alg:likelihoodratiotest_timeinvariant} and \ref{alg:pairwise_timeinvariant} correctly identify the $k$ anomalous random variables with high probability, with $O\left(\frac{k \log(n)}{\underset{1\leq v,w \leq l, v\neq w}{\min} C(p_v, p_w)}\right)$ mixed measurements.
Here, $l$ is the number of hypotheses, $p_v$ is the output probability distribution for measurements $Y = y$ under hypothesis $H_v$, and
\begin{align}
	C(p_v, p_w)=-\min_{0\leq \lambda \leq 1} \log\left(\int{p_v^{\lambda}(y)p_w^{1-\lambda}(y)dy}\right)
\end{align}
is the Chernoff information between two probability distributions $p_v$ and $p_w$.
\label{thm:numberofsamples}
\end{theorem}

\begin{IEEEproof}
In Algorithm \ref{alg:pairwise_timeinvariant}, for two probability distributions $p_v$ and $p_w$, we choose the probability likelihood ratio threshold of the Neyman-Pearson testing in such a way that the error probability decreases with the largest possible error exponent, namely the Chernoff information between $p_v$ and $p_w$:
\begin{equation*}
	C(p_v, p_w)=-\min_{0\leq \lambda \leq 1} \log\left(\int{p_v^{\lambda}(y)p_w^{1-\lambda}(y)dy}\right).
\end{equation*}

Overall, the smallest possible error exponent of making an error between any pair of probability distributions is
\begin{align}\label{eq:E}
	E = \min_{1\leq v,w \leq l, v\neq w} C(p_v, p_w).
\end{align}
Without loss of generality, we assume that $p_1$ is the true probability distribution for the observation data $\bY = \by$. Since the error probability $\Pb_{err}$ in the Neyman-Pearson testing scales as (the exponent of the scaling is asymptotically tight \cite{coverbook})
\begin{align*}
	\Pb_{err} \leq 2^{-m C(p_v, p_w)} \leq 2^{-mE},
\end{align*}
where $m$ is the number of measurements \cite[Chapter 11.9]{coverbook}. By the union bound over the $l-1$ possible pairs $(p_1, p_w)$, the probability that $p_1$ is not correctly identified as the true probability distribution scales at most as $l\times 2^{-mE} := \epsilon $, where $l=\binom{n}{k}$. From the upper and lower bounds on binomial coefficients $\binom{n}{k}$,
\begin{align*}
	\bigg( \frac{n}{k}\bigg)^k \leq \binom{n}{k} \leq \bigg( \frac{en}{k}\bigg)^k,
\end{align*}
where $e$ is the natural number, and $1 \leq k \leq n$, for the failure probability, we have
\begin{align*}
	\bigg( \frac{n}{k}\bigg)^k 2^ {-mE}\leq \binom{n}{k}2^{-mE} := \epsilon \leq \bigg( \frac{en}{k}\bigg)^k 2^ {-mE}.
\end{align*}
Thus, for the number of measurements, we have
\begin{align} \label{eq:m}
	\bigg( k \log(n) - k \log(k) - \log(\epsilon) \bigg)E^{-1} \leq m \leq \bigg( k \log(n) + k \log(e/k) - \log(\epsilon) \bigg)E^{-1}.
\end{align}
Therefore, $\Theta(k \log(n)E^{-1})$, where $E$ is introduced in \eqref{eq:E},  samples are enough for identifying the $k$ anomalous samples with high probability.
\end{IEEEproof}

Each random variable among the $n$ numbers of random variables has the same probability of being an abnormal random variable. Thus, possible locations of the $k$ different random variables out of $n$ follow uniform prior distribution; namely, every hypothesis has the same prior probability to occur. Algorithm \ref{alg:likelihoodratiotest_timeinvariant} is based on maximum likelihood detection, which is known to provide the minimum error probability with uniform prior \cite{levy2008principles}. Additionally, since the Likelihood Ratio Test (LRT) can provide the same result as the maximum likelihood estimation when the threshold value is one, Algorithm \ref{alg:pairwise_timeinvariant}, which is an LRT algorithm, can provide the same result as Algorithm \ref{alg:likelihoodratiotest_timeinvariant} with a properly chosen threshold value in the Neyman-Pearson test.

We also remark that the error exponent (Chernoff information) for the Neyman-Person test is tight, in the sense that a lower bound on the error probability for pairwise Neyman-Person test scales with the same exponent.


If we are allowed to use time-varying sketching functions, we may need fewer samples. In the next subsection, we discuss the performance of time-varying mixed measurements for this problem.

\subsection{Random Time-Varying Measurements}
\label{sec:random_timevarying}

Inspired by compressed sensing where random measurements often provide desirable sparse recovery performance \cite{candes2005decoding, DonohoTanner}, we consider random time-varying measurements. In particular, we assume that each measurement is the inner product between the random vector $\bX \in \R^{n \times 1}$ and one independent realization $\ba^j=[a_1^j,a_2^j,..., a_{n}^j]^T$ of a random sensing vector $\bA \in \R^{n \times 1}$ at time $j$. Namely, each observation is given by
$$Y^j= \langle \bA^j,\bX^j \rangle = \sum\limits_{i=1}^{n} a_i^j X_i^j, \; 1\leq j \leq m,$$
where $\ba^j=[a_1^j,a_2^j,..., a_{n}^j]^T$ is a realization of the random sensing vector $\bA$ with the PDF $p_{\bA}(\ba)$ at time $j$. We assume that the realizations $\ba^j$'s of $\bA$ are independent across different time indices.

We propose the maximum likelihood estimate method with random time-varying measurements over $\binom{n}{k}$ hypotheses in Algorithm \ref{alg:likelihoodratiotest_timevarying}. For the purpose of analyzing the error probability of the maximum likelihood estimation, we further propose a hypothesis testing algorithm based on pairwise comparison in Algorithm \ref{alg:timevaryingalg2}. The number of samples required to find the abnormal random variables is stated in Theorem \ref{thm:numberofsamples_randomtimevaring}. Before we introduce our theorem for hypothesis testing with random time-varying measurements, we newly introduce the Chernoff information between two conditional probability density functions, named it as the inner conditional Chernoff information, in Definition \ref{def:IC}.

\begin{algorithm}[t]
    \LinesNumbered
    \KwData{Observation and sensing vector at time from $1$ to $m$: $(\bA^1 = \ba^1, Y^1 = y^1)$, $(\bA^2 = \ba^2,Y^2 = y^2)$, ..., $(\bA^m = \ba^m, Y^m = y^m)$}
    \KwResult{$k$ anomalous random variables}

    \nlset{1} For each hypothesis $H_v$, $1\leq v\leq l$, $l = \binom{n}{k}$, calculate the likelihood $p_{\bA,Y|H_v}(\ba^j,y^j,\;j=1,...,m \;|\;H_v)$.\\
    \nlset{2} Choose the hypothesis with the maximum likelihood.\\
    \nlset{3} Decide the corresponding $k$ random variables as the anomalous random variables.

    \caption{Maximum likelihood estimation with random time-varying measurements}
    \label{alg:likelihoodratiotest_timevarying}
\end{algorithm}

\begin{algorithm}[t]
    \LinesNumbered
    \KwData{Observation and sensing vector at time from $1$ to $m$: $(\bA^1 = \ba^1, Y^1 = y^1)$, $(\bA^2 = \ba^2, Y^2 = y^2)$, ..., $(\bA^m = \ba^m, Y^m = y^m)$}
    \KwResult{$k$ anomalous random variables}
    \nlset{1} For all pairs of hypotheses $H_v$ and $H_w$ ($1\leq v, w \leq l$ and $v\neq w$, $l = \binom{n}{k}$), perform Neyman-Pearson testing of the following two hypotheses:
    \begin{itemize}
        \item $(\bA^1,Y^1)$, $(\bA^2,Y^2)$, ..., $(\bA^m, Y^m)$ follow probability distribution $p_{\bA,Y|H_v}(\ba,y|H_v)$
        \item $(\bA^1,Y^1)$, $(\bA^2,Y^2)$, ..., $(\bA^m, Y^m)$ follow probability distribution $p_{\bA,Y|H_w}(\ba,y|H_w)$
    \end{itemize}
    \nlset{2} Find a certain $w^*$ such that $H_{w^*}$ is the winning hypothesis, whenever it is involved in a pairwise hypothesis testing.\\
    \nlset{3} Declare the $k$ random variables producing $H_{w^*}$ as anomalous random variables.\\
    \nlset{4} If there is no such $w^*$ in Step 2, declare a failure in finding the $k$ anomalous random variables.
    \caption{Likelihood ratio test with random time-varying measurements}
    \label{alg:timevaryingalg2}
\end{algorithm}

\begin{definition}[Inner Conditional Chernoff Information] \label{def:IC}
Let $p_{\bA,Y |H_v}(\ba,y)$ be an output probability distribution for measurements $Y$'s and random sensing vectors $\bA$'s under hypothesis $H_v$. Then, for two hypotheses $H_v$ and $H_w$, the \emph{inner conditional Chernoff information} between two hypotheses $H_v$ and $H_w$ is defined as 
\par\noindent\small
\begin{align}
IC(p_{\bA,Y|H_v}, p_{\bA,Y|H_w}) :=-\min_{0\leq \lambda \leq 1} \log\left(\int{ \int{p_{\bA,Y|H_v}^{\lambda}(\ba,y|H_v)p_{\bA,Y|H_w}^{1-\lambda}(\ba,y|H_w)dy} \;d\ba}\right).
\end{align}
\normalsize
\end{definition}

With the definition of the inner conditional Chernoff information, we give our theorem on the sample complexity of our algorithms as follows.
\begin{theorem}
Consider time-varying random measurements $Y^j= y^j$, $1\leq j \leq m$, for $n$ random variables $X_1$, $X_2$, ...,  and $X_n$. Algorithms \ref{alg:likelihoodratiotest_timevarying} and \ref{alg:timevaryingalg2} correctly identify the $k$ anomalous random variables with high probability, in $O\left(\frac{k \log(n)}{\min\limits_{1\leq v,w \leq l, v\neq w} IC(p_{\bA,Y | H_v}, p_{\bA,Y | H_w})}\right)$ random time-varying measurements. Here, $l = \binom{n}{k}$ is the number of hypotheses, $p_{\bA,Y |H_v}$ is the output probability distribution for measurements $Y$'s and random sensing vectors $\bA$'s under hypothesis $H_v$, and $IC(p_{\bA,Y | H_v}, p_{\bA,Y | H_w})$ is the inner conditional Chernoff information defined in Definition \ref{def:IC}. Moreover, the inner conditional Chernoff information is equivalent to
\par\noindent\small
\begin{align}\label{eq:IC_EA}
IC(p_{\bA,Y|H_v}, p_{\bA,Y|H_w})
	& =-\min_{0\leq \lambda \leq 1} \log\left( \E_{\bA} \left(\int{p_{Y|\bA,H_v}^{\lambda}(y|\ba, H_v)p_{Y|\bA,H_w}^{1-\lambda}(y|\ba,H_w)\,dy}\right)\right).
\end{align}
\normalsize
\label{thm:numberofsamples_randomtimevaring}
\end{theorem}

\begin{IEEEproof}
In Algorithm \ref{alg:timevaryingalg2}, for two different hypotheses $H_v$ and $H_w$, we choose the probability likelihood ratio threshold of the Neyman-Pearson testing in such a way that the hypothesis testing error probability decreases with the largest error exponent, which is the Chernoff information between $p_{\bA,Y|H_v}(\ba,y|H_v)$ and $p_{\bA,Y|H_w}(\ba,y|H_w)$. Since the random time-varying sensing vectors are independent of random variable $X$ and the hypothesis $H_v$ or $H_w$, we have
\begin{align}\label{eq:conditional_prob}
    & p_{\bA,Y|H_v}(\ba,y|H_v)=p(\ba|H_v) p(y|H_v,\ba)=p(\ba) p(y|H_v,\ba), \nonumber \\
    & p_{\bA,Y|H_w}(\ba,y|H_w)=p(\ba|H_w) p(y|H_w,\ba)=p(\ba) p(y|H_w,\ba).
\end{align}

Then, the Chernoff information between $p_{\bA,Y|H_v}$ and $p_{\bA,Y|H_w}$, denoted by  $IC(p_{\bA,Y|H_v}, p_{\bA,Y|H_w}) $, is obtained as follows:
\par\noindent\small
\begin{align*}
IC(p_{\bA,Y|H_v}, p_{\bA,Y|H_w}) 	&=-\min_{0\leq \lambda \leq 1} \log\left(\int{ \int{p^{\lambda}(\ba,y|H_v)p^{1-\lambda}(\ba,y|H_w)\;dy} \;d\ba}\right)  \nonumber \\
 	&  \overset{\eqref{eq:conditional_prob}}{=}-\min_{0\leq \lambda \leq 1} \log\left(\int p_{\bA}(\ba)\left(\int{p^{\lambda}(y|\ba,H_v)p^{1-\lambda}(y|\ba,H_w)\,dy}\right) \,d\ba\right) \nonumber \\
	& \;\;=-\min_{0\leq \lambda \leq 1} \log\left( \E_{\bA} \bigg[ \int{p^{\lambda}(y|\ba,H_v)p^{1-\lambda}(y|\ba,H_w)\,dy}\bigg]\right).
\end{align*}
\normalsize
Here \eqref{eq:IC_EA} is obtained. By further working on \eqref{eq:IC_EA}, we have
\par\noindent\small
\begin{align}
	\eqref{eq:IC_EA} & =-\min_{\ba'} \min_{0\leq \lambda \leq 1} \log\bigg(  p(\ba')\bigg[ \int{p^{\lambda}(y|\ba',H_v)p^{1-\lambda}(y|\ba',H_w)\,dy} \bigg] +  \int_{\ba \neq \ba'} p(\ba) \bigg[ \underbrace{ \int{p^{\lambda}(y|\ba,H_v)p^{1-\lambda}(y|\ba,H_w)\,dy} }_{\leq 1 \;(\because \text{\;Holder's inequality})}\bigg] d\ba \bigg) \nonumber \\	
	& \geq -\min_{\ba'}\min_{0\leq \lambda \leq 1} \log\bigg(  p(\ba')\bigg[ \int{p^{\lambda}(y|\ba',H_v)p^{1-\lambda}(y|\ba',H_w)\,dy} \bigg]  +  \underbrace{\int_{\ba \neq \ba'} p(\ba) d\ba}_{=\;1-p(\ba')} \bigg)  \nonumber \\		
	& = -\min_{\ba'}\min_{0\leq \lambda \leq 1} \log\bigg(  1 - p(\ba') + p(\ba')\bigg[ \int{p^{\lambda}(y|\ba',H_v)p^{1-\lambda}(y|\ba',H_w)\,dy} \bigg]  \bigg)  \nonumber \\		
	& = -\min_{\ba'} \bigg[ \min_{0\leq \lambda \leq 1} \log\bigg(  1- p(\ba')+ p(\ba')\exp{ \bigg( \log \bigg[ \int{p^{\lambda}(y|\ba',H_v)p^{1-\lambda}(y|\ba',H_w)\,dy} \bigg]\bigg) \bigg) } \bigg]  \nonumber \\							
	& = -\min_{\ba'} \log\bigg(  1- p(\ba')+ p(\ba') e^{ - C( p_{Y|\bA',H_v},\; p_{Y|\bA',H_w})}\bigg), 			
\end{align}
\normalsize
where
\par\noindent\small
\begin{align}\label{eq:IC_C}
C( p_{Y|\bA',H_v},\; p_{Y|\bA',H_w}) = -\min_{0\leq \lambda \leq 1} \log \bigg[ \int{p^{\lambda}(y|\ba',H_v)p^{1-\lambda}(y|\ba',H_w)\,dy} \bigg],
\end{align}
\normalsize
and the first equation holds  for any realization vector $\ba'$ in the domain of $\ba$. We take the minimization over $\ba'$ in order to have the tightest lower bound of the inner conditional Chernoff information. Notice that due to the Holder's inequality, for any probability density functions $f(x)$ and $g(x)$, we have
\begin{align}\label{eq:holder_inequality}
\int f(x)^{\lambda}g(x)^{1-\lambda}dx \leq \bigg( \int f(x) dx \bigg)^{\lambda} \bigg( \int g(x) dx\bigg)^{1-\lambda} = 1.
\end{align}

In conclusion, we obtain
\par\noindent\small
\begin{align}
IC(p_{\bA,Y|H_v}, p_{\bA,Y|H_w}) \geq -\min_{\ba}\log\left(1-p_\bA(\ba)+p_\bA(\ba)e^{-C(p_{Y|\bA, H_v}, p_{Y|\bA, H_w})} \right),
\end{align}
\normalsize
where
$C(p_{Y|\bA, H_v}, p_{Y|\bA, H_w})$ is the well-known (regular) Chernoff information between $p_{Y|\bA, H_v}$ and $p_{Y|\bA, H_w}$ shown in \eqref{eq:IC_C}. As long as there exist sensing vectors $\bA$'s of a positive probability, such that the regular Chernoff information is positive, then the inner conditional Chernoff information $IC(p_{\bA,Y|H_v}, p_{\bA,Y|H_w})$ will also be positive.

Overall, the smallest possible error exponent between any pair of hypotheses is
\begin{align}\label{eq:E_time_varying}
E  = \min_{1\leq v,w \leq l, v\neq w} IC(p_{\bA,Y|H_v}, p_{\bA,Y|H_w}).
\end{align}
Without loss of generality, we assume $H_1$ is the true hypothesis. Since the error probability $\Pb_{err}$ in the Neyman-Pearson testing satisfies
\begin{align}
	\Pb_{err} \leq 2^{-m ( IC(p_{\bA,Y|H_v}, p_{\bA,Y|H_w}))} \leq 2^{-mE},
\end{align}
by the union bound over the $l-1$ possible pairs $(H_1, H_w)$, where $l=\binom{n}{k}$, the probability that $H_1$ is not correctly identified as the true hypothesis is upper-bounded by $l\times 2^{-mE}$ in terms of scaling. Hence, as shown in the proof of Theorem \ref{thm:fixed_time_invariant_measurements}, $m=\Theta(k \log(n)E^{-1})$, where $E$ is introduced in \eqref{eq:E_time_varying}, samples are enough for identifying the $k$ anomalous samples with high probability.  
\end{IEEEproof}

\subsection{Deterministic Time-Varying Measurements}
 \label{sec:deterministic_timevarying}
In this subsection, we consider mixed measurements which are varied over time. However, each sensing vector is predetermined. Hence, for exactly $p(\bA=\ba)m$ (assuming that $p(\bA=\ba)m$ are integers) measurements, a realized sensing vector $\ba$ is used. In contrast, in random time-varying measurements, each sensing vector $\bA$ is taken randomly, and thus the number of measurements taking realization $\ba$ is random. We define the predetermined sensing vector at time $j$ as $\ba^j$.

For deterministic time-varying measurements, we introduce the maximum likelihood estimate method among $l = \binom{n}{k}$ hypotheses in Algorithm \ref{alg:likelihoodratiotestdeterministic}. To analyze the error probability, we consider another hypothesis testing method based on pairwise comparison with deterministic time-varying measurements in Algorithm \ref{alg:timevaryingalg2deterministic}. Before introducing the sample complexity of hypothesis testing with deterministic time-varying measurements, we define the outer Chernoff information between two probability density functions given hypotheses and a sensing vector in Definition \ref{def:OC}.

\begin{algorithm}[t]
    \LinesNumbered
    \KwData{Observation data $\by = [y^1, y^2, ..., y^m]^T  \in \R^{m \times 1}$, deterministic sensing vectors $[{\ba^1}, {\ba^2}, ..., {\ba^m}]^T \in \R^{m \times n}$ }
    \KwResult{$k$ anomalous random variables}
    \nlset{1} For each hypothesis $H_v$ ($1\leq v\leq l$, $l= \binom{n}{k}$), calculate the likelihood $p_{Y|H_v,\bA}(y|H_v,\ba)$.\\
    \nlset{2} Choose the hypothesis with the maximum likelihood.\\
    \nlset{3} Declare the corresponding $k$ random variables as the anomalous random variables.\\
    \caption{Maximum likelihood estimation with deterministic time-varying measurements}
    \label{alg:likelihoodratiotestdeterministic}
\end{algorithm}

\begin{algorithm}[t]
    \LinesNumbered
    \KwData{observation data $\by = [y^1, y^2, ..., y^m]^T \in \R^{m \times 1}$, deterministic sensing vectors $[{\ba^1}, {\ba^2}, ..., {\ba^m}]^T \in \R^{m \times n}$}
    \KwResult{$k$ anomalous random variables}
    \nlset{1} For all pairs of hypotheses $H_v$ and $H_w$ ($1\leq v, w \leq l$, $l= \binom{n}{k}$, and $v\neq w$), perform Neyman-Pearson testing of the following two hypotheses:
    \begin{itemize}
        \item $Y^1$, $Y^2$, ..., $Y^m$ follow the probability distribution $p_{Y|H_v,\bA}(y|H_v,\ba)$
        \item $Y^1$, $Y^2$, ..., $Y^m$ follow probability distribution $p_{Y|H_w,\bA}(y|H_w,\ba)$
    \end{itemize}
    \nlset{2} Find a certain $w^*$, such that $H_{w^*}$ is the winning hypothesis, whenever it is involved in a pairwise hypothesis testing.\\
    \nlset{3} Declare the $k$ random variables producing $H_{w^*}$ as anomalous random variables.\\
    \nlset{4} If there is no such $w^*$ in Step 2, declare a failure in finding the $k$ anomalous random variables.
    \caption{Likelihood ratio test with deterministic time-varying measurements}
    \label{alg:timevaryingalg2deterministic}
\end{algorithm}

\begin{definition}[Outer Conditional Chernoff Information]\label{def:OC}
For $\lambda\in[0,1]$, two hypotheses $H_v$ and $H_w$ ($1\leq v ,w \leq l$), and a sensing vector $\bA$, define
\begin{align*}
p_{\lambda} (y| \ba, H_v, H_w) :=\frac{p^\lambda_{Y| \bA, H_v}(y|\ba,H_v)p^{1-\lambda}_{Y|\bA,H_w}(y|\ba,H_w)}{\int p^\lambda_{Y|\bA,H_v}(y|\ba,H_v)p^{1-\lambda}_{Y|\bA,H_w}(y|\ba,H_w)dy},
\end{align*}
\begin{align*}
Q_{\lambda, v\rightarrow w}:=\E_{\bA} \bigg[ D \bigg(     p_{\lambda} (y|\ba,H_v,H_w)~||~p (y|\ba,H_v)               \bigg) \bigg],
\end{align*}
\begin{align*}
Q_{\lambda, w\rightarrow v}:=\E_{\bA} \bigg[ D \bigg(     p_{\lambda} (y|\ba,H_v,H_w)~||~p (y|\ba,H_w)                \bigg) \bigg],
\end{align*}
where $D(P||Q) := \sum_{y \in \chi} P(y) \log \frac{P(y)}{Q(y)}$ is the relative entropy or Kullback-Leibler distance between two probability mass functions $P$ and $Q$. Then, the outer conditional Chernoff information $OC(p_{Y|\bA,H_v}, p_{Y|\bA,H_w})$ between $H_v$ and $H_w$, under deterministic time-varying sensing vector $\bA$, is defined as
\begin{align}\label{eq:minimum_error_exponent}
OC(p_{Y|\bA,H_v}, p_{Y|\bA,H_w}):=Q_{\lambda, v\rightarrow w}=Q_{\lambda, w\rightarrow v},
\end{align}
where $\lambda$ is chosen such that $Q_{\lambda, v\rightarrow w}=Q_{\lambda, w\rightarrow v}$.
\end{definition}

With this definition, the following theorem describes the sample complexity of our algorithms in deterministic time varying measurements.
\begin{theorem}\label{thm:numberofsamples_deterministic_timevarying}
Consider time-varying deterministic observations $Y^j = y^j$, $1\leq j \leq m$, for $n$ random variables $X_1$, $X_2$, ..., $X_n$. $l = \binom{n}{k}$ is the number of hypotheses for the distribution of the vector $\bX = [X_1,X_2,...,X_n]^T$. Then with $O \bigg(\frac{k \log(n)}{\min\limits_{1\leq v,w \leq l, v\neq w} OC(p_{Y|\bA,H_v}, p_{Y|\bA,H_w})} \bigg)$ random time-varying measurements, Algorithms \ref{alg:likelihoodratiotestdeterministic} and  \ref{alg:timevaryingalg2deterministic} correctly identify the $k$ anomalous random variables with high probability.
Here $l$ is the number of hypotheses, $p_{Y|\bA,H_v}(y|\ba,H_v)$, $1\leq v \leq l$ is the output probability distribution for observations $Y$'s under hypothesis $H_v$ and sensing vector $\bA$, and $OC(p_{Y|\bA,H_v}, p_{Y|\bA,H_w})$ is the outer conditional Chernoff information defined in Definition \ref{def:OC}. Moreover, the outer conditional Chernoff information is equivalent to
\par\noindent\small
\begin{align} \label{eq:outer_CI}
OC(p_{Y|\bA,H_v}, p_{Y|\bA,H_w}) &=-\min_{0\leq \lambda \leq 1} \int p_{\bA}(\ba)\log \left(\int{p_{Y|\bA,H_v}^{\lambda}(y|\ba,H_v)p_{Y|\bA,H_w}^{1-\lambda}(y|\ba,H_w)\,dy}\right) \,d\ba \nonumber\\
&=-\min_{0\leq \lambda \leq 1}  \E_{\bA} \left(\log \left(\int{p_{Y|\bA,H_v}^{\lambda}(y|\ba,H_v)p_{Y|\bA,H_w}^{1-\lambda}(y|\ba,H_w)\,dy}\right)\right).
\end{align}
\normalsize
\end{theorem}
For readability, we place the proof of  Theorem \ref{thm:numberofsamples_deterministic_timevarying}
 in Appendix \ref{apx:proof_deter_time_varying}. It is noteworthy that from the Jensen's inequality, the outer Chernoff information introduced in \eqref{eq:outer_CI} is greater than or equal to the inner Chernoff information introduced in \eqref{eq:IC_EA}.

\section{Compressed Hypothesis Testing in the Regime of Undersampling}
\label{sec:undersampling}
Compressed hypothesis testing is especially effective when the number of samples allowed is  in the subsampling regime, where the number of samples is small, sometimes even smaller than the number of random variables. Now we give a lower bound on the error probability of determining the set of anomalous random variables when the number of individual samples is smaller than the number of random variables.

\begin{theorem}\label{thm:lowerboundforseparateobservations}
Consider $n$ independent random variables, among which $(n-k)$ random variables follow a known probability distribution $f_1(\cdot)$; while the other $k$ random variables follow another known probability distribution $f_2(\cdot)$. Suppose that we take $m<n$ samples of individual random variables (no mixing). Then the probability of misidentifying the $k$ abnormal random variables is at least
$$\Pb_{err} \geq   1- \frac{\sum_{i=\max{(0, k+m-n)}}^{\min{(k, m)}} \binom{m}{i}  }{\binom{n}{k}}.$$
\end{theorem}

\begin{IEEEproof}
To lower bound the error probability, we assume that whenever a random variable is ever sampled, the decoder knows correctly what distribution that random variable follows. Consider the case where $m$ samples are observed. Suppose that $i$ (where $i\geq\max{(0, k+m-n)}$ and $i \leq \min{(k, m)}$)  random variables that follow the abnormal distribution $f_2(\cdot)$  are observed in these $m$ samples. Then there are at least
$(n-m)$ random variables are never sampled, and among them there are $(k-i)$ random variables that follow the abnormal distribution $f_2(\cdot)$. Determining correctly the  $(k-i)$ random variables that follow the abnormal distribution $f_2(\cdot)$ is at most of probability $\frac{1}{\binom{n-m}{k-i}}$.
So the probability of correctly identifying the $k$ abnormal random variables is at most

$$ \frac{ \sum_{i=\max{(0, k+m-n)}}^{\min{(k, m)}} \binom{m}{i}  \times \binom{n-m}{k-i} \times \frac{1}{\binom{n-m}{k-i}} }{\binom{n}{k}}.$$

This proves the lower bound on misidentifying the $k$ abnormal random variables in separate measurements, where $m < n$.
\end{IEEEproof}

As we can see, if $m>k$ and $m \ll n$, the error probability can be very close to $1$.  In contrast, compressed hypothesis testing can potentially greatly lower the error probability even with $m \ll n$ samples. In fact, following the proof of Theorem \ref{thm:numberofsamples_deterministic_timevarying}, we have the following results on the error probability for deterministic time-varying measurements.

\begin{theorem}\label{thm:errorprobability_undersampling_mixedmeasurements}
Consider time-varying deterministic observations $Y^j = y^j$, $1\leq j \leq m$, for $n$ random variables $X_1$, $X_2$, ..., $X_n$. $l = \binom{n}{k}$ is the number of hypotheses for the distribution of the vector $\bX = [X_1,X_2,...,X_n]^T$. Then with $m$ time-varying measurements, using Algorithms \ref{alg:likelihoodratiotestdeterministic} and  \ref{alg:timevaryingalg2deterministic}, the probability of incorrectly identifying the $k$ anomalous random variables is upper bounded by
$$  \binom{n}{k} e^{-m \;\times \; \big( \min\limits_{1\leq v,w \leq l, v\neq w} OC(p_{Y|\bA,H_v}, p_{Y|\bA,H_w}) \big) }.$$
Here $m$ is the number of measurements and can be smaller than $n$,  $l$ is the number of hypotheses, $p_{Y|\bA,H_v}(y|\ba,H_v)$, $1\leq v \leq l$ is the output probability distribution for observations $Y$'s under hypothesis $H_v$ and sensing vector $\bA$, and $OC(p_{Y|\bA,H_v}, p_{Y|\bA,H_w})$ is the outer conditional Chernoff information defined in Definition \ref{def:OC}. 
\normalsize
\end{theorem}

If the outer Chernoff information for the mixed measurements is sufficiently small, the error probability can indeed be made close to 0 even if the number of measurements is smaller than the problem dimension $n$. In contrast, according to Theorem \ref{thm:lowerboundforseparateobservations}, for small $m$, the error probability of separate observations is lower bounded by a number close to 1, and it is impossible at all for separate observations to achieve low error probability. We remark that, compared with earlier theorems which deal mostly with the error exponent (where the number of samples $m$ is large and goes to infinity),  Theorem \ref{thm:errorprobability_undersampling_mixedmeasurements} is for the undersampling regime where the number of samples is smaller than the number of variables.
Through various numerical experiments in Section \ref{sec:simulation}, we validate the observations above.


\section{Examples of Compressed Hypothesis Testing}
\label{sec:example}

In this section, we provide simple examples in which smaller error probability can be achieved in hypothesis testing through mixed observations than the traditional individual sampling approach, with the same number of measurements. Especially, we consider Gaussian probability distributions in our examples.

\subsection{Example 1: two Gaussian random variables}
\label{sec:example-exam1}
In this example, we consider $n=2$, and $k=1$. We group the two independent random variables $X_1$ and $X_2$ in a random vector $[X_1, X_2]^T$. Suppose that there are two hypotheses for a $2$-dimensional random vector $[X_1, X_2]^T$, where $X_1$ and $X_2$ are independent:\\[-10pt]
\begin{itemize}
     \item $H_1$: $X_1 \sim \cN(A, \sigma^2)$ and $X_2 \sim \cN(B, \sigma^2)$,
     \item $H_2$: $X_1 \sim \cN(B, \sigma^2)$ and $X_2 \sim \cN(A, \sigma^2)$. \\[-10pt]
\end{itemize}
Here $A$ and $B$ are two distinct constants, and $\sigma^2$ is the variance of the two Gaussian random variables. At each time index, only one observation is allowed, and the observation is restricted to a linear mixing of $X_1$ and $X_2$. Namely,
$$Y^j=a_1 X_1+a_2 X_2.$$
We assume that the sensing vector $[a_1, a_2]^T$ does not change over time. Clearly, when $a_1 \neq 0$ and $a_2=0$, the sensing vector reduces to a separate observation of $X_1$; and when $a_1 = 0$ and $a_2 \neq 0$, it reduces to a separate observation of $X_2$. In these cases, the observation follows distribution $\cN(A,\sigma^2)$ for one hypothesis, and follows distribution $\cN(B, \sigma^2)$ for the other hypothesis. The Chernoff information between these two distributions is
\par\noindent\small
\begin{align}\label{eq:CI_Gauss_case1}
	C(\cN(A, \sigma^2),\cN(B, \sigma^2))
	& =-\min_{0\leq \lambda \leq 1} \log \int{p_{\cN(A,\sigma^2)}^{\lambda}(x)p_{\cN(B,\sigma^2)}^{1-\lambda}(x)\,dx} \nonumber\\
    & =-\min_{0\leq \lambda \leq 1} \log \int{\bigg(\frac{1}{2\pi \sigma^2}\bigg)^{\lambda/2} e^{-\frac{\lambda (x-A)^2}{2\sigma^2}} \bigg(\frac{1}{2\pi \sigma^2}\bigg)^{(1-\lambda)/2} e^{-\frac{(1-\lambda) (x-B)^2}{2\sigma^2}}\,dx} \nonumber\\
    & =-\min_{0\leq \lambda \leq 1} \log \bigg(\frac{1}{2\pi \sigma^2}\bigg)^{1/2} \int{ e^{-\frac{\lambda (x-A)^2 + (1-\lambda)(x-B)^2}{2\sigma^2}} \,dx} \nonumber\\
    & =-\log \bigg(\frac{1}{2\pi \sigma^2}\bigg)^{1/2} -\min_{0\leq \lambda \leq 1} \log \int{ e^{-\frac{ (x - (A\lambda - B\lambda +B))^2}{2\sigma^2}}  e^{- \frac{\lambda A^2 + B^2 - \lambda B^2 - (A\lambda - B\lambda +B)^2 }{2\sigma^2}}  \,dx} \nonumber\\
    & =-\log \bigg(\frac{1}{2\pi \sigma^2}\bigg)^{1/2} -\min_{0\leq \lambda \leq 1} \bigg( - \frac{\lambda A^2 + B^2 - \lambda B^2 - (A\lambda - B\lambda +B)^2 }{2\sigma^2} +  \log \underbrace{ \int{ e^{-\frac{ (x - (A\lambda - B\lambda +B))^2}{2\sigma^2}} \,dx}}_{=\sqrt{2\pi \sigma^2}} \bigg) \nonumber\\
    & =-\log \bigg(\frac{1}{2\pi \sigma^2}\bigg)^{1/2} - \log\sqrt{2\pi \sigma^2} -\min_{0\leq \lambda \leq 1} - \frac{\lambda A^2 + B^2 - \lambda B^2 - (A\lambda - B\lambda +B)^2 }{2\sigma^2}  \nonumber\\
    & =\max_{0\leq \lambda \leq 1} \frac{\lambda A^2 + B^2 - \lambda B^2 - (A\lambda - B\lambda +B)^2 }{2\sigma^2} \nonumber \\
	&=\frac{(A-B)^2}{8\sigma^2}.  \quad\quad\quad\quad (\text{by plugging the optimal $\lambda$, i.e., 1/2})
\end{align}
\normalsize

For the mixed measurements, when the hypothesis $H_1$ holds, the observation $Y^j$ follows Gaussian distribution $\cN(a_1 A+a_2 B, (a_1^2+a_2^2)\sigma^2 )$. Similarly, when the hypothesis $H_2$ holds, the observation $Y^j$ follows Gaussian distribution $\cN(a_1 B+a_2 A, (a_1^2+a_2^2)\sigma^2 )$.  The Chernoff information between these two Gaussian distributions $\cN(a_1 A+a_2 B, (a_1^2+a_2^2)\sigma^2 )$ and $\cN(a_1 B+a_2 A, (a_1^2+a_2^2)\sigma^2 )$ is given by
\begin{align}\label{eq:mix_Chrenoff_ex}
\frac{[(a_1 A+a_2 B)-(a_1 B+a_2 A)]^2}{8(a_1^2+a_2^2)\sigma^2} &=\frac{[(a_1-a_2)^2(A-B)^2]}{8(a_1^2+a_2^2)\sigma^2} = \frac{2(A-B)^2}{8\sigma^2},
\end{align}
where the last equality is obtained by taking the measurement vector $[a_1, a_2]^T = [a_1, -a_1]^T$. Therefore, with mixed measurements, we can double the Chernoff information. This shows that linear mixed observations can offer strict improvement in terms of reducing the error probability in hypothesis testing by increasing the error exponent.


\subsection{Example 2: Gaussian random variables with different means}
In this example, we consider the mixed observations for two hypotheses of Gaussian random vectors. In general, suppose that there are two hypotheses for an $n$-dimensional random vector $[X_1, X_2,...,X_{n}]^T$,\\[-5pt]
\begin{itemize}
     \item $H_1$: $[X_1, X_2,...,X_{n}]$ follow jointly Gaussian distributions $\cN(\bmu_1, \bSigma_1)$,
     \item $H_2$: $[X_1, X_2,...,X_{n}]$ follow jointly Gaussian distributions $\cN(\bmu_2, \bSigma_2)$.\\[-5pt]
\end{itemize}
Here $\bSigma_1$ and $\bSigma_2$ are both $n \times n$ covariance matrices.

Suppose at each time instant, only one observation is allowed, and the observation is restricted to a time-invariant sensing vector $\bA \in \R^{n \times 1}$; namely
\begin{align*}
	Y^j= \langle \bA, \bX^j \rangle,
\end{align*}
Under these conditions, the observation follows distribution $\cN(\bA^T \bmu_1, \bA^T \bSigma_1 \bA)$ for hypothesis $H_1$, and follows distribution $\cN(\bA^T \bmu_2, \bA^T \bSigma_2 \bA)$ for the other hypothesis $H_2$. We would like to choose a sensing vector $\bA$ which maximizes the Chernoff information between the two possible univariate Gaussian distributions, namely
$$\max_{\bA} \;C(\cN(\bA^T \bmu_1, \bA^T \bSigma_1 \bA),\cN(\bA^T \bmu_2, \bA^T \bSigma_2 \bA)).$$

In fact, from \cite{ChernoffInformation}, the Chernoff information between these two distributions is
\begin{align*}
& C(\cN(\bA^T \bmu_1, \bA^T \bSigma_1 \bA),\cN(\bA^T \bmu_2, \bA^T \bSigma_2 \bA) ) \\
& = \max_{0\leq \lambda \leq 1} \left[ \frac{1}{2} \log{\left( \frac{\bA^T (\lambda \bSigma_1+(1-\lambda) \bSigma_2) \bA}{(\bA^T \bSigma_1 \bA)^{\lambda} (\bA^T \bSigma_2 \bA)^{1-\lambda}} \right)} +\frac{\lambda (1-\lambda) (\bA^T (\bmu_1-\bmu_2))^2}{2 \bA^T (\lambda \bSigma_1+(1-\lambda) \bSigma_2) \bA} \right].
\end{align*}

We first look at a special case when $\bSigma=\bSigma_1=\bSigma_2$. Under this condition, the maximum Chernoff information is given by
$$\max_{\bA}\max_{0\leq \lambda \leq 1} \frac{\lambda(1-\lambda) [\bA^T (\bmu_1-\bmu_2)]^2}{2\bA^T \bSigma \bA}.$$
Taking $\bA'=\bSigma^{\frac{1}{2}} \bA$, then this reduces to
$$\max_{\bA'}\max_{0\leq \lambda \leq 1} \frac{\lambda(1-\lambda) [(\bA')^T \bSigma^{-\frac{1}{2}}(\bmu_1-\bmu_2)]^2}{2(\bA')^T \bA'}.$$
From Cauchy-Schwarz inequality, it is easy to see that the optimal $\lambda=\frac{1}{2}$, $\bA'=\bSigma^{-\frac{1}{2}}(\bmu_1-\bmu_2)$, and $\bA=\bSigma^{-1}(\bmu_1-\bmu_2)$. Under these conditions, the maximum Chernoff information is given by
\begin{align}
	\frac{1}{8} (\bmu_1-\bmu_2)^T \bSigma^{-1} (\bmu_1-\bmu_2).
\end{align}
Note that in general, $\bA'=\bSigma^{-\frac{1}{2}}(\bmu_1-\bmu_2)$ is not a separate observation of a certain individual random variable, but rather a linear mixing of the $n$ random variables. Therefore, a mixed measurement can maximize the Chernoff information.

\subsection{Example 3: Gaussian random variables with different variances}
\label{subsec:Example3}
Let's look at mixed observations for Gaussian random variables with different variances. Consider the same setting in Example 2, except that we now look at another special case when $\bmu=\bmu_1=\bmu_2$. We will study the optimal sensing vector under this scenario. Then the Chernoff information becomes
\begin{align}
& C(\cN(\bA^T \bmu, \bA^T \bSigma_1 \bA),\cN(\bA^T \bmu, \bA^T \bSigma_2 \bA) ) \nonumber\\
&= \max_{0\leq \lambda \leq 1} \frac{1}{2} \log{ \left( \frac{ \bA^T (\lambda \bSigma_1+(1-\lambda) \bSigma_2) \bA }{ (\bA^T \bSigma_1 \bA)^{\lambda} (\bA^T \bSigma_2 \bA)^{1-\lambda}} \right)  }.
\end{align}

To find the optimal sensing vector $\bA$, we are solving this optimization problem
\begin{align}
	\max_{\bA}\max_{0\leq \lambda \leq 1} \frac{1}{2} \log{\left( \frac{\bA^T (\lambda \bSigma_1+(1-\lambda) \bSigma_2) \bA}{(\bA^T \bSigma_1 \bA)^{\lambda} (\bA^T \bSigma_2 \bA)^{1-\lambda}} \right)}.
\end{align}
For a certain $\bA$, we define
\begin{align*}
	B :=\frac{\max{(\bA^T \bSigma_1 \bA, \bA^T \bSigma_2 \bA) }}{\min{(\bA^T \bSigma_1 \bA, \bA^T \bSigma_2 \bA)  }}.
\end{align*}
Note that $B \geq 1$. By symmetry over $\lambda$ and $1-\lambda$, maximizing the Chernoff information can always be reduced to
\begin{align}
\label{eq:findingOptA}
\max_{B\geq 1} \max_{0\leq \lambda \leq 1} \frac{1}{2} \log{\left( \frac{\lambda + (1-\lambda) B}{B^{1-\lambda}} \right)}.
\end{align}
The optimal $\lambda$, denoted by $\lambda^{\star}$, is obtained by finding the point which makes the first order differential equation to zero as follows:
\begin{align*}
    \lambda^{\star}=\frac{-(B-1)+B\log(B)}{(B-1) \log(B)}.
\end{align*}
By plugging $\lambda^{\star}$ to (\ref{eq:findingOptA}), we obtain the following optimization problem:
\begin{align}
\label{eq:OptAObjFunc}
\max_{B \geq 1} \frac{1}{2} \left \{ -1 + \frac{B}{B-1} \log(B) + \log \bigg( \frac{B-1}{B \log(B)}\bigg)   \right\}.
\end{align}
We note that the objective function is an increasing function in $B$, when $B \geq 1$, which is proven in Lemma \ref{lemma:monotonic}.

\begin{lemma}\label{lemma:monotonic}
The optimal objective value of the following optimization problem
$$\max_{0\leq \lambda \leq 1} \frac{1}{2} \log{\left( \frac{\lambda + (1-\lambda) B}{B^{1-\lambda}} \right)},$$
is an increasing function in $B \geq 1$.
\end{lemma}

\begin{IEEEproof}
 We only need to show that for any $\lambda \in [0,1]$, $\left( \frac{\lambda + (1-\lambda) B}{B^{1-\lambda}} \right)$ is an increasing function in $B\geq 1$.
In fact, the derivative of it with respect to $\lambda$ is
$$\lambda(1-\lambda) (B^{\lambda-1}-B^{\lambda-2})\geq 0.$$
Then the conclusion of this lemma immediately follows.
\end{IEEEproof}

This means we need to maximize $B$, in order to maximize the Chernoff information. Hence, to find the optimal $\bA$ maximizing $B$, we solve the following two optimization problems:
\begin{equation}
       \max_{\bA}{\bA^T \bSigma_1 \bA } ~~~~~\text{subject~to}~~\bA^T \bSigma_2 \bA \leq 1,
\label{eq:twoquadraticprogram1}
\end{equation}
and
\begin{equation}
       \max_{\bA}{\bA^T \bSigma_2 \bA } ~~~~~\text{subject~to}~~\bA^T \bSigma_1 \bA \leq 1.
\label{eq:twoquadraticprogram2}
\end{equation}
Then the maximum of the two optimal objective values is equal to the optimal objective value of optimizing $B$, and the corresponding $\bA$ is the optimal sensing vector maximizing the Chernoff information. These two optimization problems are not convex optimization programs. However, they still hold zero duality gap from the S-procedure \cite[Appendix B]{boyd}. In fact, they are respectively equivalent to the following two semidefinite programming optimization problems:
\begin{equation}
\begin{aligned}
& \underset{\gamma, \;\lambda}{\text{minimize}}
& & -\gamma \\
& \text{subject to}
& & \lambda \geq 0 \\
&&& {\left( \begin{array}{cc}
-\bSigma_1+\lambda\bSigma_2 & 0\\
0 & -\lambda-\gamma  \end{array} \right) \succeq 0},
\end{aligned}
\label{eq:sdpfor2quadratic1}
\end{equation}
and
\begin{equation}
\begin{aligned}
& \underset{\gamma,\; \lambda}{\text{minimize}}
& & -\gamma \\
& \text{subject to}
& & \lambda \geq 0 \\
&&& {\left( \begin{array}{cc}
-\bSigma_2+\lambda\bSigma_1 & 0\\
0 & -\lambda-\gamma  \end{array} \right) \succeq 0}.
\end{aligned}
\label{eq:sdpfor2quadratic2}
\end{equation}
Thus, they can be efficiently solved via a generic optimization solver.

For example, when $\bSigma_1$ and $\bSigma_2$ are given as follows:
\begin{align*}
    \bSigma_1 = \begin{bmatrix} 1&0&0\\0&1&0\\0&0&100 \end{bmatrix},\quad\quad \bSigma_2 = \begin{bmatrix} 100&0&0\\0&1&0\\0&0&1 \end{bmatrix},
\end{align*}
we obtain $[0,0,-1]^T$ for the optimal sensing vector $\bA$. This is because the diagonal elements in the covariance matrices $\bSigma_1$ and $\bSigma_2$ represent the variance of random variables $[X_1,X_2,X_3]^T$, and the biggest difference in variance is shown in random variable $X_1$ or $X_3$. Therefore, checking the random variable $X_3$ (or $X_1$) through realizations is an optimal way to figure out whether the random variables follows $\bSigma_1$ or $\bSigma_2$. On the other hand, if the random variables are dependent, and $\bSigma_1$ and $\bSigma_2$ are given as follows:
\begin{align*}
    \bSigma_1 = \begin{bmatrix} 1&0.5&0.5\\0.5&1&0.5\\0.5&0.5&100 \end{bmatrix},\quad\quad \bSigma_2 = \begin{bmatrix} 100&0.5&0.5\\0.5&1&0.5\\0.5&0.5&1 \end{bmatrix},
\end{align*}
then, we obtain $[0.4463,0.0022,-0.8949]^T$ for the optimal sensing vector $\bA$ which maximizes the Chernoff information. Namely, mixed measurements can maximize the Chernoff information, which can achieve the best error exponent.

\subsection{Example 4: $k=1$ anomalous random variable among $n=7$ random variables }

Let us introduce a specific example to help understanding before considering more general examples on the compressed hypothesis testing. In this example, we have six random variables following the distribution $\cN (0, 1)$, and the other random variable following the distribution $\cN (0, \sigma^2)$, where $\sigma^2 > 1$. We assume that all random variables $X_1$, $X_2$,..., and $X_7$ are independent. So overall, there are seven hypotheses:\\[-5pt]
\begin{itemize}
\item $H_1$: ($X_1$, $X_2$, ..., $X_7$) $\sim$ ($\cN(0, \sigma^2)$, $\cN(0, 1)$, ..., $\cN(0, 1)$),
\item $H_2$: ($X_1$, $X_2$, ..., $X_7$) $\sim$ ($\cN(0, 1)$, $\cN(0, \sigma^2)$, ..., $\cN(0, 1)$),
\item ......
\item $H_7$: ($X_1$, $X_2$, ..., $X_7$) $\sim$ ($\cN(0, 1)$, $\cN(0, 1)$, ..., $\cN(0, \sigma^2)$).\\[-5pt]
\end{itemize}
In this example, we will show that the Chernoff information with separate measurements is smaller than the Chernoff information with mixed measurements. We first calculate the Chernoff information between any two hypotheses with separate measurements. In separate measurements, for a hypothesis $H_v$, the probability distribution for the output is $\cN(0, \sigma^2)$ only when the random variable $X_v$ is observed. Otherwise, the output distribution follows $\cN(0, 1)$. Then, for any pair of hypotheses $H_v$ and $H_w$, when the random variable $X_v$ is observed, the probability distributions for the output are $\cN(0, \sigma^2)$ and $\cN(0, 1)$ respectively. Similarly, for hypotheses $H_v$ and $H_w$, when the random variable $X_w$ is observed, the probability distributions for the output are $\cN(0, 1)$ and $\cN(0, \sigma^2)$ respectively. For the separate measurements, seven sensing vectors from $\ba_1^T$ to $\ba_7^T$ for seven hypotheses are predetermined as follows:
\begin{align*}
    \begin{bmatrix} {\ba^1}^T \\ {\ba^2}^T \\{\ba^3}^T \\{\ba^4}^T \\{\ba^5}^T \\{\ba^6}^T \\{\ba^7}^T\end{bmatrix} = \begin{bmatrix}
        1 & 0 & 0 & 0 & 0 & 0 & 0 \\
        0 & 1 & 0 & 0 & 0 & 0 & 0 \\
        0 & 0 & 1 & 0 & 0 & 0 & 0 \\
        0 & 0 & 0 & 1 & 0 & 0 & 0 \\
        0 & 0 & 0 & 0 & 1 & 0 & 0 \\
        0 & 0 & 0 & 0 & 0 & 1 & 0 \\
        0 & 0 & 0 & 0 & 0 & 0 & 1
      \end{bmatrix}.
\end{align*}
And we take deterministic time-varying measurements using these seven sensing vectors.

The Chernoff information between any two hypotheses, e.g., $H_v = H_1$ and $H_w = H_2$, with separate measurements is calculated as follows:
\small
\begin{align}
&OC(p_{Y|\bA,H_v}, p_{Y|\bA, H_w}) \nonumber \\
&=-\min_{0\leq \lambda \leq 1}  \E_{\bA} \left( \log \left(\int{p_{Y|\bA,H_v}^{\lambda}(y|\ba,H_v)p_{Y|\bA,H_w}^{1-\lambda}(y|\ba, H_w)\,dy}\right) \right)\nonumber \\
&=-\min_{0\leq \lambda \leq 1}  \bigg[\frac{1}{7} \bigg(  \underbrace{  \log \left(\int{p_{\mathcal{N}(0,\sigma^2)}^{\lambda}(y)p_{\mathcal{N}(0,1)}^{1-\lambda}(y)\,dy}\right) }_{ \text{from measuremet vector}\;\ba_1^T}  +  \underbrace{ \log \left(\int{p_{\mathcal{N}(0,1)}^{\lambda}(y)p_{\mathcal{N}(0,\sigma^2)}^{1-\lambda}(y)\,dy}\right) }_{\text{from measuremet vector}\;\ba_2^T} +  \underbrace{  \cancelto{0}{ \log \left( \int{p_{\mathcal{N}(0,1)}^{\lambda}(y)p_{\mathcal{N}(0,1)}^{1-\lambda}(y)\,dy }\right) } }_{\text{from measuremet vectors}\;\ba_3^T} \nonumber \\
    & \quad\quad\quad\quad\quad +  \underbrace{ \cancelto{0}{  \log \left(\int{p_{\mathcal{N}(0,1)}^{\lambda}(y)p_{\mathcal{N}(0,1)}^{1-\lambda}(y)\,dy}  \right) } }_{\text{from measuremet vectors}\;\ba_4^T} + ... +  \underbrace{  \cancelto{0}{ \log \left(\int{p_{\mathcal{N}(0,1)}^{\lambda}(y)p_{\mathcal{N}(0,1)}^{1-\lambda}(y)\,dy}\right) } }_{\text{from measuremet vectors}\;\ba_7^T} \bigg)\bigg] \nonumber \\
&=-\min_{0\leq \lambda \leq 1}  \left [\frac{1}{7} \left( \log \left(\int{p_{\cN(0,1)}^{\lambda}(y)p_{\cN(0,\sigma^2)}^{1-\lambda}(y)\,dy}\right) \right) +\frac{1}{7} \left( \log \left(\int{p_{\cN(0,\sigma^2)}^{\lambda}(y)p_{\cN(0,1)}^{1-\lambda}(y)\,dy}\right) \right)\right]  \nonumber
\end{align}
\begin{align}\label{eq:differentvariancesSeparateObservations}
& =-\frac{1}{7} \min_{0\leq \lambda \leq 1} \bigg[ \log \int{\bigg(\frac{1}{2\pi \sigma^2}\bigg)^{(1-\lambda)/2} e^{-\frac{(1-\lambda) x^2}{2\sigma^2}} \bigg(\frac{1}{2\pi}\bigg)^{\lambda/2} e^{-\frac{\lambda x^2}{2}}\,dx} + \log \int{\bigg(\frac{1}{2\pi \sigma^2}\bigg)^{\lambda/2} e^{-\frac{\lambda x^2}{2\sigma^2}} \bigg(\frac{1}{2\pi}\bigg)^{(1-\lambda)/2} e^{-\frac{(1-\lambda) x^2}{2}}\,dx} \bigg] \nonumber\\
    & =-\frac{1}{7} \min_{0\leq \lambda \leq 1} \bigg[ \log{ \bigg(\frac{1}{2\pi \sigma^2}\bigg)^{(1-\lambda)/2} \bigg(\frac{1}{2\pi}\bigg)^{\lambda/2} + \log \bigg(\frac{1}{2\pi \sigma^2}\bigg)^{\lambda/2} \bigg(\frac{1}{2\pi}\bigg)^{(1-\lambda)/2}  + \log \int e^{-\frac{(1-\lambda) x^2}{2\sigma^2}}  e^{-\frac{\lambda x^2}{2}}\,dx} + \log \int{ e^{-\frac{\lambda x^2}{2\sigma^2}} e^{-\frac{(1-\lambda) x^2}{2}}\,dx} \bigg] \nonumber\\
    & =-\frac{1}{7} \min_{0\leq \lambda \leq 1} \bigg[ \log{ \bigg(\frac{1}{2\pi \sigma^2}\bigg)^{(1-\lambda)/2} \bigg(\frac{1}{2\pi}\bigg)^{\lambda/2} + \log \bigg(\frac{1}{2\pi \sigma^2}\bigg)^{\lambda/2} \bigg(\frac{1}{2\pi}\bigg)^{(1-\lambda)/2}  + \log \int e^{-\frac{ ((1-\lambda) + \lambda \sigma^2) x^2}{2\sigma^2}} \,dx} + \log \int{ e^{-\frac{ (\lambda + (1-\lambda)\sigma^2) x^2}{2\sigma^2}} \,dx} \bigg] \nonumber\\
    & =-\frac{1}{7} \min_{0\leq \lambda \leq 1} \bigg[ \log \bigg(\frac{1}{2\pi \sigma^2}\bigg)^{(1-\lambda)/2} \bigg(\frac{1}{2\pi}\bigg)^{\lambda/2} + \log \bigg(\frac{1}{2\pi \sigma^2}\bigg)^{\lambda/2} \bigg(\frac{1}{2\pi}\bigg)^{(1-\lambda)/2}  + \log  \frac{ \sqrt{2\pi \sigma^2} }{ \sqrt{ (1-\lambda) + \lambda \sigma^2 }} + \log  \frac{\sqrt{ 2\pi \sigma^2}}{ \sqrt{ \lambda + (1-\lambda)\sigma^2 }} \bigg]  \nonumber\\
    & =-\frac{1}{7} \min_{0\leq \lambda \leq 1} \bigg[ - \frac{\lambda}{2} \log 2\pi -\frac{1-\lambda}{2} \log 2\pi -\frac{1}{2} \log (1-\lambda + \lambda \sigma^2) + \frac{1}{2} \log 2\pi \sigma^2 -\frac{1}{2} \log ( \lambda + (1-\lambda)\sigma^2) \bigg] \nonumber\\
    & =-\frac{1}{7} \min_{0\leq \lambda \leq 1}  \bigg[ -\frac{1}{2} \log (1-\lambda + \lambda \sigma^2) + \log \sigma -\frac{1}{2} \log ( \lambda + (1-\lambda)\sigma^2) \bigg]  \nonumber\\
    & =-\frac{1}{7} \bigg[  \log \sigma + \min_{0\leq \lambda \leq 1}  -\frac{1}{2}  \log \bigg( (1-\lambda + \lambda \sigma^2)( \lambda + (1-\lambda)\sigma^2) \bigg) \bigg]\nonumber\\
    & =-\frac{1}{7} \bigg[  \log \sigma + \min_{0\leq \lambda \leq 1}  -\frac{1}{2}  \log \bigg( -(\sigma^2 -1)^2 (\lambda - \frac{1}{2})^2 + \frac{(\sigma^2 - 1)^2}{4} + \sigma^2 \bigg) \bigg]    \quad\quad\quad (\text{By applying optimal $\lambda = \frac{1}{2}$}) \nonumber\\
    & = \frac{1}{7} \log \frac{1 + \sigma^2}{2\sigma}
\end{align}
\normalsize

Let us then calculate the Chernoff information between hypotheses $H_v$ and $H_w$ with mixed measurements. For the mixed measurements, we consider using the parity check matrix of $(7,4)$ Hamming codes as follows:
\begin{align*}
    \begin{bmatrix} {\ba^1}^T \\ {\ba^2}^T \\{\ba^3}^T \end{bmatrix} = \begin{bmatrix}
        1 & 0 & 0 & 1 & 1 & 0 & 1 \\
        0 & 1 & 0 & 1 & 0 & 1 & 1 \\
        0 & 0 & 1 & 0 & 1 & 1 & 1
      \end{bmatrix}.
\end{align*}
We use each row vector of the parity check matrix as a sensing vector for a deterministic time-varying measurement. For example, the $i$-th row vector is used for the $j$-th mixed measurement, where $i=(j\; \text{mod}\; 3) +1$. Thus, we use total three sensing vectors for mixed measurements repeatedly.

For the pair of hypotheses $H_v$ and $H_w$, we can have total $21 (= \binom{7}{2}) $ cases in the calculation of the outer Chernoff information. For that, we have the following lemma:
\begin{lemma}\label{lemma:OCI_minimum}
Given mixed measurements in the parity check Hamming code matrix, for a pair of hypotheses, the outer Chernoff information between $H_1$ and $H_4$ (or $H_4$ and $H_7$) is the minimum one out of $21 (= \binom{7}{2})$ combination cases.
\end{lemma}
\begin{IEEEproof}
From the definition of the outer Chernoff information introduced in \eqref{eq:outer_CI}, we can calculate the outer Chernoff information as follows:
\begin{align*}
OC(p_{Y|\bA,H_v}, p_{Y|\bA, H_w})
&=-\min_{0\leq \lambda \leq 1}  \E_{\bA} \left( \log \left(\int{p_{Y|\bA,H_v}^{\lambda}(y|\ba,H_v)p_{Y|\bA,H_w}^{1-\lambda}(y|\ba,H_w)\,dy}\right) \right).
\end{align*}
For example, for a pair of hypotheses $H_1$ and $H_2$, we have
\begin{align*}
OC(p_{Y|\bA,H_v}, p_{Y|\bA, H_w})
&=-\min_{0\leq \lambda \leq 1}  \bigg[\frac{1}{3} \bigg(  \underbrace{  \log \left(\int{p_{\mathcal{N}(0,\sigma^2+3)}(y|\ba_1,H_1)^{\lambda}(y)p_{\mathcal{N}(0,4)}(y|\ba_1,H_2)^{1-\lambda}(y)\,dy}\right) }_{ \text{from measuremet vector}\;\ba_1^T}  \\
    & \quad\quad\quad\quad\quad\quad +  \underbrace{ \log \left(\int{p_{\mathcal{N}(0,4)}(y|\ba_2,H_1)^{\lambda}(y)p_{\mathcal{N}(0,\sigma^2+3)}(y|\ba_2,H_2)^{1-\lambda}(y)\,dy}\right) }_{\text{from measuremet vector}\;\ba_2^T} \\
     & \quad\quad\quad\quad\quad\quad +  \underbrace{ \cancelto{0}{\log \left(\int{p_{\mathcal{N}(0,4)}(y|\ba_3,H_1)^{\lambda}(y)p_{\mathcal{N}(0,4)}(y|\ba_3,H_2)^{1-\lambda}(y)\,dy}\right)} }_{\text{from measuremet vector}\;\ba_3^T}  \bigg)\bigg].
\end{align*}
For another pair of hypotheses $H_1$ and $H_4$, we have
\begin{align*}
OC(p_{Y|\bA,H_v}, p_{Y|\bA, H_w})
    &=-\min_{0\leq \lambda \leq 1}  \bigg[\frac{1}{3} \bigg(  \underbrace{ \cancelto{0}{ \log \left(\int{p_{\mathcal{N}(0,\sigma^2+3)}(y|\ba_1,H_1)^{\lambda}(y)p_{\mathcal{N}(0,\sigma^2 + 3)}(y|\ba_1,H_4)^{1-\lambda}(y)\,dy}\right) } }_{ \text{from measuremet vector}\;\ba_1^T}  \nonumber \\
    & \quad\quad\quad\quad\quad\quad +  \underbrace{ \log \left(\int{p_{\mathcal{N}(0,4)}(y|\ba_2,H_1)^{\lambda}(y)p_{\mathcal{N}(0,\sigma^2+3)}(y|\ba_2,H_4)^{1-\lambda}(y)\,dy}\right) }_{\text{from measuremet vector}\;\ba_2^T}  \nonumber\\
     & \quad\quad\quad\quad\quad\quad +  \underbrace{ \cancelto{0}{\log \left(\int{p_{\mathcal{N}(0,4)}(y|\ba_3,H_1)^{\lambda}(y)p_{\mathcal{N}(0,4)}(y|\ba_3,H_4)^{1-\lambda}(y)\,dy}\right)} }_{\text{from measuremet vector}\;\ba_3^T}  \bigg)\bigg].
\end{align*}
Among 21 cases, the outer Chernoff information between $H_1$ and $H_4$ (or $H_4$ and $H_7$) has only one remaining term in the calculation. Then, to complete the proof of this lemma, let us introduce the following lemma:
\begin{lemma}\label{lemma:holder_inequality}
For any probability density functions $f_i(x)$, $i=1,2,...,m$, where $m$ is an even positive number, the following inequality holds:
\begin{align}
	& \underset{0 \leq \lambda \leq 1}{\text{minimize}}\;  \bigg[ \log \int f_1^{\lambda}(x) f_2^{1-\lambda}(x) dx + \log \int f_3^{\lambda}(x) f_4^{1-\lambda}(x) dx  + ... + \log \int f_{m-1}^{\lambda}(x) f_m^{1-\lambda}(x) dx \bigg] \label{OCI}\\
	& \leq 	\underset{0 \leq \lambda \leq 1}{\text{minimize}}\; \log \int f_{2j-1}^{\lambda}(x) f_{2j}^{1-\lambda}(x) dx, \nonumber
\end{align}
where $j$ can be any number between 1 and $m/2$.
\end{lemma}
\begin{IEEEproof}
The lemma can be proven by showing that each term of the objective function in \eqref{OCI} is non-positive over $0 \leq \lambda \leq 1$. From the Holder's inequality shown in \eqref{eq:holder_inequality}, $\int f_{2j-1}^{\lambda}(x) f_{2j}^{1- \lambda} dx \leq 1$, $j=1,2,...,m/2$. By taking logarithm, each term of the objective function in \eqref{OCI} becomes non-positive. Therefore, minimizing the sum of all non-positive terms provides smaller or equal optimal value than that of minimizing only one term.
\end{IEEEproof}
From Lemma \ref{lemma:holder_inequality}, we can conclude that the outer Chernoff information between $H_1$ and $H_4$ (or $H_4$ and $H_7$), which has only one remaining term, is the minimum one out of 21 cases.
\end{IEEEproof}

Then, let us calculate the outer Chernoff information between $H_v = H_1$ and $H_w = H_4$ as follows:
\begin{align}\label{eq:differentvariancesHamming}
OC(p_{Y|\bA,H_v}, p_{Y|\bA, H_w})
&=-\min_{0\leq \lambda \leq 1}  \E_{\bA} \left( \log \left(\int{p_{Y|\bA,H_v}^{\lambda}(y|\ba,H_v)p_{Y|\bA,H_w}^{1-\lambda}(y|\ba,H_w)\,dy}\right) \right) \nonumber \\
    &=-\min_{0\leq \lambda \leq 1}  \bigg[\frac{1}{3} \bigg(  \underbrace{ \cancelto{0}{ \log \left(\int{p_{\mathcal{N}(0,\sigma^2+3)}(y|\ba_1,H_1)^{\lambda}(y)p_{\mathcal{N}(0,\sigma^2 + 3)}(y|\ba_1,H_4)^{1-\lambda}(y)\,dy}\right) } }_{ \text{from measuremet vector}\;\ba_1^T}  \nonumber \\
    & \quad\quad\quad\quad\quad\quad +  \underbrace{ \log \left(\int{p_{\mathcal{N}(0,4)}(y|\ba_2,H_1)^{\lambda}(y)p_{\mathcal{N}(0,\sigma^2+3)}(y|\ba_2,H_4)^{1-\lambda}(y)\,dy}\right) }_{\text{from measuremet vector}\;\ba_2^T}  \nonumber\\
     & \quad\quad\quad\quad\quad\quad +  \underbrace{ \cancelto{0}{\log \left(\int{p_{\mathcal{N}(0,4)}(y|\ba_3,H_1)^{\lambda}(y)p_{\mathcal{N}(0,4)}(y|\ba_3,H_4)^{1-\lambda}(y)\,dy}\right)} }_{\text{from measuremet vector}\;\ba_3^T}  \bigg)\bigg]  \nonumber\\
     &=-\min_{0\leq \lambda \leq 1}  \frac{1}{3} \log \left(\int{p_{\mathcal{N}(0,4)}(y|\ba_2,H_1)^{\lambda}(y)p_{\mathcal{N}(0,\sigma^2+3)}(y|\ba_2,H_4)^{1-\lambda}(y)\,dy}\right)   \nonumber\\
& = \frac{1}{6} \left( -1 + \frac{B}{B-1} \log(B) + \log \bigg( \frac{B-1}{B \log(B)}\bigg)   \right),
\end{align}
where $B=\frac{\max{(\sigma^2+3,\; 4)}}{\min{(\sigma^2+3,\; 4)}}.$ The final inequality is obtained from the result in Section \ref{subsec:Example3}, where $\bA$ in Example 3 is $\ba_2$ here, $\bmu_1$ and $\bmu_2$ are zero vectors here, and $\bSigma_1$ and $\bSigma_2$ in Example 3 are $\diag(\sigma^2+3,1,1,...,1)$ and $\diag(1,1,1,\sigma^2+3,1,...,1)$ respectively. Here, $\diag(\sigma^2+3,1,1,...,1)$ represents a diagonal matrix having elements $\sigma^2+3$, $1$, ..., $1$ in diagonal.

\begin{figure}
\begin{center}
\includegraphics[width=0.60\textwidth]{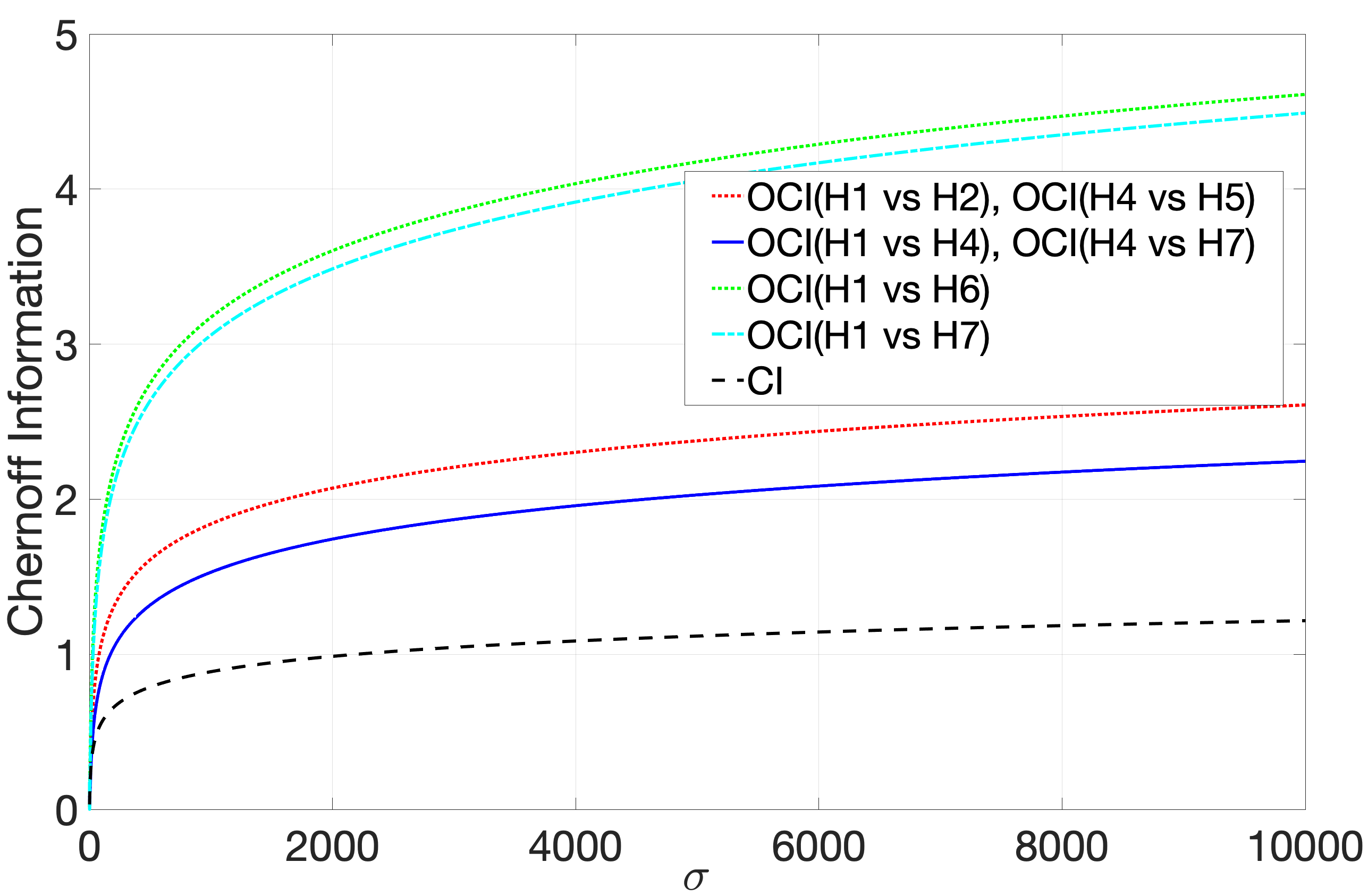}
\end{center}
\caption{Comparison of the Chernoff information with mixed measurements against the Chernoff information with separate measurements by varying the variance in Example 4. OCI represents the outer Chernoff information, i.e., Chernoff information with mixed measurements, and CI represents the Chernoff information with separate measurements. }
\label{fig:OCI_comparison_different_variance}
\end{figure}

When $\sigma^2 \gg 1$, for separate observations, we have
\begin{align*}
    OC(p_{Y|\bA,H_v}, p_{Y|\bA, H_w}) = \frac{1}{7} \log \bigg(\frac{\sigma^2+1}{2\sigma} \bigg) \approx \frac{1}{7} \log \bigg(\frac{\sigma}{2} \bigg),
\end{align*}
while for mixed measurements through the parity-check matrix of Hamming codes, $OC(p_{Y|\bA,H_v}, p_{Y|\bA, H_w})$ asymptotically satisfies
\begin{align*}
    OC(p_{Y|\bA,H_v}, p_{Y|\bA, H_w}) & = \frac{1}{6} \left( -1 + \frac{ \sigma^2 + 3}{ \sigma^2 -1} \log(\frac{\sigma^2 + 3}{4}) + \log \bigg( \frac{\sigma^2 -1}{ (\sigma^2 + 3) \log(\frac{\sigma^2 + 3}{4})}\bigg)   \right) \\
    &  \approx \frac{1}{6} \left( -1 + \log(\frac{\sigma^2}{4}) + \log \bigg( \frac{1}{ \log(\frac{\sigma^2}{4})}\bigg)   \right) \\
    & =  \frac{1}{6} \log \bigg( \frac{\sigma^2}{ 8e \cdot \log(\sigma/2)} \bigg),
\end{align*}
where $e$ is the natural number. For the comparison in the Chernoff information between separate measurements and mixed measurements, let us subtract the values in logarithm and check whether the subtracted result is positive or negative. For large enough $\sigma$, the following condition holds
\begin{align*}
	\frac{\sigma^2}{8e \cdot \log(\sigma/2)} - \frac{\sigma}{2} > 0 \Leftrightarrow e^{\sigma} > \bigg( \frac{\sigma}{2} \bigg)^{4e}.
\end{align*}
Therefore, we can conclude that for large enough $\sigma$, the Chernoff information with mixed measurements, i.e., the outer Chernoff information, becomes greater than that with separate measurements. Fig. \ref{fig:OCI_comparison_different_variance} shows the outer Chernoff information, denoted by OCI, and the Chernoff information with separate measurements, denoted by CI, in the figure. From Fig. \ref{fig:OCI_comparison_different_variance}, it is clearly shown that the Chernoff information with mixed measurements can be larger than the Chernoff information with separate measurements, and the outer Chernoff information between $H_1$ and $H_4$ (or $H_4$ and $H_7$) is the minimum one among other cases. For simplicity of the figure, we present the outer Chernoff information results for only a few cases of hypotheses pair in Fig. \ref{fig:OCI_comparison_different_variance}.

Additionally, the inner Chernoff information between $H_v = H_1$ and $H_w = H_4$ can be calculated as follows:
\begin{align*}
IC(p_{Y, \bA|H_v}, p_{Y, \bA | H_w}) &=-\min_{0\leq \lambda \leq 1}    \log \bigg[ \E_{\bA} \bigg( \int{p_{Y|\bA, H_v}^{\lambda}(y|\ba, H_v)p_{Y|\bA, H_w}^{1-\lambda}(y|\ba, H_w)\,dy} \bigg) \bigg]  \nonumber \\
    &=-\min_{0\leq \lambda \leq 1}  \log \bigg[\frac{1}{3} \underbrace{ \cancelto{1}{ \int{p_{\mathcal{N}(0,\sigma^2+3)}(y|\ba_1,H_1)^{\lambda}(y)p_{\mathcal{N}(0,\sigma^2+3)}(y|\ba_1,H_4)^{1-\lambda}(y)\,dy}} }_{ \text{from measuremet vector}\;\ba_1^T}  \\
    & \quad\quad\quad\quad\quad\quad +  \frac{1}{3} \underbrace{  \int{p_{\mathcal{N}(0,4)}(y|\ba_2,H_1)^{\lambda}(y)p_{\mathcal{N}(0,\sigma^2+3)}(y|\ba_2,H_4)^{1-\lambda}(y)\,dy}  }_{\text{from measuremet vector}\;\ba_2^T} \\
     & \quad\quad\quad\quad\quad\quad +  \frac{1}{3} \underbrace{ \cancelto{1}{ \int{p_{\mathcal{N}(0,4)}(y|\ba_3,H_1)^{\lambda}(y)p_{\mathcal{N}(0,4)}(y|\ba_3,H_4)^{1-\lambda}(y)\,dy} } }_{\text{from measuremet vector}\;\ba_3^T}  \bigg]\\
    &=-\min_{0\leq \lambda \leq 1}  \log \bigg[ \frac{1}{3}  \int{  \bigg(\frac{1}{2 \pi \cdot 4} \bigg)^{\lambda/2} e^{-\frac{\lambda x^2}{2 \cdot 4} } \cdot \bigg(\frac{1}{2 \pi (\sigma^2 + 3)} \bigg)^{(1-\lambda)/2} e^{-\frac{(1-\lambda) x^2}{2 (\sigma^2 +3)} }  \,dx}  \; + \; \frac{2}{3}  \bigg]
\end{align*}
\begin{align*}
    &=-\min_{0\leq \lambda \leq 1}  \log \bigg[ \frac{1}{3} \bigg(\frac{1}{2 \pi \cdot 4} \bigg)^{\lambda/2}  \bigg(\frac{1}{2 \pi (\sigma^2 + 3)} \bigg)^{(1-\lambda)/2}  \int{ e^{-\frac{\lambda x^2}{2 \cdot 4} } \cdot  e^{-\frac{(1-\lambda) x^2}{2 (\sigma^2 +3)} }  \,dx}  \; + \; \frac{2}{3}  \bigg]\\
    &=-\min_{0\leq \lambda \leq 1}  \log \bigg[ \frac{1}{3} \bigg(\frac{1}{2 \pi \cdot 4} \bigg)^{\lambda/2}  \bigg(\frac{1}{2 \pi (\sigma^2 + 3)} \bigg)^{(1-\lambda)/2}  \int{ e^{-\frac{\lambda (\sigma^2+3) x^2 + 4(1-\lambda)x^2 }{2 (\sigma^2+3) \cdot 4} }   \,dx}  \; + \; \frac{2}{3}  \bigg]\\
    &=-\min_{0\leq \lambda \leq 1}  \log \bigg[\frac{1}{3} \bigg(\frac{1}{2 \pi \cdot 4} \bigg)^{\lambda/2}  \bigg(\frac{1}{2 \pi (\sigma^2 + 3)} \bigg)^{(1-\lambda)/2}  \int{ e^{-\frac{ (\lambda (\sigma^2+3) + 4(1-\lambda))x^2 }{2 (\sigma^2+3) \cdot 4} }   \,dx}  \; + \; \frac{2}{3}  \bigg]\\
  & \quad( \sqrt{ \lambda(\sigma^2+3) + 4(1-\lambda)} \;x  = y \;\; \rightarrow  \sqrt{ \lambda(\sigma^2+3) + 4(1-\lambda)}  \;dx  = dy) \\
      &=-\min_{0\leq \lambda \leq 1}  \log \bigg[\frac{1}{3} \bigg(\frac{1}{2 \pi \cdot 4} \bigg)^{\lambda/2}  \bigg(\frac{1}{2 \pi (\sigma^2 + 3)} \bigg)^{(1-\lambda)/2}  \int{ e^{-\frac{ y^2 }{2 (\sigma^2+3)\cdot 4}} \cdot \frac{1}{\sqrt{\lambda (\sigma^2+3) + 4(1-\lambda)} }   \,dy}  + \frac{2}{3}  \bigg]\\
     &=-\min_{0\leq \lambda \leq 1}  \log \bigg[ \frac{1}{3} \bigg(\frac{1}{2 \pi \cdot 4} \bigg)^{\lambda/2}  \bigg(\frac{1}{2 \pi (\sigma^2 + 3)} \bigg)^{(1-\lambda)/2} \cdot \frac{1}{\sqrt{\lambda (\sigma^2+3) + 4(1-\lambda)} } \int{ e^{-\frac{ y^2 }{2 (\sigma^2+3)\cdot 4}}    \,dy}  +  \frac{2}{3}  \bigg]\\
    &=-\min_{0\leq \lambda \leq 1}  \log \bigg[  \frac{1}{3} \bigg(\frac{1}{2 \pi \cdot 4} \bigg)^{\lambda/2}  \bigg(\frac{1}{2 \pi (\sigma^2 + 3)} \bigg)^{(1-\lambda)/2} \cdot \bigg( \frac{ 2 \pi (\sigma^2 +3) \cdot 4 }{ \lambda (\sigma^2+3) + 4(1-\lambda)} \bigg)^{1/2}  + \frac{2}{3}  \bigg]\\
    &=-\min_{0\leq \lambda \leq 1}  \log \bigg[ \frac{1}{3} \bigg(\frac{1}{4} \bigg)^{\lambda/2}  \bigg(\frac{1}{ \sigma^2 + 3} \bigg)^{(1-\lambda)/2} \cdot \bigg( \frac{ (\sigma^2 +3) \cdot 4 }{ \lambda (\sigma^2+3) + 4(1-\lambda)} \bigg)^{1/2} +  \frac{2}{3}  \bigg]\\
    &=-\min_{0\leq \lambda \leq 1}  \log \bigg[  \frac{1}{3} \bigg(\frac{ \sigma^2 +3 }{4} \bigg)^{\lambda/2}  \cdot \bigg( \frac{  4 }{ \lambda (\sigma^2+3) + 4(1-\lambda)} \bigg)^{1/2} + \frac{2}{3}  \bigg]
\end{align*}
By considering the first-order condition, we have the following critical point, which is between 0 and 1; namely,
\begin{align} \label{eq:lambda_star_H1vsH4}
\lambda^{\star} = -\dfrac{4\log\left({\sigma}^2+3\right)-{\sigma}^2-4\log\left(4\right)+1}{\left({\sigma}^2-1\right)\log\left({\sigma}^2+3\right)-\log\left(4\right){\sigma}^2+\log\left(4\right)}.
\end{align}
Therefore, the inner Chernoff information is obtained as
\begin{align*}
   IC(p_{Y, \bA|H_v}, p_{Y, \bA | H_w}) = - \log\left(\dfrac{\left({\sigma}^2+3\right)^\frac{\lambda^{\star}}{2}{\cdot}2^{1-\lambda^{\star}}}{3\sqrt{\left({\sigma}^2+3\right)\lambda^{\star}+4\left(1-\lambda^{\star}\right)}}+\dfrac{2}{3}\right),
\end{align*}
where $\lambda^{\star}$ is in \eqref{eq:lambda_star_H1vsH4}.

\subsection{Example 5: $k=1$ anomalous random variable among $n$ random variables with different means}
For the benefit of mixed measurements in hypothesis testing, we take into account a general example with $n$ random variables having different means. More specifically, only one random variable follows distribution $\cN (\mu, 1)$, where $\mu \neq 0$, and the rest $n-1$ random variables follow the distribution $\cN (0, 1)$. Under the assumption that all random variables $X_1$, $X_2$,..., and $X_{n}$ are independent, we have $n$ hypotheses as follows:\\[-5pt]
\begin{itemize}
 \item $H_1$: ($X_1$, $X_2$, ..., $X_{n}$) $\sim$ ($\cN(\mu, 1)$, $\cN(0, 1)$, ..., $\cN(0,1)$),
 \item $H_2$: ($X_1$, $X_2$, ..., $X_{n}$) $\sim$ ($\cN(0, 1)$, $\cN(\mu,1)$, ..., $\cN(0, 1)$),
 \item ......
 \item $H_{n}$: ($X_1$, $X_2$, ..., $X_{n}$) $\sim$ ($\cN(0, 1)$, $\cN(0, 1)$, ..., $\cN(\mu, 1)$).\\[-5pt]
\end{itemize}
In this example, we will show that the Chernoff information with mixed measurements can be larger than that with separate measurements.

For a hypothesis $H_v$, the probability distribution for the output with separate measurements, is $\cN(\mu, 1)$ only when the random variable $X_v$ is observed. Otherwise, the output distribution follows $\cN(0, 1)$. Then, for any pair of hypotheses $H_v$ and $H_w$, when the random variable $X_v$ is observed, the probability distributions for the output are $\cN(\mu, 1)$ and $\cN(0, 1)$ respectively. Similarly, for hypotheses $H_v$ and $H_w$, when the random variable $X_w$ is observed, the probability distributions for the output are $\cN(\mu, 1)$ and $\cN(0, 1)$ respectively. From \eqref{eq:outer_CI}, the Chernoff information between $H_v$ and $H_w$ with separate measurements is obtained by
\begin{align*}
&OC(p_{Y|\bA,H_v}, p_{Y|\bA, H_w}) \nonumber \\
&=-\min_{0\leq \lambda \leq 1}  \E_{\bA} \left( \log \left(\int{p_{Y|\bA,H_v}^{\lambda}(y|\ba,H_v)p_{Y|\bA,H_w}^{1-\lambda}(y|\ba, H_w)\,dy}\right) \right)\nonumber \\
&=-\min_{0\leq \lambda \leq 1}  \left [\frac{1}{n} \left( \log \left(\int{p_{\cN(\mu,1)}^{\lambda}(y)p_{\cN(0,1)}^{1-\lambda}(y)\,dy}\right) \right) +\frac{1}{n} \left( \log \left(\int{p_{\cN(0,1)}^{\lambda}(y)p_{\cN(\mu,1)}^{1-\lambda}(y)\,dy}\right) \right)\right] \nonumber \\
    & =-\frac{1}{n} \min_{0\leq \lambda \leq 1} \bigg[ \log \int{\bigg(\frac{1}{2\pi}\bigg)^{(1-\lambda)/2} e^{-\frac{(1-\lambda) x^2}{2}} \bigg(\frac{1}{2\pi}\bigg)^{\lambda/2} e^{-\frac{\lambda (x-\mu)^2}{2}}\,dx} + \log \int{\bigg(\frac{1}{2\pi}\bigg)^{\lambda/2} e^{-\frac{\lambda x^2}{2}} \bigg(\frac{1}{2\pi}\bigg)^{(1-\lambda)/2} e^{-\frac{(1-\lambda) (x-\mu)^2}{2}}\,dx}\bigg] \nonumber \\
    & =-\frac{1}{n} \min_{0\leq \lambda \leq 1} \bigg[  \log \bigg(\frac{1}{2\pi}\bigg)^{1/2} \int{ e^{-\frac{(1-\lambda) x^2 + \lambda(x - \mu)^2}{2}} \,dx} + \log \bigg(\frac{1}{2\pi}\bigg)^{1/2} \int{ e^{-\frac{\lambda x^2 + (1-\lambda)(x - \mu)^2}{2}} \,dx} \bigg] \nonumber \\
    & =-\frac{1}{n} \min_{0\leq \lambda \leq 1} \bigg[ \log \bigg(\frac{1}{2\pi}\bigg)^{1/2} \int{ e^{-\frac{ (x- \lambda \mu)^2 }{2}} \cdot e^{-\frac{ \lambda \mu^2 (1- \lambda) }{2}}  \,dx} + \log \bigg(\frac{1}{2\pi}\bigg)^{1/2} \int{ e^{-\frac{ (x - \mu(1-\lambda))^2 }{2}} \cdot e^{-\frac{ \lambda \mu^2 (1 - \lambda)}{2}} \,dx} \bigg]\nonumber \\
    & =-\frac{1}{n} \min_{0\leq \lambda \leq 1} \bigg[ \log \bigg(\frac{1}{2\pi}\bigg)^{1/2} \cdot e^{-\frac{ \lambda \mu^2 (1- \lambda) }{2}} \int{ e^{-\frac{ (x- \lambda \mu)^2 }{2}}   \,dx} + \log \bigg(\frac{1}{2\pi}\bigg)^{1/2} \cdot e^{-\frac{ \lambda \mu^2 (1 - \lambda)}{2}} \int{ e^{-\frac{ (x - \mu(1-\lambda))^2 }{2}} \,dx} \bigg]\nonumber \\
    & =-\frac{1}{n} \min_{0\leq \lambda \leq 1} \bigg[ \log \bigg(\frac{1}{2\pi}\bigg)^{1/2} \cdot e^{-\frac{ \lambda \mu^2 (1- \lambda) }{2}} \cdot (2 \pi)^{1/2} + \log \bigg(\frac{1}{2\pi}\bigg)^{1/2} \cdot e^{-\frac{ \lambda \mu^2 (1 - \lambda)}{2}} \cdot (2 \pi)^{1/2} \bigg] \nonumber \\
     & =-\frac{1}{n} \min_{0\leq \lambda \leq 1}\bigg[  \log  e^{-\frac{ \lambda \mu^2 (1- \lambda) }{2}}  + \log  e^{-\frac{ \lambda \mu^2 (1 - \lambda)}{2}} \bigg]\nonumber \\
     & =-\frac{1}{n} \min_{0\leq \lambda \leq 1}  \bigg[ - \lambda \mu^2 (1- \lambda) \bigg]\nonumber  \\
     & =-\frac{1}{n} \min_{0\leq \lambda \leq 1} \bigg[  \mu^2 ( \lambda^2 - \lambda)  \bigg]\nonumber \\
     & =-\frac{1}{n} \min_{0\leq \lambda \leq 1} \bigg[  \mu^2 ( \lambda^2 - \lambda + \frac{1}{4} - \frac{1}{4}) \bigg] \nonumber \\
     & =-\frac{1}{n} \min_{0\leq \lambda \leq 1}  \bigg[ \mu^2 ( \lambda - \frac{1}{2})^2 - \frac{\mu^2}{4} \bigg]  \nonumber \\
     & = \frac{\mu^2}{4n},
\end{align*}
where the final equality is obtained by plugging in the optimal solution, i.e., $\lambda^{\star}=\frac{1}{2}$. Therefore, we have the Chernoff information with separate measurements as follows:
\begin{align}\label{eq:differentMeansSeparateObservations_finalresults}
OC(p_{Y|\bA,H_v}, p_{Y|\bA, H_w})= \frac{\mu^2}{4 n }.
\end{align}

The Chernoff information between hypotheses $H_v$ and $H_w$ with mixed measurements is calculated as follows. For the mixed measurements, we consider a matrix, size in $m \times n$, with the number of ones in a row is $n_r$, where each row vector is used for a sensing vector as a deterministic time-varying measurement.  For any pair of hypotheses $H_v$ and $H_w$, we assume that there is a sensing vector among $m$ sensing vectors which measures one and only one of $X_v$ and $X_w$. Without loss of generality, we further assume that a sensing vector measures $X_v$ but not $X_w$.
Suppose the hypothesis $H_v$ is true. Then, the output probability distribution for that measurement is $\cN(\mu, n_r)$. If the hypothesis $H_w$ is true,  the output probability distribution is expressed as $\cN(0, n_r)$. The Chernoff information between $H_v$ and $H_w$ with mixed measurements is calculated as
\begin{align}\label{eq:differentMeanOCI}
OC(p_{Y|\bA,H_v}, p_{Y|\bA, H_w})
&=-\min_{0\leq \lambda \leq 1}  \E_{\bA} \left( \log \left(\int{p_{Y|\bA,H_v}^{\lambda}(y|\ba,H_v)p_{Y|\bA,H_w}^{1-\lambda}(y|\ba,H_w)\,dy}\right) \right) \nonumber \\
&= -\min_{0\leq \lambda \leq 1}  \frac{1}{m}  \log \left(\int{p_{\cN(0, n_r)}^{\lambda}(y)p_{\cN(\mu,n_r)}^{1-\lambda}(y)\,dy}\right) \nonumber \\
    & =-\frac{1}{m} \min_{0\leq \lambda \leq 1} \log \int{\bigg(\frac{1}{2\pi n_r}\bigg)^{1/2} e^{-\frac{(1-\lambda) (x - \mu)^2}{2\cdot n_r}} e^{-\frac{\lambda x^2}{2 \cdot n_r}}\,dx} \nonumber \\
     & =-\frac{1}{m} \min_{0\leq \lambda \leq 1} \log \bigg(\frac{1}{2\pi n_r}\bigg)^{1/2}  \int{e^{-\frac{(1-\lambda) (x - \mu)^2}{2\cdot4}} e^{-\frac{\lambda x^2}{2 \cdot n_r}}\,dx}  \nonumber \\
     & =-\frac{1}{m} \min_{0\leq \lambda \leq 1} \bigg[ \log \bigg(\frac{1}{2\pi n_r}\bigg)^{1/2} + \log \int{e^{-\frac{(1-\lambda) (x - \mu)^2}{2\cdot n_r}} e^{-\frac{\lambda x^2}{2 \cdot n_r}}\,dx}  \bigg]\nonumber \\
     & =-\frac{1}{m} \min_{0\leq \lambda \leq 1} \bigg[ \log \bigg(\frac{1}{2\pi n_r}\bigg)^{1/2} + \log \int{e^{-\frac{(1-\lambda) (x - \mu)^2 + \lambda x^2}{2\cdot n_r}} \,dx}  \bigg]\nonumber \\
     & =-\frac{1}{m} \min_{0\leq \lambda \leq 1} \bigg[ \log \bigg(\frac{1}{2\pi n_r}\bigg)^{1/2} + \log \int{e^{-\frac{(x - (1-\lambda)\mu)^2 + \lambda \mu^2 (1-\lambda)}{2\cdot n_r}} \,dx}  \bigg]\nonumber\\
          & =-\frac{1}{m} \min_{0\leq \lambda \leq 1} \bigg[ \log \bigg(\frac{1}{2\pi n_r}\bigg)^{1/2} + \log \int{e^{-\frac{(x - (1-\lambda)\mu)^2}{2 \cdot n_r}} e^{-\frac{\lambda \mu^2 (1-\lambda)}{2\cdot n_r}} \,dx}  \bigg]\nonumber\\
     & =-\frac{1}{m} \min_{0\leq \lambda \leq 1} \bigg[ \log \bigg(\frac{1}{2\pi n_r}\bigg)^{1/2} + \log e^{-\frac{\lambda \mu^2 (1-\lambda)}{2\cdot n_r}} \int{e^{-\frac{(x - (1-\lambda)\mu)^2}{2 \cdot n_r}}  \,dx}  \bigg]\nonumber \\
     & =-\frac{1}{m} \min_{0\leq \lambda \leq 1} \bigg[ \log \bigg(\frac{1}{2\pi n_r}\bigg)^{1/2} + \log e^{-\frac{\lambda \mu^2 (1-\lambda)}{2\cdot n_r}} \sqrt{2\pi n_r}  \bigg]\nonumber \\
     &  =-\frac{1}{m} \min_{0\leq \lambda \leq 1}  \bigg[  -\frac{\lambda \mu^2 (1-\lambda)}{2 \cdot n_r}  \bigg]\nonumber \\
     &  =-\frac{\mu^2}{2mn_r} \min_{0\leq \lambda \leq 1}  \bigg[  \lambda^2  - \lambda  \bigg]\nonumber \\
     &  =-\frac{\mu^2}{2mn_r} \min_{0\leq \lambda \leq 1}  \bigg[  \bigg( \lambda  - \frac{1}{2} \bigg)^2 - \frac{1}{4} \bigg]\nonumber \\
     &  =\frac{\mu^2}{8mn_r},
\end{align}
where the last equality is obtained by taking the optimal solution, i.e., $\lambda^{\star}=\frac{1}{2}$.  Therefore, as long as $n> 2mn_r$, where $n_r$ is the number of ones in a sensing vector, the Chernoff information with mixed measurements can be larger than that with separate measurements.

Additionally, if we consider random time varying measurements, then the mixed measurements will be randomly obtained among $m$ numbers of sensing vectors. In this case, the inner Chernoff information is calculated as
\begin{align}\label{eq:differentMeanICI}
IC(p_{Y, \bA|H_v}, p_{Y, \bA | H_w})
&=-\min_{0\leq \lambda \leq 1}    \log \left( \int{p_{Y,\bA |H_v}^{\lambda}(y,\ba|H_v)p_{Y, \bA|H_w}^{1-\lambda}(y,\ba|H_w)\,dy} \right) \nonumber \\
&=-\min_{0\leq \lambda \leq 1}    \log \left( \int p_{\bA}(\ba)  \bigg( \int{p_{Y |\bA, H_v}^{\lambda}(y|\ba,H_v)p_{Y| \bA, H_w}^{1-\lambda}(y|\ba,H_w)\,dy}  \bigg) d\ba \right) \nonumber \\
&= -\min_{0\leq \lambda \leq 1}  \log \left(  \frac{1}{m} \int{p_{\cN(\mu , n_r)}^{\lambda}(y)p_{\cN(0,n_r)}^{1-\lambda}(y)\,dy}   + \frac{m-1}{m} \right)  \nonumber \\
&= -\min_{0\leq \lambda \leq 1}  \log \left(  \frac{1}{m} \int{ \bigg(\frac{1}{2 \pi n_r} \bigg)^{1/2} e^{ - \frac{\lambda (x - \mu)^2 + (1-\lambda)x^2}{2 n_r} } \,dx}  + \frac{m-1}{m} \right)  \nonumber \\
&= -\min_{0\leq \lambda \leq 1}  \log \left(  \frac{1}{m} \bigg(\frac{1}{2 \pi n_r} \bigg)^{1/2}  \cdot e^{ - \frac{ \lambda \mu^2 (1 - \lambda)}{2 n_r} } \int{ e^{ - \frac{ (x - \lambda \mu)^2}{2 n_r} }\,dx}   + \frac{m-1}{m} \right)  \nonumber \\
&= -\min_{0\leq \lambda \leq 1}  \log \left(  \frac{1}{m} \bigg(\frac{1}{2 \pi n_r} \bigg)^{1/2}  \cdot e^{ - \frac{ \lambda \mu^2 (1 - \lambda)}{2 n_r} } (2 \pi n_r)^{1/2}  + \frac{m-1}{m} \right)  \nonumber \\
&= -\min_{0\leq \lambda \leq 1}  \log \left(  \frac{1}{m}  \cdot e^{ - \frac{ \lambda \mu^2 (1 - \lambda)}{2 n_r} }  + \frac{m-1}{m} \right)  \nonumber \\
&= -\min_{0\leq \lambda \leq 1}  \log \left(  \frac{1}{m}  \cdot e^{  \frac{ \mu^2}{2 n_r} (\lambda^2 - \lambda)  } + \frac{m-1}{m} \right) \nonumber \\
&= -\min_{0\leq \lambda \leq 1}  \log \left(  \frac{1}{m} \cdot e^{- \frac{\mu^2}{8 n_r} } \cdot e^{  \frac{ \mu^2}{2 n_r} (\lambda - \frac{1}{2})^2 } + \frac{m-1}{m} \right)  \nonumber \\
&= - \log \left(  \frac{1}{m} \cdot e^{- \frac{\mu^2}{8 n_r} } + \frac{m-1}{m} \right)
\end{align}
where the last equality is obtained by plugging the optimal Chernoff coefficient to $\lambda$, i.e., $\lambda^{\star}=1/2$.

\begin{figure}[t]
    \centering
    \includegraphics[scale=0.4]{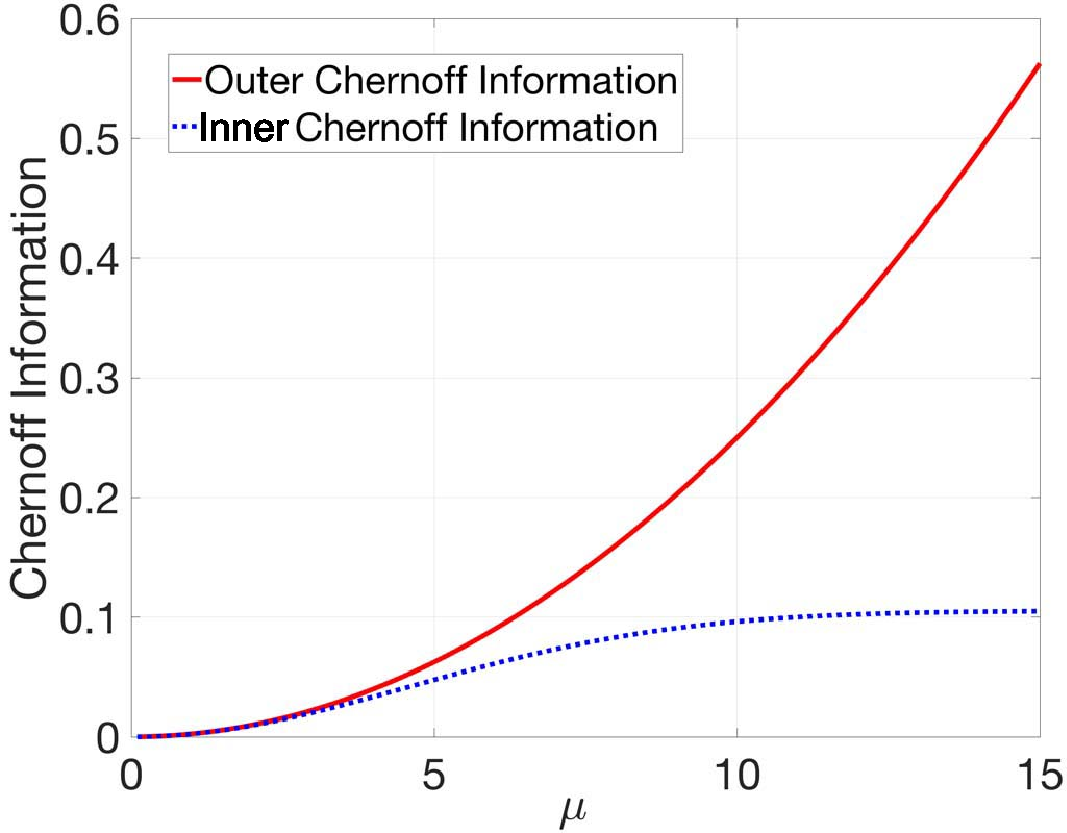}
    \caption{Comparison between the outer Chernoff information and the inner Chernoff information. The parameters $m$ and $n_r$ are set to 10 and 5 respectively.}
    \label{fig:OC_vs_IC}
\end{figure}

Fig. \ref{fig:OC_vs_IC} shows the comparison between the outer Chernoff information in \eqref{eq:differentMeanOCI} and the inner Chernoff information in \eqref{eq:differentMeanICI}. As mentioned in the Section \ref{sec:deterministic_timevarying}, it is demonstrated that the outer Chernoff information is greater than or equal to the inner Chernoff information.

\subsection{Example 6: $k=1$ anomalous random variable among $n$ random variables with different variances}
In order to clearly see the benefit of mixed measurements in hypothesis testing, let us consider another example for general $n$ with $k=1$. Here, only one random variable follows distribution $\cN (0, \sigma^2)$, and the rest $n-1$ random variables follow the distribution $\cN (0, 1)$. For simplicity, we consider the case when $\sigma^2 \gg 1$, and assume that all random variables $X_1$, $X_2$,..., and $X_{n}$ are independent. Then, we have $n$ hypotheses as follows:\\[-5pt]
\begin{itemize}
 \item $H_1$: ($X_1$, $X_2$, ..., $X_{n}$) $\sim$ ($\cN(0, \sigma^2)$, $\cN(0, 1)$, ..., $\cN(0, 1)$),
 \item $H_2$: ($X_1$, $X_2$, ..., $X_{n}$) $\sim$ ($\cN(0, 1)$, $\cN(0, \sigma^2)$, ..., $\cN(0, 1)$),
 \item ......
 \item $H_{n}$: ($X_1$, $X_2$, ..., $X_{n}$) $\sim$ ($\cN(0, 1)$, $\cN(0, 1)$, ..., $\cN(0, \sigma^2)$).\\[-5pt]
\end{itemize}
Through this example, we will show that the Chernoff information with mixed measurements can be larger than that with separate measurements.

With separate measurements, for a hypothesis $H_v$, the probability distribution for the output is $\cN(0, \sigma^2)$ only when the random variable $X_v$ is observed. Otherwise, the output distribution follows $\cN(0, 1)$. Then, for any pair of hypotheses $H_v$ and $H_w$, when the random variable $X_v$ is observed, the probability distributions for the output are $\cN(0, \sigma^2)$ and $\cN(0, 1)$ respectively. Similarly, for hypotheses $H_v$ and $H_w$, when the random variable $X_w$ is observed, the probability distributions for the output are $\cN(0, 1)$ and $\cN(0, \sigma^2)$ respectively.
From \eqref{eq:outer_CI}, the Chernoff information between $H_v$ and $H_w$ with separate measurements is given by
\begin{align}\label{eq:differentvariancesSeparateObservations100}
&OC(p_{Y|\bA,H_v}, p_{Y|\bA, H_w}) \nonumber \\
&=-\min_{0\leq \lambda \leq 1}  \E_{\bA} \left( \log \left(\int{p_{Y|\bA,H_v}^{\lambda}(y|\ba,H_v)p_{Y|\bA,H_w}^{1-\lambda}(y|\ba, H_w)\,dy}\right) \right)\nonumber \\
&=-\min_{0\leq \lambda \leq 1}  \left [\frac{1}{n} \left( \log \left(\int{p_{\cN(0,1)}^{\lambda}(y)p_{\cN(0,\sigma^2)}^{1-\lambda}(y)\,dy}\right) \right) +\frac{1}{n} \left( \log \left(\int{p_{\cN(0,\sigma^2)}^{\lambda}(y)p_{\cN(0,1)}^{1-\lambda}(y)\,dy}\right) \right)\right].
\end{align}
Due to the symmetry in the terms of the objective function, by plugging the optimal solution, i.e., $\lambda^{\star}=\frac{1}{2}$, in \eqref{eq:differentvariancesSeparateObservations100}, we have
\begin{align}\label{eq:differentvariancesSeparateObservations_finalresults100}
OC(p_{Y|\bA,H_v}, p_{Y|\bA, H_w}) = \frac{1}{n} \log(\frac{\sigma^2+1}{2\sigma}) \approx \frac{1}{n} \log(\frac{\sigma}{2}),
\end{align}
where the approximation is obtained from the condition $\sigma^2 \gg 1$.

With mixed measurements, the Chernoff information between hypotheses $H_v$ and $H_w$ is calculated as follows. For the mixed measurements, we consider a matrix, size in $m \times n$, with the number of ones in a row is $n_r$, where each row vector is used for a sensing vector as a deterministic time-varying measurement.  For any pair of hypotheses $H_v$ and $H_w$, we assume that there is a sensing vector among $m$ sensing vectors which measures one and only one of $X_v$ and $X_w$. Without loss of generality, we further assume that a sensing vector measures $X_v$ but not $X_w$.
Suppose the hypothesis $H_v$ is true. Then, the output probability distribution for that measurement is $\cN(0, \sigma^2+n_r-1)$. If the hypothesis $H_w$ is true,  the output probability distribution is expressed as $\cN(0, n_r)$. The Chernoff information between $H_v$ and $H_w$ with mixed measurements can be calculated as follows:
\begin{align}\label{eq:differentvariances100}
OC(p_{Y|\bA,H_v}, p_{Y|\bA, H_w})
&=-\min_{0\leq \lambda \leq 1}  \E_{\bA} \left( \log \left(\int{p_{Y|\bA,H_v}^{\lambda}(y|\ba,H_v)p_{Y|\bA,H_w}^{1-\lambda}(y|\ba,H_w)\,dy}\right) \right) \nonumber \\
&= -\min_{0\leq \lambda \leq 1}  \frac{1}{m}  \log \left(\int{p_{\cN(0,\sigma^2+n_r-1)}^{\lambda}(y)p_{\cN(0,n_r)}^{1-\lambda}(y)\,dy}\right), \nonumber \\
& = \frac{1}{2m} \left( -1 + \frac{B}{B-1} \log(B) + \log \bigg( \frac{B-1}{B \log(B)}\bigg)   \right),
\end{align}
where $B=\frac{\max{(\sigma^2+n_r - 1 ,\; n_r)}}{\min{(\sigma^2+n_r - 1,\; n_r)}}$. We obtain the final result from the same reason that we obtain \eqref{eq:differentvariancesHamming}.

If we assume that $\sigma^2 \gg n_r$, then the outer Chernoff information can be further approximated as follows:
\begin{align}
    OC(p_{Y|\bA,H_v}, p_{Y|\bA, H_w})
    & =  \frac{1}{2m} \left( -1 + \frac{\sigma^2 + n_r - 1}{\sigma^2 - 1} \log( \frac{\sigma^2 + n_r - 1}{n_r}) + \log \bigg( \frac{\sigma^2 - 1}{ (\sigma^2 + n_r -1) \log( \frac{\sigma^2 + n_r - 1}{ n_r} )}\bigg)   \right) \nonumber \\
    & \approx \frac{1}{2m} \left( -1 +  \log( \frac{\sigma^2}{n_r} ) + \log \bigg( \frac{1}{ \log( \frac{\sigma^2}{ n_r})}\bigg)   \right) \nonumber\\
    & = \frac{1}{2m}  \log \bigg( \frac{\sigma^2}{ e \cdot n_r \cdot \log( \frac{\sigma^2}{ n_r})}\bigg)
     \label{eq:differentvariancesHamming_scaling},
\end{align}
where $e$ is the natural number. For the comparison between separate measurement and mixed measurement in Chernoff information, let us subtract the values in logarithm and check whether the subtracted result is positive or negative. For large enough $\sigma$, the following condition holds
\begin{align}
\frac{\sigma^2}{ e \cdot n_r \cdot \log( \frac{\sigma^2}{ n_r})} - \frac{\sigma}{2} > 0\; \Leftrightarrow \;e^{\sigma} > \bigg( \frac{\sigma}{\sqrt{n_r}}\bigg)^{e\cdot n_r}.
\end{align}
Therefore, as long as $n > 2m$ and $\sigma$ is large enough, the Chernoff information with mixed measurements can be larger than that with separate measurements. For example, if we use the Hamming code parity check matrix for our measurement matrix, then, the size of the parity check matrix is $m \times n$, where $n=2^m-1$. Therefore, when $n$ is large, the Chernoff information with separate measurements becomes much smaller than that with mixed measurements.

\color{black}
\section{Characterization of Optimal Sensing Vector Maximizing the Error Exponent}
\label{sec:optimalmixing}

In this section, we derive a characterization of the optimal deterministic time-varying measurements which maximize the error exponent of hypothesis testing. We further explicitly design the optimal measurements for some simple examples. We begin with the following lemma about the error exponent of hypothesis testing.
\begin{lemma}
Suppose that there are overall $l=\binom{n}{k}$ hypotheses. For any fixed $k$ and $n$, the error exponent of the error probability of hypothesis testing is given by
\begin{equation*}
    E=\min_{1\leq v,w \leq l, v\neq w} OC(p_{Y|\bA,H_v}, p_{Y|\bA,H_w}).
\end{equation*}
\label{lem:exactexponent}
\end{lemma}

\begin{IEEEproof}
We first consider an upper bound on the error probability of hypothesis testing. Without loss of generality, let us assume $H_1$ is the true hypothesis. The error probability $\Pb_{err}$ in the Neyman-Pearson testing is stated as follows (the error exponent is tight for Neyman-Person testing as $m \rightarrow \infty$ \cite{coverbook}):
\begin{align*}
    \Pb_{err} \leq 2^{-m OC(p_{Y|\bA,H_v}, p_{Y|\bA, H_w})} \leq 2^{-mE}.
\end{align*}
By the union bound over the $l-1$ possible pairs $(H_1, H_w)$, the probability that $H_1$ is not correctly identified as the true hypothesis is upper bounded by $l\times 2^{-mE}$ in terms of scaling.

Now we take into account a lower bound on the error probability of hypothesis testing. Without loss of generality, we assume that $E$ is achieved between the hypothesis $H_1$ and the hypothesis $H_2$, namely,
\begin{equation*}
    E=OC(p_{Y|\bA,H_1}, p_{Y|\bA,H_2}).
\end{equation*}

Suppose that we are given the prior information that either hypothesis $H_1$ or $H_2$ is true. Knowing this prior information will not increase the error probability. Under this prior information, the error probability behaves asymptotically as $2^{-m E}$ as $m\rightarrow \infty$.  This shows that the error exponent of hypothesis testing is no bigger than $E$.
\end{IEEEproof}

The following theorem gives a simple characterization of the optimal probability density function $p_{\bA}(\ba)$ for the sensing vector. This enables us to explicitly find the optimal sensing vectors, under certain special cases of Gaussian random variables.
\begin{theorem}\label{thm:optimal_sensing_vector}
In order to maximize the error exponent in hypothesis testing, the optimal sensing vectors have a distribution $p^{*}_{\bA}(\ba)$ which maximizes the minimum of the outer conditional Chernoff information between different hypotheses:
\begin{align*}
    p^{*}_\bA(\ba)=\underset{p_\bA(\ba)}{\argmax} \min_{1\leq v,w \leq l, v\neq w} OC(p_{Y|\bA,H_v}, p_{Y|\bA,H_w}).
\end{align*}

When $k=1$ and the $n$ random variables of interest are independent Gaussian random variables with the same variances, the optimal $p^{*}_\bA(\ba)$ admits a discrete probability distribution:
\begin{equation*}
    p^{*}_\bA(\ba)= \sum_{\sigma~\mbox{as a permutation} } \frac{1}{n!}\delta(\ba-\sigma(\ba^*)),
\end{equation*}
where $\ba^*$ is a constant $n$-dimensional vector such that
\begin{align}\label{thm:optimal_k=1_Gassuaindifferent means}
    \ba^*=\underset{\ba}{\argmax} \sum\limits_{1\leq v,w \leq l, v\neq w} C(p_{Y|\bA,H_v}, p_{Y|\bA,H_w}).
\end{align}
Here  $C(p_{Y|\bA,H_v}, p_{Y|\bA,H_w})$ represents the (regular) Chernoff information between $p_{Y|\bA, H_v}$, and $p_{Y|\bA, H_w}$.
\end{theorem}

\begin{IEEEproof}
The first statement follows from Lemma \ref{lem:exactexponent}.  Hence, we only need to prove the optimal sensing vectors for Gaussian random variables with the same variance under $k=1$. We let $l=\binom{n}{k}$.
For any $p_\bA(\ba)$, we have
\begin{eqnarray*}
     && \hspace{-3em}\min_{1\leq v,w \leq l, v\neq w} OC(p_{Y|\bA,H_v}, p_{Y|\bA,H_w}) \\
     &\leq& \frac{1}{\binom{n}{2}}\sum\limits_{1\leq v,w \leq l, v\neq w} OC(p_{Y|\bA,H_v}, p_{Y|\bA,H_w})\\
     & \overset{\eqref{eq:outer_CI} }{=}&  \frac{1}{\binom{n}{2}}\sum\limits_{1\leq v,w \leq l, v\neq w} \bigg[  - \min_{0\leq \lambda \leq 1} \int p_{\bA}(\ba)\log \left(\int{p_{Y|\bA,H_v}^{\lambda}(y|\ba,H_v)p_{Y|\bA,H_w}^{1-\lambda}(y|\ba,H_w)\,dy}\right) \,d\ba \bigg]  \\
     &\leq& \frac{1}{\binom{n}{2}}\sum\limits_{1\leq v,w \leq l, v\neq w}  \int p_{\bA}(\ba) \bigg[ - \min_{0\leq \lambda \leq 1} \log \left(\int{p_{Y|\bA,H_v}^{\lambda}(y|\ba,H_v)p_{Y|\bA,H_w}^{1-\lambda}(y|\ba,H_w)\,dy}\right) \bigg] \,d\ba  \\
     &=& \frac{1}{\binom{n}{2}}\sum\limits_{1\leq v,w \leq l, v\neq w} \int{p_\bA(\ba) C(p_{Y|\bA,H_v}, p_{Y|\bA,H_w})}\,d\ba \\
     &=& \frac{1}{\binom{n}{2}}\int{p_\bA(\ba) \sum\limits_{1\leq v,w \leq l, v\neq w}  C(p_{Y|\bA,H_v}, p_{Y|\bA,H_w})}\,d \ba \\
     &\overset{\eqref{thm:optimal_k=1_Gassuaindifferent means} }{\leq}& \frac{1}{\binom{n}{2}}\sum\limits_{1\leq v,w \leq l, v\neq w} C(p_{Y|\bA^*,H_v}, p_{Y|\bA^*,H_w}),
\end{eqnarray*}
where the second inequality follows from the fact that for any functions $f_i(\lambda)$, $i=1,...,m$, and for any $\lambda$ in the intersection domain of all the $f_i(\lambda)$'s, $\min_{\lambda} ( f_1(\lambda) +  f_2(\lambda) + ... + f_m(\lambda) ) \geq \min_{\lambda} f_1(\lambda) + \min_{\lambda} f_2(\lambda) + ... + \min_{\lambda} f_m(\lambda)$, and $\bA^*$ is the same expression for $\ba^*$ in the form of random variable.

On the other hand, for two Gaussian distributions with the same variances, the optimal $\lambda$ in \eqref{eq:outer_CI} is always equal to $\frac{1}{2}$, no matter what
$p_\bA(\ba)$ is chosen as shown in Examples 1 and 2 in Section \ref{sec:example}.  By symmetry, when
\begin{equation*}
p^{*}_\bA(\ba)= \sum_{\sigma~\mbox{as a permutation} } \frac{1}{n!}\delta(\ba-\sigma(\ba^*)),
\end{equation*}
and $\lambda = \frac{1}{2}$, for any two different hypotheses $H_v$ and $H_w$,
\begin{eqnarray*}
     && \hspace{-3em} OC(p_{Y|\bA,H_v}, p_{Y|\bA,H_w}) \\
     &\overset{\eqref{eq:outer_CI} }{=}&- \min_{0\leq \lambda \leq 1} \int p^{*}_{\bA}(\ba)\log \left(\int{p_{Y|\bA,H_v}^{\lambda}(y|\ba,H_v)p_{Y|\bA,H_w}^{1-\lambda}(y|\ba,H_w)\,dy}\right) \,d\ba\\
     &=&  \int p^{*}_{\bA}(\ba) \bigg[ - \min_{0\leq \lambda \leq 1} \log \left(\int{p_{Y|\bA,H_v}^{\lambda}(y|\ba,H_v)p_{Y|\bA,H_w}^{1-\lambda}(y|\ba,H_w)\,dy}\right) \bigg] \,d\ba  \\
     &=&\int{p^{*}_\bA(\ba) C(p_{Y|\bA,H_v}, p_{Y|\bA,H_w})}\,d \ba \\
     &=&\frac{1}{\binom{n}{2}}\int{p^{*}_\bA(\ba) \sum\limits_{1\leq v,w \leq l, v\neq w}  C(p_{Y|\bA,H_v}, p_{Y|\bA,H_w})}\,d \ba \\
     &=&\int{p^{*}_\bA(\ba) \frac{1}{\binom{n}{2}} \sum\limits_{1\leq v,w \leq l, v\neq w}  C(p_{Y|\bA^*,H_v}, p_{Y|\bA^*,H_w})}\,d \ba \\
     &=&\frac{1}{\binom{n}{2}} \sum\limits_{1\leq v,w \leq l, v\neq w}  C(p_{Y|\bA^*,H_v}, p_{Y|\bA^*,H_w}),
\end{eqnarray*}
where the second equality is from the fact that $\lambda=\frac{1}{2}$ is the common maximizer for Chernoff information between any two probability distribution having the same variances, i.e.,  for any functions $f_i(\lambda)$, $i=1,...,m$, $\min_{\lambda} ( f_1(\lambda) +  f_2(\lambda) + ... + f_m(\lambda) ) = \min_{\lambda} f_1(\lambda) + \min_{\lambda} f_2(\lambda) + ... + \min_{\lambda} f_m(\lambda)$ due to the common solution $\lambda$, the fourth equality is from the permutation symmetry of $p^{*}_\bA(\ba)$, the fifth equality is again from the generation of $p^{*}_\bA(\ba)$ from permutations of $\ba^*$, and the last equality follows from  $\int p^{*}_\bA(\ba)\,d \ba=1$.

This means that under $p^{*}_\bA(\ba)$,
\begin{eqnarray*}
    &&\hspace{-3em}\min_{1\leq v,w \leq l, v\neq w} OC(p_{Y|\bA,H_v}, p_{Y|\bA,H_w})\\
    &=&\frac{1}{\binom{n}{2}} \sum\limits_{1\leq v,w \leq l, v\neq w}  C(p_{Y|\bA^*,H_v}, p_{Y|\bA^*,H_w}).
\end{eqnarray*}
This further implies that the upper bound on $\min\limits_{1\leq v,w \leq l, v\neq w} OC(p_{Y|\bA,H_v}, p_{Y|\bA,H_w})$ is achieved, and we can conclude that $p^{*}_\bA(\ba)$ is the optimal distribution.
\end{IEEEproof}

We can consider Theorem \ref{thm:optimal_sensing_vector} to calculate explicitly the optimal sensing vectors for $n$ independent Gaussian random variables of the same variance $\sigma^2$, among which $k=1$ random variable has a mean $\mu_1$ and the mean of the other $(n-1)$ random variables is $\mu_2$. To obtain the optimal measurements maximizing the error exponent $E$, we first need to calculate a constant vector $\ba^*$ such that
    $$\ba^*=\underset{\ba}{\argmax} \sum\limits_{1\leq v,w \leq l, v\neq w} C(p_{Y|\bA,H_v}, p_{Y|\bA,H_w}).$$
After a simple calculation, $\ba^*$ is the optimal solution to
\begin{equation}
    \begin{aligned}
    & \underset{\ba}{\text{maximize}}
    & & \frac{1}{\binom{n}{2}} \sum\limits_{1\leq i,j \leq n, i\neq j}  {(a_i-a_j)^2} \\
    & \text{subject to}
    & & \sum\limits_{i=1}^{n}a_i^2 \leq 1,
    \end{aligned}
\label{eq:optimization_fork=1_Guassianmean}
\end{equation}
where $a_i$ represents the $i$-th element of vector $\ba$, and the corresponding optimal error exponent is
\begin{align}
	\frac{1}{\binom{n}{2}} \sum\limits_{1\leq i,j \leq n, i\neq j}  \frac{(a_i^*-a_j^*)^2(\mu_1-\mu_2)^2}{8\sigma^2 \sum\limits_{i=1}^{n}(a_i^*)^2}.
\end{align}
See \eqref{eq:mix_Chrenoff_ex} for a simple example.

This optimization problem is not a convex optimization program, however, it still admits zero duality gap from the S-procedure \cite[Appendix B]{boyd}, and can be efficiently solved via a generic solver. In fact, we obtain the optimal solution $\ba^*=[\frac{1}{\sqrt{2}},\frac{-1}{\sqrt{2}},0,...,0]^T$. Then an optimal distribution $p^{*}_\bA(\ba)$ for the sensing vector is
\begin{equation*}
    p^{*}_\bA(\ba)= \sum_{\sigma~\mbox{as a permutation} } \frac{1}{n!}\delta(\ba-\sigma(\ba^*)).
\end{equation*}
Namely, this optimal sensing vector is to uniformly choose two random variables, say $X_1$ and $X_2$, and take their weighted sum
$\frac{1}{\sqrt{2}} X_1-\frac{1}{\sqrt{2}} X_2$. Correspondingly, the optimal error exponent is
\begin{align*}
    \frac{(\mu_1-\mu_2)^2}{8\sigma^2} \frac{n}{\binom{n}{2}}= \frac{(\mu_1-\mu_2)^2}{4 \sigma^2 (n-1)}.
\end{align*}
In contrast, if we perform separate observations of $n$ random variables individually, the error exponent will be
\begin{align*}
    \frac{(\mu_1-\mu_2)^2}{8\sigma^2} \frac{n-1}{\binom{n}{2}}=\frac{(\mu_1-\mu_2)^2}{4\sigma^2 n}.
\end{align*}
In fact, linear mixed observations increase the error exponent by $\frac{n}{n-1}$ times. When $n$ is small, the improvement is significant. For example, when $n=2$, the error exponent is doubled.

From this example of Gaussian random variables with different means, we can make some interesting observations.
On the one hand, quite surprisingly, separate observations of random variables are \emph{not} optimal in maximizing the error exponent. On the other hand, in the optimal measurement scheme, each measurement takes a linear mixing of only two random variables, instead of mixing all the random variables. It is interesting to see whether these observations hold for other types of random variables.

\section{Efficient Algorithms for Hypothesis Testing with Mixed Observations}
\label{sec:efficient_algo}
In Section \ref{sec:compressed_hypothesis_testing}, we introduce the maximum likelihood estimate methods for three different types of mixed observations. Even though the maximum likelihood estimate method is the optimal test achieving the smallest hypothesis testing error probability, conducting the maximum likelihood estimator over $\binom{n}{k}$ hypotheses is computationally challenging. Especially, when $n$ and $k$ are large, finding $k$ abnormal random variables out of $n$ random variables is computationally intractable by using the maximum likelihood estimator. To overcome this drawback, we further design efficient algorithms to find $k$ abnormal random variables among $n$ random variables for large $n$ and $k$ by Least Absolute Shrinkage and Selection Operator (LASSO) based algorithm, and Message Passing (MP) based algorithm.

\subsection{LASSO based hypothesis testing algorithm}
The LASSO, also known as Basis Pursuit denoising \cite{Chen1998atomic,Tibshirani1996regreesion,baraldi2019basis}, is a well-known sparse regression tool in statistics, and has been successfully applied to various fields such as signal processing \cite{Chen1995examples,ekanadham2011recovery}, machine learning \cite{Sra2011optimization}, and control system \cite{Kukreja2006least,cho2022iterative}. We consider to use the LASSO algorithm to detect $k$ anomalous random variables when they have different means from the other $n-k$ random variables.

Suppose that each of the abnormal random variables has a non-zero mean, while each of the other $(n-k)$ random variables has  a zero mean. For a sensing matrix $\bA \in R^{m \times n}$, and observation $\by \in R^{m}$,  the LASSO optimization formulation is given as follows:
\begin{align}
\label{fun:LASSO}
    \underset{\bx}{\text{minimize}}\;\;\frac{1}{2} ||\by - \bA\bx||_2^2 + \lambda||\bx||_1,
\end{align}
where $\bx$ is a variable, and $\lambda$ is a parameter for the penalty term, i.e., $\ell_1$ norm of $\bx$, which makes more elements of $\bx$ driven to zero as $\lambda$ increases.

We use the LASSO formulation (\ref{fun:LASSO}) to obtain a solution $\hat{\bx}$, by taking $\by=[Y^1, ..., Y^m]^T$ and $\bA$ as an $m \times n$ sensing matrix whose $j$-th row vector equals to the sensing vector ${\ba^j}^T$ introduced in \eqref{eq:observation}. Since we are interested in finding $k$ abnormal random variables, we solve (\ref{fun:LASSO}) and select $k$ largest elements of $\hat{\bx}$ in amplitude. We decide the corresponding $k$ random variables as the $k$ abnormal random variables.

\subsection{MP based hypothesis testing algorithm}
We further design a message passing based method to discover general abnormal random variables,  even if the abnormal random variables have the same mean as the regular random variables. Our message passing based algorithm uses \textit{bipartite graph} to perform statistical inference. We remark that message passing algorithms have been successfully applied to decoding for error correcting code \cite{mackay2003information}, including Low Density Parity Check (LDPC) code. Unlike sparse bipartite graphs already used in LDPC codes and compressed sensing \cite{ExpanderCode, XHExpander, IndykSparse}, a novelty in this paper is that our sparse bipartite graphs are allowed to have more measurement nodes than variable nodes, namely $m \geq n$. Also, in our problem setup, we consider random variables which have different realizations at different time. Thus, it is a different problem setup from ones in \cite{ExpanderCode, XHExpander, IndykSparse}.

\begin{figure}[t]
    \centering
    \includegraphics[scale=0.4]{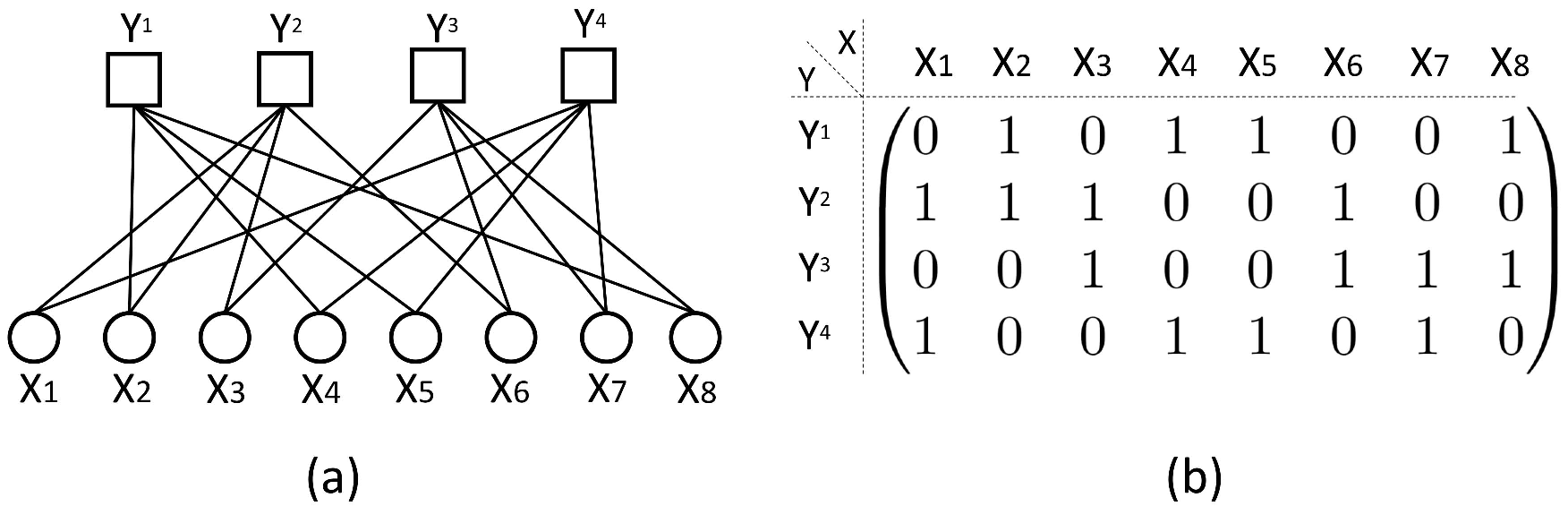}
    \caption{Illustration of a factor graph (a) from a matrix (b). A random variable $X_i$ and $Y^j$ are considered as a variable node and a check node in the graph respectively.}
    \label{fig:factor_graph}
\end{figure}

Let us denote the $j$-th linear mixed function by ${\ba^j}^T \bX$, where $\bX$ is a random variable vector, and ${\ba^j}$ is the $j$-th sensing vector for a linear mixed observation. The observation random variable at time $j$, i.e. $Y^j$, is simply represented as ${\ba^j}^T \bX^j$ introduced in (\ref{eq:observation}). Note that over time indices $j$'s, the random variable vector $\bX$ takes independent realizations. Now we define a bipartite factor graph as follows. We will represent a random variable $X_i$ using a \textit{variable node} on the bottom, and  represent an observation $Y^j$ as a \textit{check node} on the top. Variable nodes and check nodes appear on two sides of the bipartite factor graph. A line is drawn between a node $X_i$ and a node $Y^j$ if and only if that random variable $X_i$ is involved in the observation $Y^j$. We call the observation $Y^j$ linked with $X_i$ as a \textit{neighbor} of $X_i$. We will denote the set of neighbors of $X_i$ by $\cQ(X_i)$. Similarly, the set of random variable $X_i$'s linked with a random variable $Y^j$ is denoted by $\cQ(Y^j)$. Fig. \ref{fig:factor_graph}(a) is an example of the factor graph for a given sensing matrix $\bA$ in Fig. \ref{fig:factor_graph}(b).

\begin{figure}[t]
    \centering
    \includegraphics[scale=1.7]{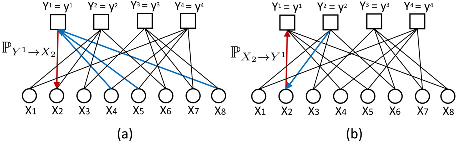}
    \caption{(a) Message sent from a check node to a variable node. The message sent from a check node $Y^1$ to a variable node $X_2$ (red arrow) is expressed as the probability $\Pb_{Y^1 \rightarrow X_2} (X_2~ \text{is abnormal} \;|\; Y^1 = y^1)$ by considering probabilities $\Pb_{X_i \rightarrow Y^1}$, $i=4,5,8$ (blue arrow). (b) Message sent from a variable node to a check node. The message sent from a variable node $X_2$ to a check node $Y^1$ (red arrow) is expressed as the probability $\Pb_{X_2 \rightarrow Y^1} (X_2~ \text{is abnormal})$ by considering probability $\Pb_{Y^2 \rightarrow X_2}$.}
    \label{fig:message}
\end{figure}

In our message passing based algorithm, messages are exchanged between variable nodes and check nodes. The messages are the probabilities that a random variable $X_i$ is abnormal. More precisely, the message sent from a variable node $X_i$ to a check node $Y^j$ is the probability that the variable node $X_i$ is abnormal, namely $\Pb_{X_i \rightarrow Y^j}(X_i~\text{is abnormal})$, based on local information gathered at variable node $X_i$. Similarly, the message sent from a check node $Y^j$ to a variable node $X_i$ is the probability that the variable node $X_i$ is abnormal based on local information gathered at variable node $Y^j$. For example, Fig. \ref{fig:message}(a) shows the message sent from a check node $Y^1$ to a variable node $X_2$, which is the probability that $X_2$ is abnormal given observation $y^1$ and incoming messages from $X_4$ to $Y^1$, $X_5$ to $Y^1$, and $X_8$ to $Y^1$. Fig. \ref{fig:message}(b) illustrates the message from a variable node $X_2$ to a check node $Y_1$, which is the probability that $X_2$ is abnormal when we consider the incoming message from $Y^2$ to $X_2$. The message $\mathfrak{m}$ from a check node $Y^j$ to a variable node $X_i$ is expressed as a function of incoming messages from neighbors of $X_i$ and $Y^j$, i.e., $\cQ(X_i)$ and $\cQ(Y^j)$ respectively, and given by
\begin{align}
\label{eq:message_y_x}
\mathfrak{m}_{Y^j \rightarrow X_i} & = f(\mathfrak{m}_{X \in \cQ(Y^j)\setminus X_i \rightarrow Y^j},  Y^j) \nonumber \\
                                  & = \Pb_{Y_j \rightarrow X_i}(X_i \text{ is abnormal}\; |\; Y_j = y_j),
\end{align}
where $f(\cdot)$ is a function calculating the probability of $X_i$ being abnormal based on incoming messages from its neighbor variable nodes and realized observation $y^j$. The message $\mathfrak{m}$ from a variable node to a check node is expressed by
\begin{align}
\label{eq:message_x_y}
    \mathfrak{m}_{X_i \rightarrow Y^j} & = h(\mathfrak{m}_{Y\in \cQ(X_j)\setminus Y^j \rightarrow X_i}) \nonumber\\
                                      & = \Pb_{X_i \rightarrow Y^j}(X_i \text{ is abnormal}),
\end{align}
where $h(\cdot)$ is a function calculating the probability of $X_i$ being abnormal based on incoming messages from check nodes. In the same way, we calculate the probability that $X_i$ is normal.

\section{Numerical Experiments}
\label{sec:simulation}
We numerically evaluate the performance of mixed observations in hypothesis testing. We first simulate the error probability of identifying abnormal random variables through linear mixed observations. The linear mixing used in the simulation is based on sparse bipartite graphs \cite{ExpanderCode, XHExpander, IndykSparse}. In the sparse bipartite graphs, $n$ variable nodes on the bottom are used to represent the $n$ random variables, and $m$ measurement nodes on the top are used to represent the $m$ measurements as shown in Fig. \ref{fig:factor_graph}. If and only if the $i$-th random variable is nontrivially involved in the $j$-th measurement, there is an edge connecting the $i$-th variable node to the $j$-th measurement node. In this simulation, there are $6$ edges emanating from each measurement node on the top, and  there are $\frac{6m}{n}$ edges emanating from each variable node on the bottom. After a uniformly random permutation, the $6m$ edges emanating from the measurement nodes are plugged into the $6m$ edge ``sockets'' of the bottom variable nodes. If there is an edge connecting the $i$-th variable node to the $j$-th measurement node, then the linear mixing coefficient before the $i$-th random variable in the $j$-th measurement is set to $1$; otherwise that linear mixing coefficient is set to $0$. We denote the Likelihood Test (LT) with separate observations and mixed (compressed) observations as SLT and CLT respectively. We also abbreviate the message passing and LASSO based hypothesis testing method to MP and LASSO in the figures below.

\subsection{Random variables with different variances}
\begin{figure}[t]
\centering
\includegraphics[scale=0.15]{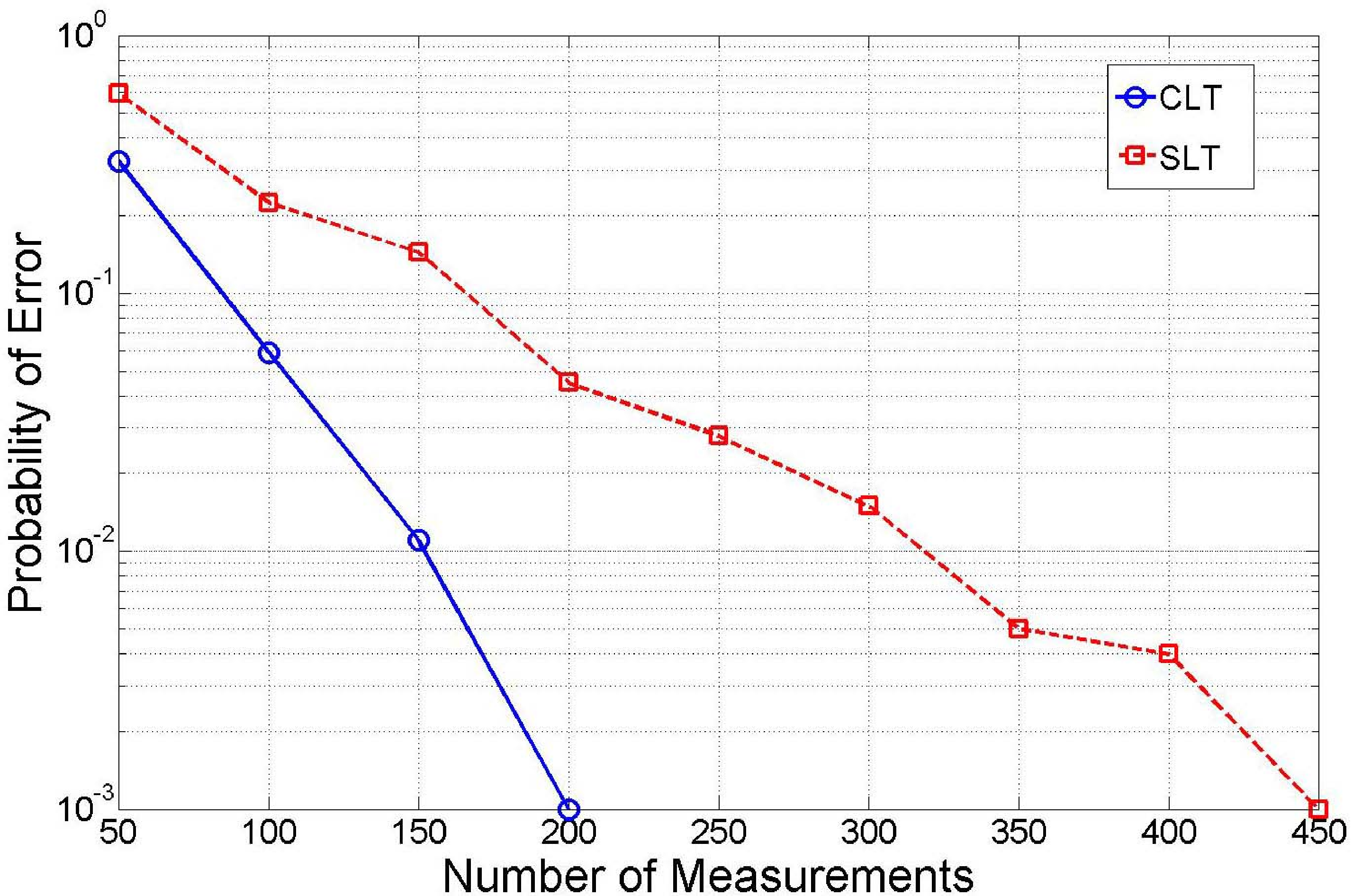}
\caption{Error probability in log scale as the number of measurements $m$ is varied, when $(n,k)=(100,1)$. The normal and abnormal random variables follow $\cN(0,1)$ and $\cN(0,100)$ respectively.}
\label{fig:N100Variance}
\end{figure}

In the first simulation example, we take $n=100$, and let $m$ vary from $50$ to $450$. We assume that $k=1$ random variable follows the Gaussian distribution $\cN(0, 100)$, and the other $(n-k)=99$ random variables follow another distribution $\cN(0,1)$. We use the maximum likelihood estimation to find the anomalous random variables through the described linear mixed observations based on sparse bipartite graphs. For comparison, we also implement the maximum likelihood estimator for separate observations of random variables, where we first take $\lfloor\frac{m}{n} \rfloor$ separate observations of each random variables, and then take an additional separate observation of $(m \mod n)$ random variables selected uniformly at random.  For each $m$, we perform $1000$ random trials, and record the number of trials failing to identify the abnormal random variables.  The error probability is plotted in Fig. \ref{fig:N100Variance}. As shown in Fig. \ref{fig:N100Variance}, the hypothesis testing with mixed observations provides significant reduction in the error probability of hypothesis testing, under the same number of observations.

\begin{figure}[t]
\centering
\includegraphics[scale=0.15]{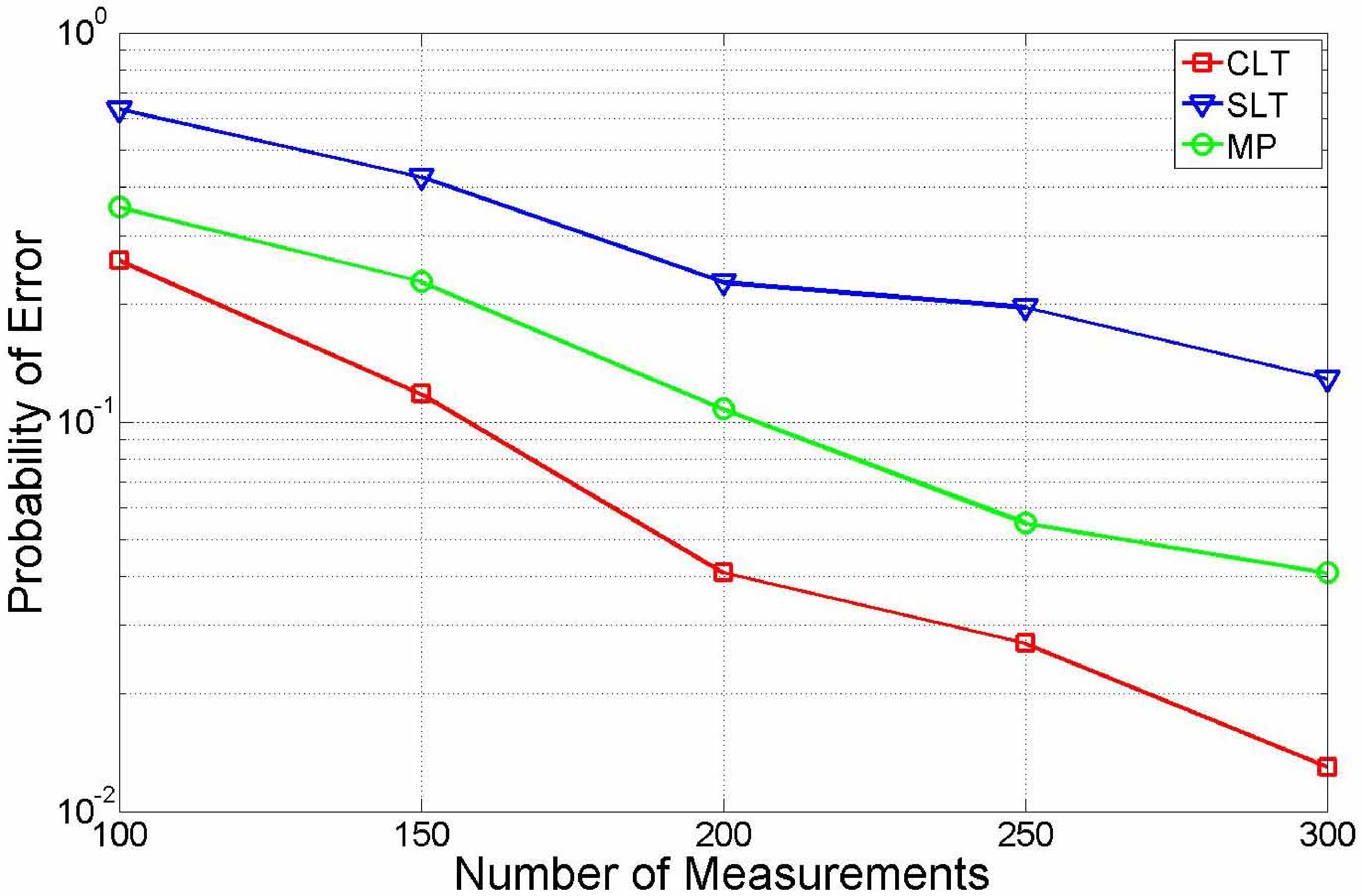}
\caption{Error probability in log scale as the number of measurements $m$ is varied, when $(n,k)=(200,1)$. The normal and abnormal random variables follow $\cN(0,1)$ and $\cN(0,100)$ respectively.}
\label{fig:comparision_LRT_MP1}
\end{figure}

\begin{figure}[t]
\centering
\includegraphics[scale=0.15]{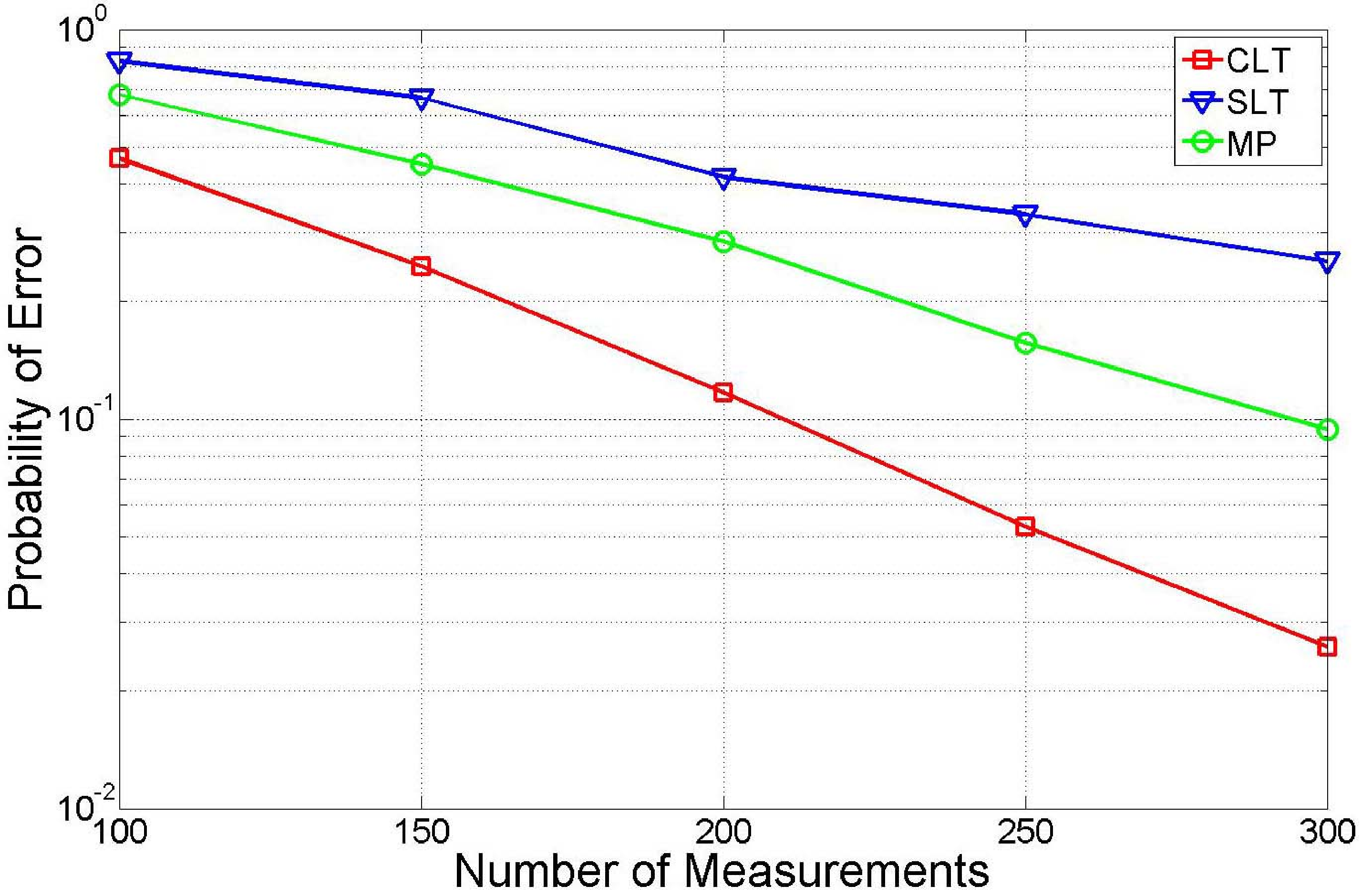}
\caption{Error probability in log scale as the number of measurements $m$ is varied, when $(n,k)=(200,2)$. The normal and abnormal random variables follow $\cN(0,1)$ and $\cN(0,100)$ respectively.}
\label{fig:comparision_LRT_MP2}
\end{figure}

We further carry out simulations for $n=200$. Figs. \ref{fig:comparision_LRT_MP1} and \ref{fig:comparision_LRT_MP2} show the error probability in hypothesis testing when normal and abnormal random variables follow the Gaussian distributions $\cN(0,1)$ and $\cN(0,100)$ on $k=1$ and $2$ respectively. The error probability in hypothesis testing is obtained from 1000 random trials. In these simulations, we use MP based hypothesis testing algorithm and compare it against LT methods. As shown in Figs. \ref{fig:comparision_LRT_MP1} and \ref{fig:comparision_LRT_MP2}, the error probability of MP based algorithm is worse than that of CLT. One reason for the worse performance of MP based algorithm is error propagation during iterations.


\begin{figure}[t]
\centering
\includegraphics[scale=0.15]{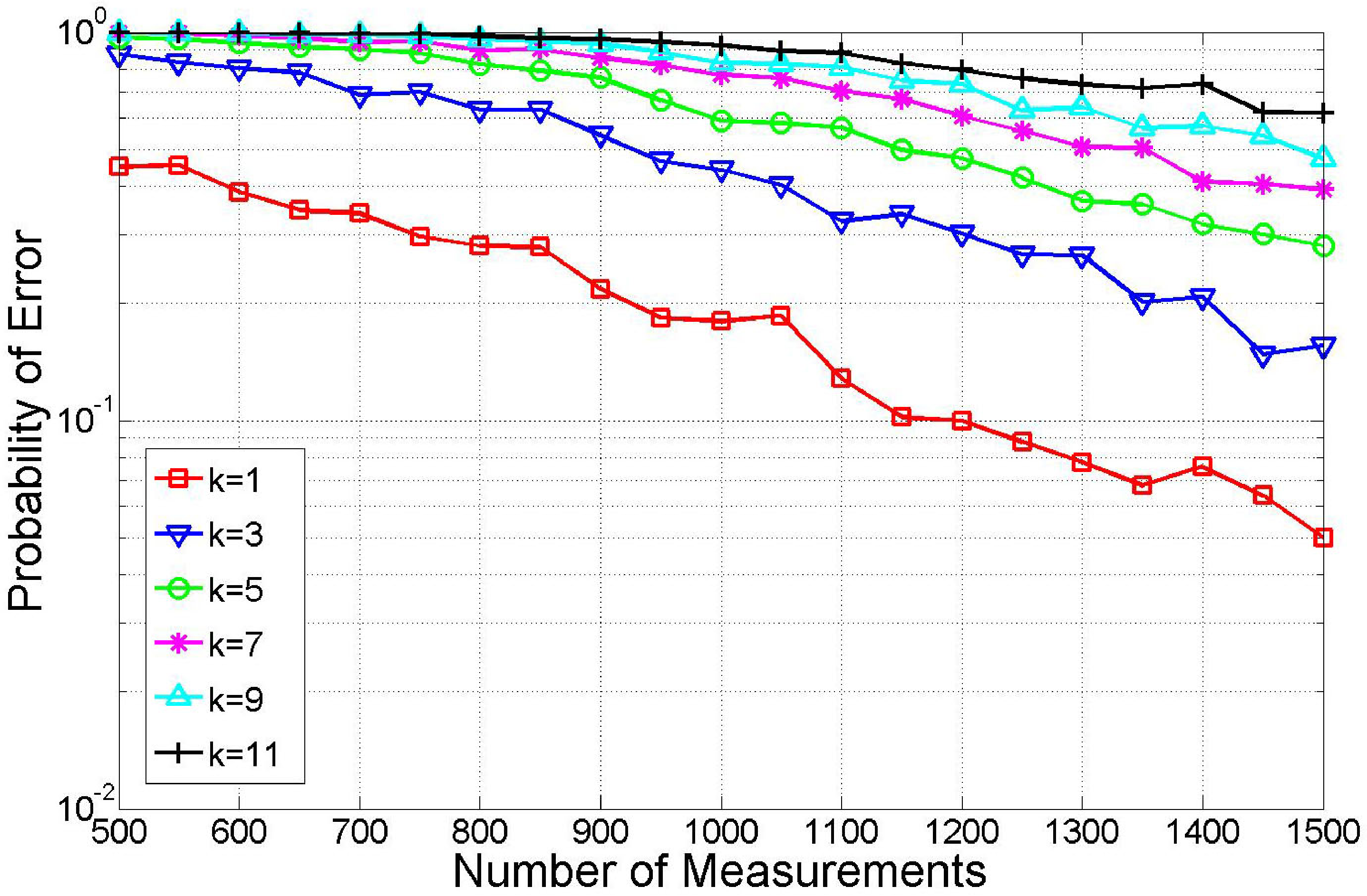}
\caption{Error probability of MP based hypothesis testing algorithm in log scale as the number of measurements $m$ is varied on various $k$ values, when $n=1000$. The normal and abnormal random variables follow $\cN(0,1)$ and $\cN(0,100)$ respectively.}
\label{fig:MP_HD_var_0_10}
\end{figure}

For large values of $n$ and $k$, Fig. \ref{fig:MP_HD_var_0_10} shows the performance of MP based hypothesis testing algorithm when the two types of random variables have different variances. We vary $k$ from $1$ to $11$ when $n = 1000$. In this parameter setup, LT methods have difficulties in finding the $k$ abnormal random variables out of $n$, since $n$ and $k$ are huge. The normal and abnormal random variables follow the Gaussian distribution $\cN(0,1)$ and $\cN(0,100)$ respectively, and the error probability is obtained from 500 random trials. Fig. \ref{fig:MP_HD_var_0_10} demonstrates that as $k$ becomes larger, more measurements are required to figure out the abnormal random variables.

\subsection{Random variables with different means}
\begin{figure}[t]
\centering
\includegraphics[scale=0.15]{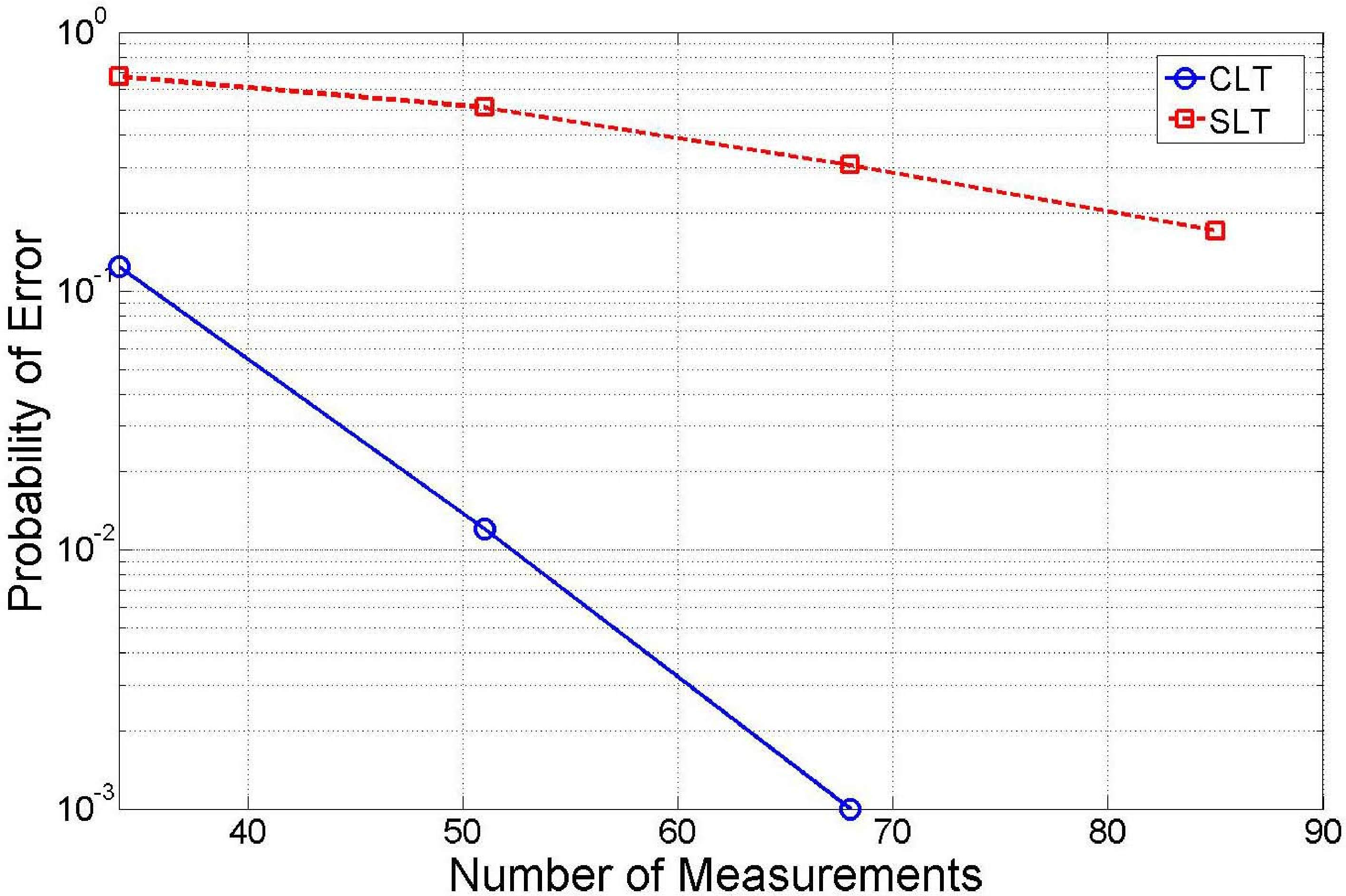}
\caption{Error probability in log scale as the number of measurements $m$ is varied, when $n=102$. The normal and abnormal random variables follow $\cN(8,1)$ and $\cN(0,1)$ respectively.}
\label{fig:N102Expectation}
\end{figure}

Under the same simulation setup as in Fig. \ref{fig:N100Variance}, we test the error probability performance of mixed observations for two Gaussian distributions: the anomalous Gaussian distribution $\cN(0, 1)$, and the normal Gaussian distribution $\cN(8,1)$. We also slightly adjust the number of total random variables to $n=102$ and $k=1$, to make sure that each random variable participates in the same integer number $\frac{6m}{n}$ of measurements. Mixed observations visibly reduce the error probability under the same number of measurements, compared with separate observations. For example, even when $m=68<n=102$, CLT correctly identifies the anomalous random variable in $999$ out of $1000$ cases by using $m$ mixed observations from the bipartite graphs. Fig. \ref{fig:N102Expectation} shows the result when the two types of random variables have different means.

\begin{figure}[t]
\centering
\includegraphics[scale=0.15]{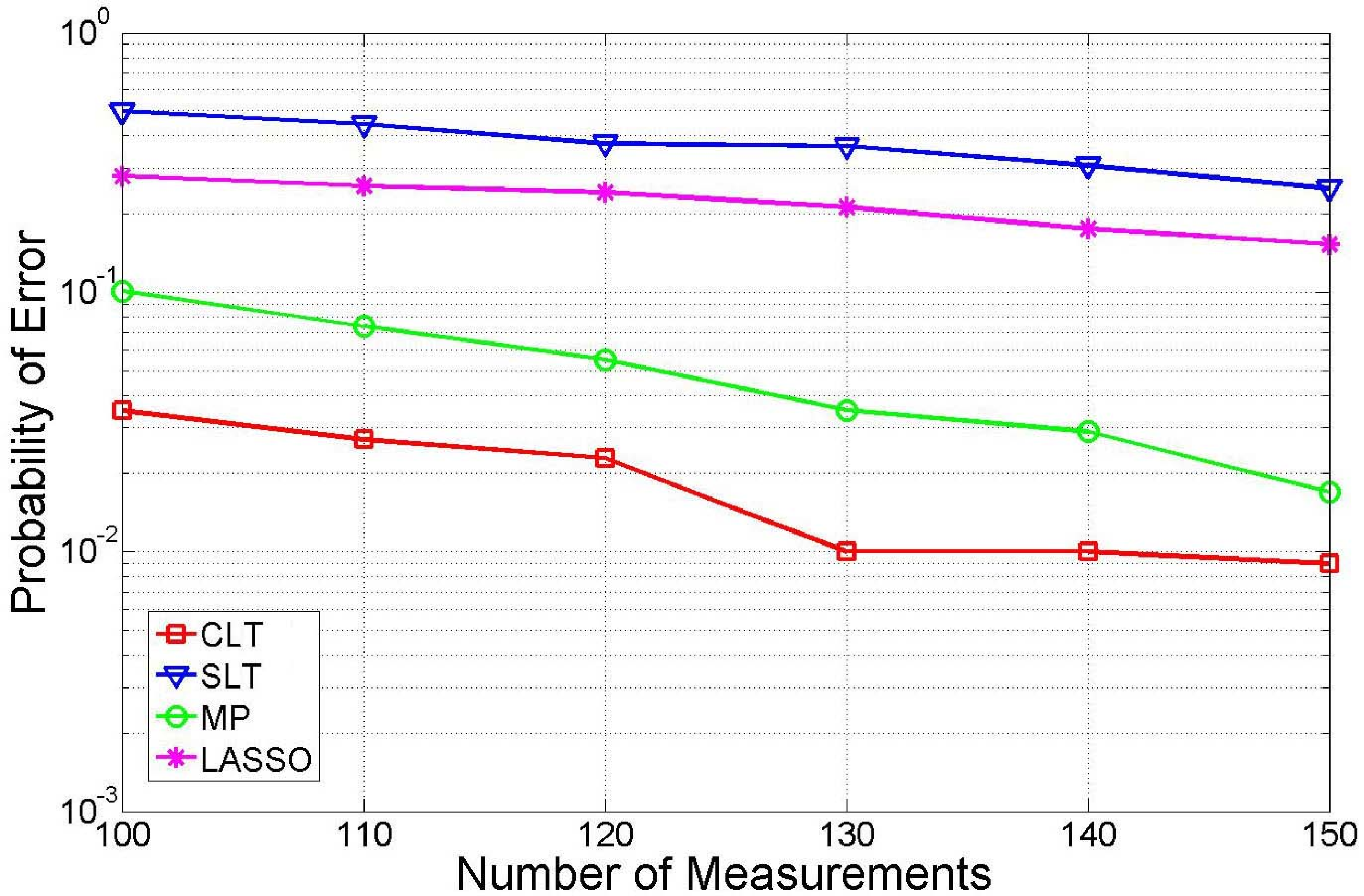}
\caption{Error probability in log scale as the number of measurements $m$ is varied, when $(n,k)=(200,1)$. The normal and abnormal random variables follow $\cN(0,1)$ and $\cN(7,1)$ respectively.}
\label{fig:comparision_LRT_MP_LASSO1}
\end{figure}

\begin{figure}[t]
\centering
\includegraphics[scale=0.15]{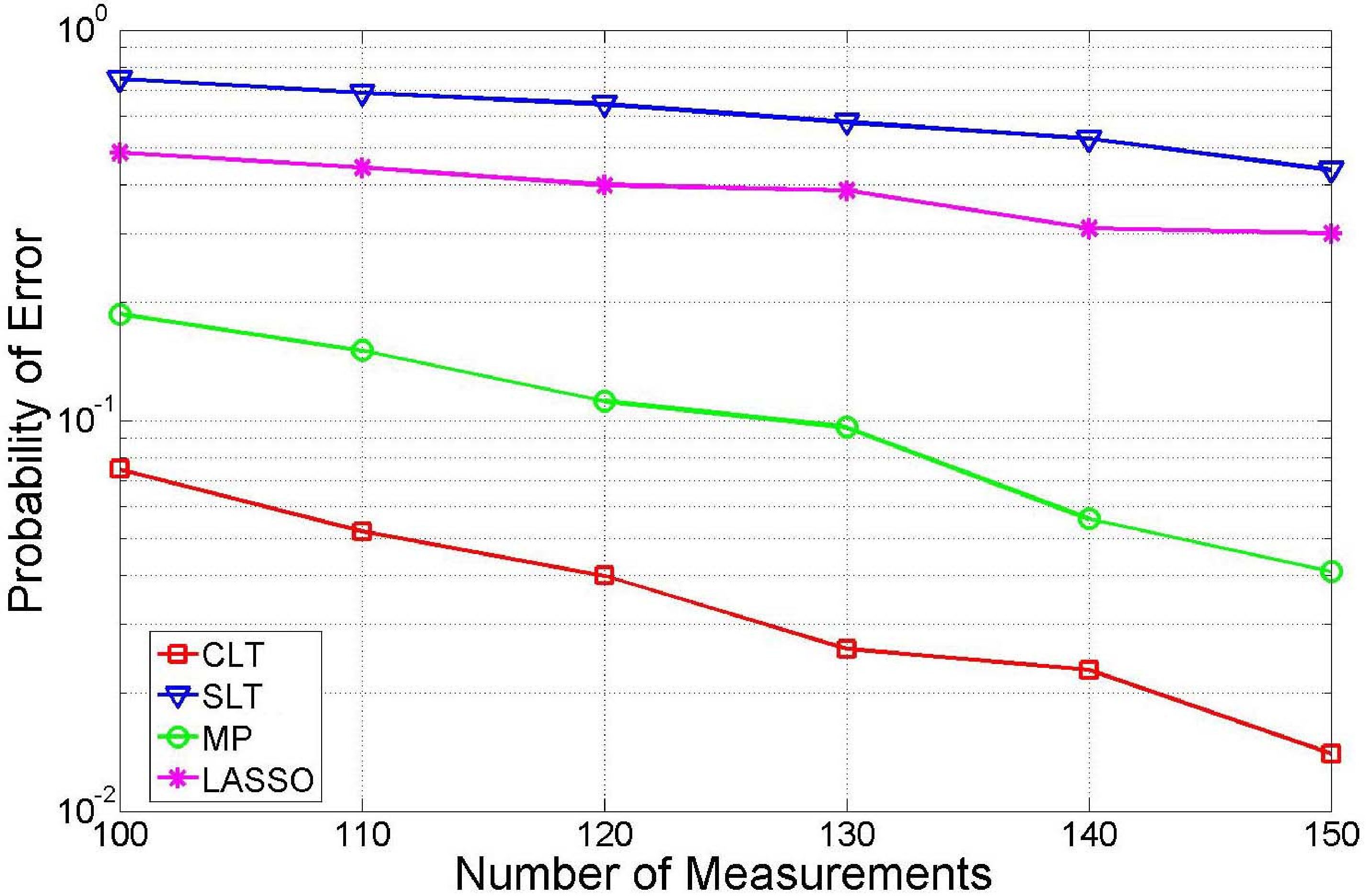}
\caption{Error probability in log scale as the number of measurements $m$ is varied, when $(n,k)=(200,2)$. The normal and abnormal random variables follow $\cN(0,1)$ and $\cN(7,1)$ respectively.}
\label{fig:comparision_LRT_MP_LASSO2}
\end{figure}

In addition, we carry out simulations to show the results from MP and LASSO based hypothesis testing methods and compare them against the results from LT methods. Figs. \ref{fig:comparision_LRT_MP_LASSO1} and  \ref{fig:comparision_LRT_MP_LASSO2} show the simulation results when the normal and abnormal random variables follow the Gaussian distribution $\cN(0,1)$ and $\cN(7,1)$ on $k=1$ and $2$ respectively. The error probability is obtained from 1000 random trials for each $m$. In this simulation setup, CLT and SLT show the best and the worst performance in error probability among the methods respectively.

\begin{figure}[t]
\centering
\includegraphics[scale=0.15]{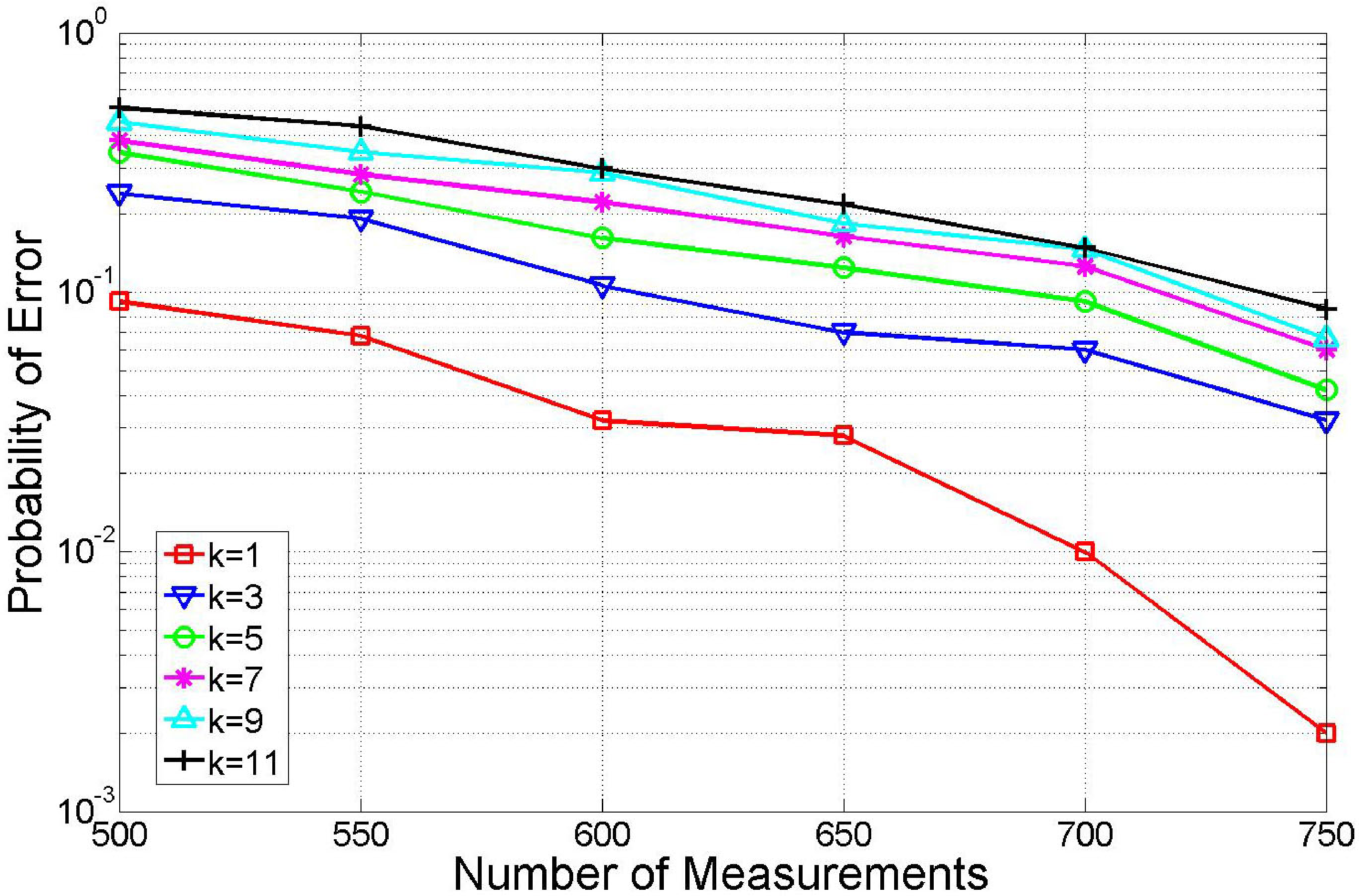}
\caption{Error probability of MP based hypothesis testing algorithm in log scale as the number of measurements $m$ is varied on various $k$ values, when $n=1000$. The normal and abnormal random variables follow $\cN(0,1)$ and $\cN(8,1)$ respectively.}
\label{fig:MP_HD_mean_0_8}
\end{figure}

\begin{figure}[t]
\centering
\includegraphics[scale=0.15]{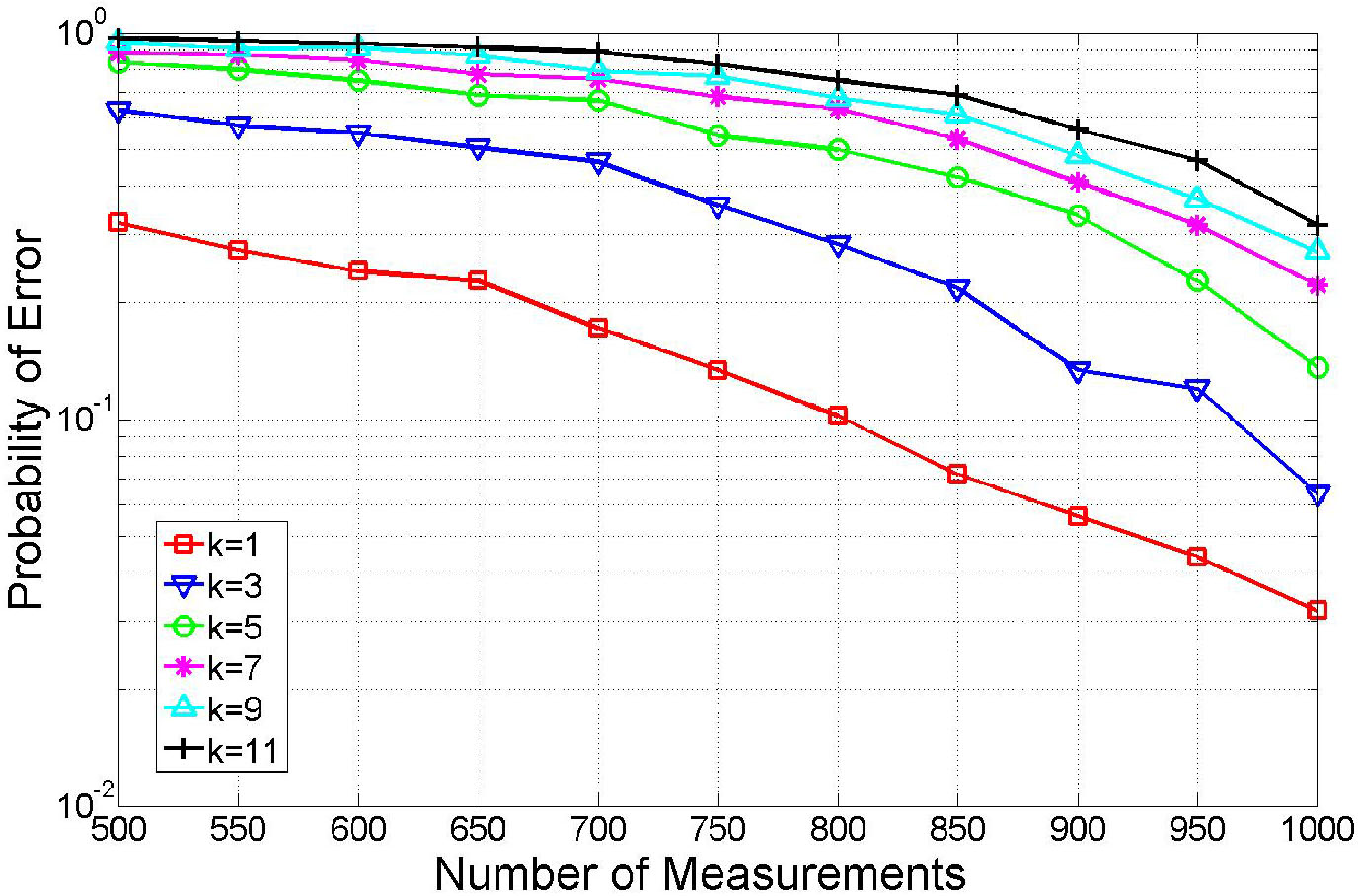}
\caption{Error probability of LASSO based hypothesis testing algorithm in log scale as the number of measurements $m$ varied on various $k$ values, when $n=1000$. The normal and abnormal random variables follow $\cN(0,1)$ and $\cN(8,1)$ respectively.}
\label{fig:LASSO_HD_mean_0_8}
\end{figure}

\begin{figure}[t]
\centering
\includegraphics[scale=0.15]{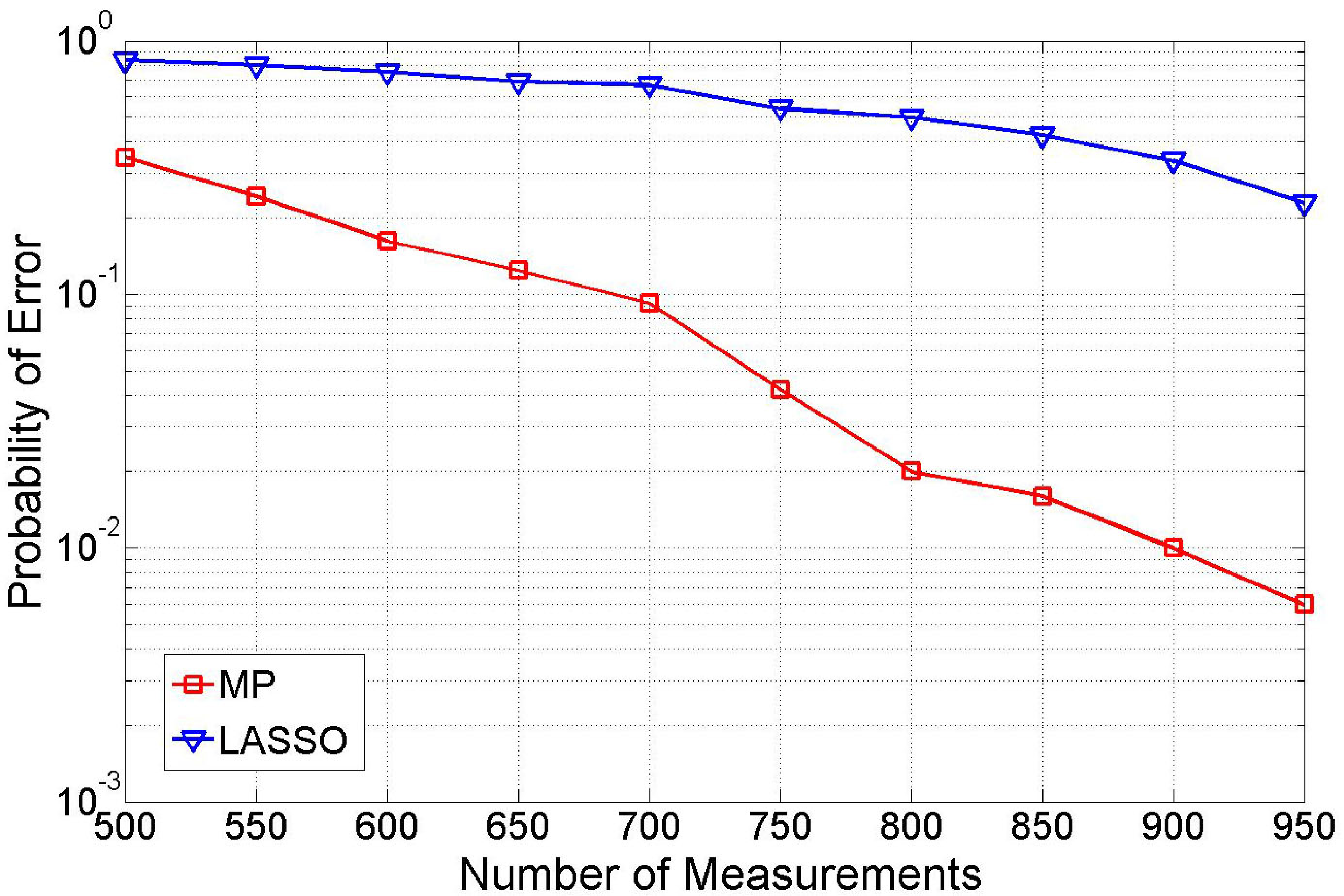}
\caption{Error probability in log scale as the number of measurements $m$ is varied on various $k$ values, when $(n,k)=(1000,5)$. The normal and abnormal random variables follow $\cN(0,1)$ and $\cN(8,1)$ respectively.}
\label{fig:MP_LASSO_HD}
\end{figure}

We compare the performance of MP and LASSO based hypothesis testing algorithms on large values of $n$ and $k$ which is the  computational challenging case for LT in Figs. \ref{fig:MP_HD_mean_0_8} and \ref{fig:LASSO_HD_mean_0_8}. For these simulations, we set $n$ to $1000$ and $k$ from $1$ to $11$. We perform $500$ random trials for each $m$ to obtain the error probability. The normal and abnormal random variables follow $\cN(0,1)$ and $\cN(8,1)$ respectively. Finally, Fig. \ref{fig:MP_LASSO_HD} shows the comparison result between MP and LASSO based hypothesis testing methods.

\subsection{Random variables with different means and variances}
\begin{figure}[t]
\centering
\includegraphics[scale=0.15]{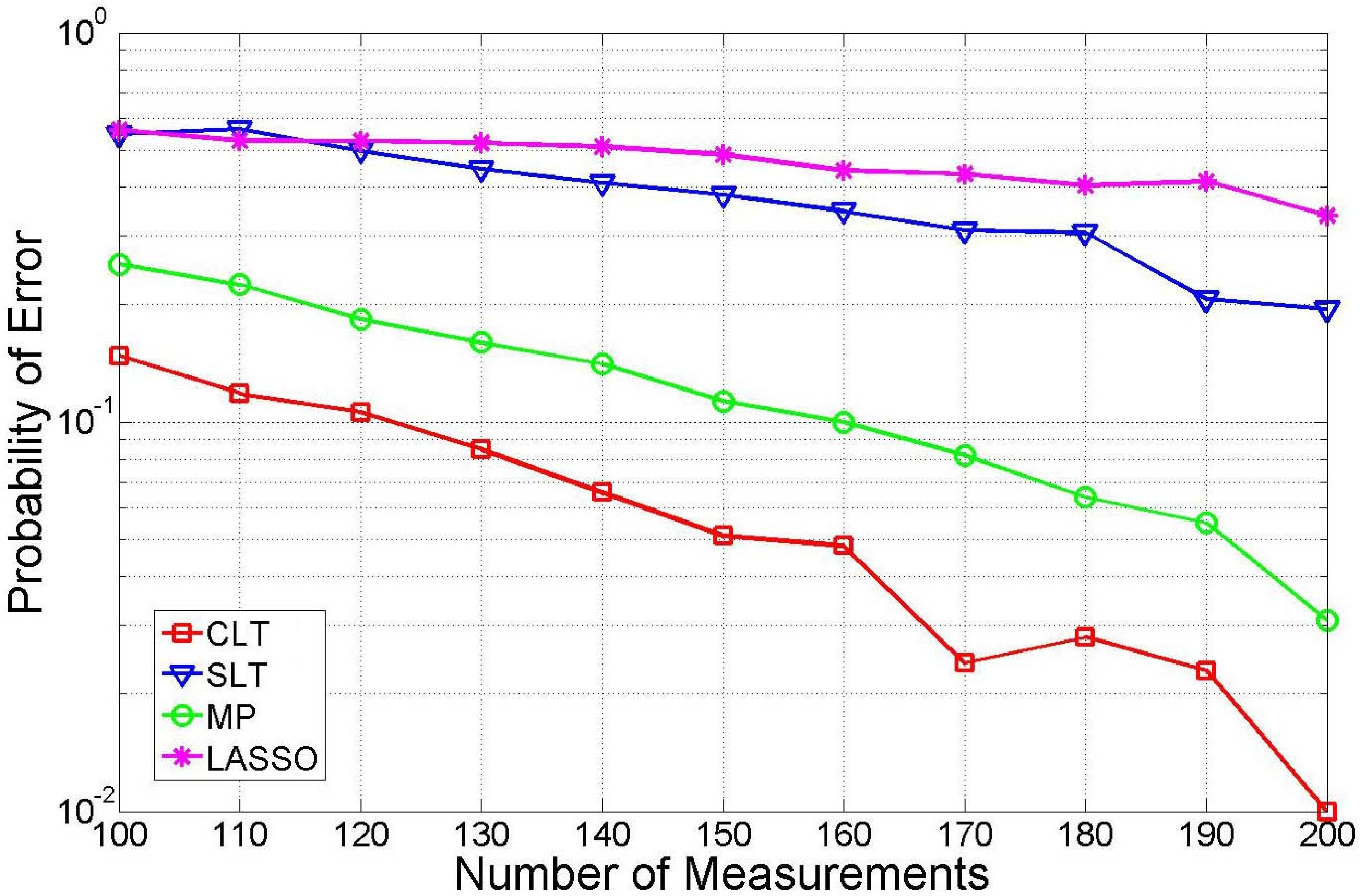}
\caption{Error probability in log scale as the number of measurements $m$ is varied, when $(n,k)=(200,1)$. The normal and abnormal random variables follow $\cN(0,1)$ and $\cN(7,100)$ respectively.}
\label{fig:comparision_LRT_MP_LASSO_diff_mean_var1}
\end{figure}

\begin{figure}[t]
\centering
\includegraphics[scale=0.15]{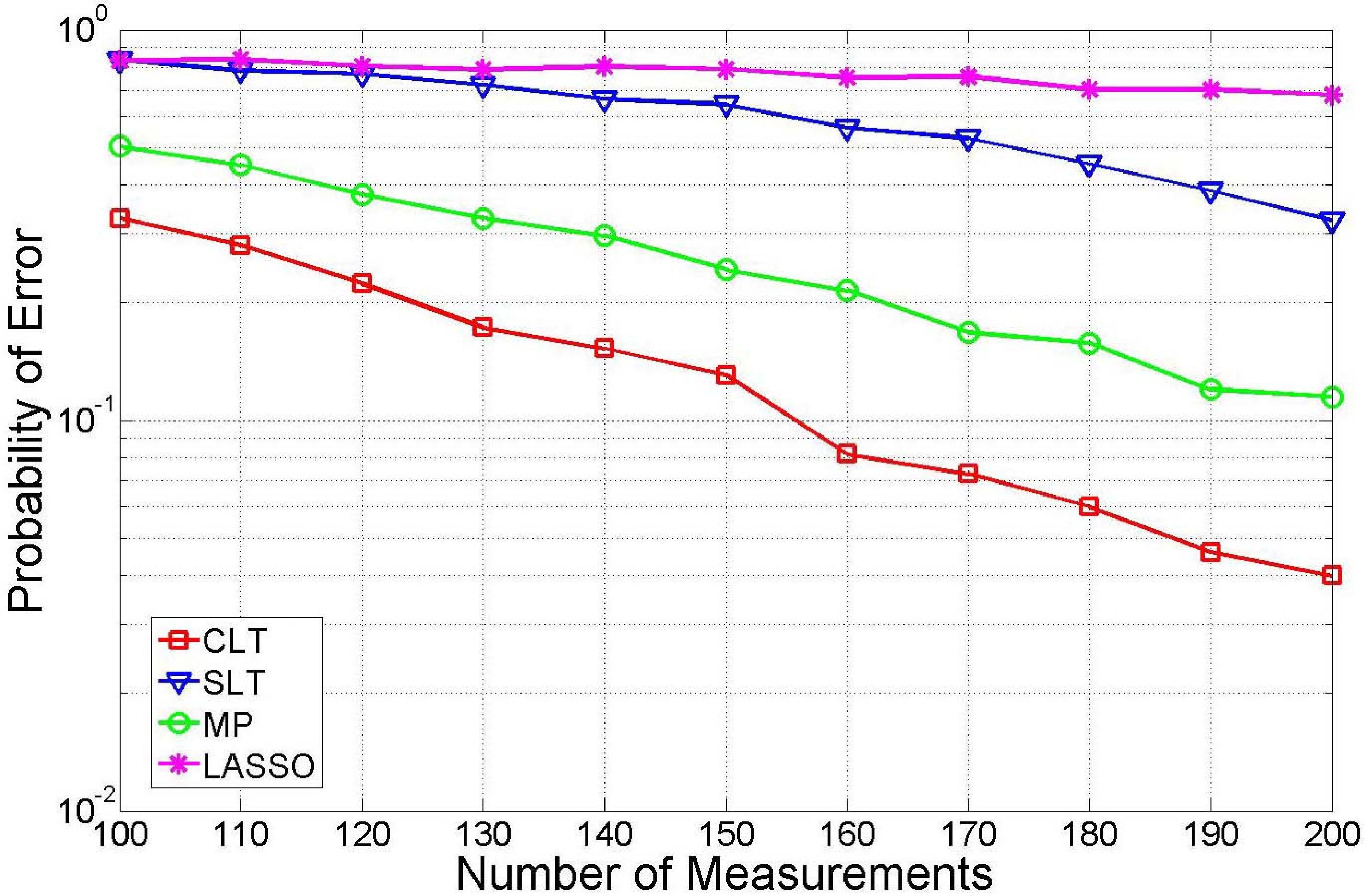}
\caption{Error probability in log scale as the number of measurements $m$ is varied, when $(n,k)=(200,2)$. The normal and abnormal random variables follow $\cN(0,1)$ and $\cN(7,100)$ respectively.}
\label{fig:comparision_LRT_MP_LASSO_diff_mean_var2}
\end{figure}

We further test the performance of mixed observations in error probability for two Gaussian distributions with different means and variances. Figs. \ref{fig:comparision_LRT_MP_LASSO_diff_mean_var1} and \ref{fig:comparision_LRT_MP_LASSO_diff_mean_var2} show the results when two types of random variables have different means and variances. The abnormal and normal random variables follow $\cN(0, 1)$, and $\cN(7,100)$ respectively. We obtain the error probability from 1000 random trials for each $m$. The simulation results show that CLT and MP still identify the abnormal random variables with fewer observations than SLT.

\section{Conclusion}\label{sec:conclusion}
In this paper, we studied the compressed hypothesis testing problem, which is finding $k$ anomalous random variables following a different probability distribution among $n$ random variables by using mixed observations of these $n$ random variables. Our analysis  showed that mixed observations, compared with separate observations of individual random variables, can reduce the number of samples required to identify the anomalous random variables accurately. Compared with conventional compressed sensing problems, in our setting, each random variable may take dramatically different realizations in different observations. Therefore, the compressed hypothesis testing problem considered in this paper is quite different from the conventional compressed sensing problem. Additionally, for large-scale hypothesis testing problems, we designed efficient algorithms - Least Absolute Shrinkage and Selection Operator (LASSO) and Message Passing (MP) based algorithms. Numerical experiments demonstrate that mixed observations can play a significant role in reducing the required samples in hypothesis testing problems.

There are some open questions remained in performing hypothesis testing from mixed observations. For example, for random variables of non-Gaussian distributions, it is not explicitly known what linear mixed observations maximize the anomaly detection error exponents. In addition, it is very interesting to explore the mixed observations for anomaly detection with random variables having unknown abnormal probability distributions.


\bibliographystyle{IEEEtran}
\bibliography{CompressedHT}

\section{Appendix}
\subsection{Proof of Theorem \ref{thm:numberofsamples_deterministic_timevarying}}
\label{apx:proof_deter_time_varying}
We provide the proof of Theorem \ref{thm:numberofsamples_deterministic_timevarying} here. The framework of this proof follows the book written by Cover and Thomas \cite[Chapter 11]{coverbook}
\begin{IEEEproof}
In Algorithm \ref{alg:timevaryingalg2deterministic}, for two different hypotheses $H_v$ and $H_w$, we choose the probability likelihood ratio threshold of the Neyman-Pearson testing in a way, such that the hypothesis testing error probability decreases with the largest error exponent. Now we focus on deriving what this largest error exponent is, under deterministic time-varying measurements.

For simplicity of presentation, we first consider a special case: there are only two possible sensing vectors $\ba_1$ and $\ba_2$; and one half of the sensing vectors are $\ba_1$ while the other half are $\ba_2$. The conclusions can be extended to general distribution $p_\bA(\ba)$ on $\bA$, in a similar way of reasoning. In addition, we assume that the observation data is over a discrete space $\chi$, which can also be generalized to a continuous space without affecting the conclusion in this theorem. Since we use the probability mass function over a discrete space, we will use upper letters for the probability mass functions to distinguish the probability density functions over a continuous space in this proof. Suppose we take $m$ measurements in total, our assumption translates to that $\frac{1}{2} m$ measurements are taken from the sensing vector $\ba_1$, and $\frac{1}{2} m$ measurements are taken from the sensing vector $\ba_2$. Without loss of generality, we consider two hypotheses denoted by $H_1$ and $H_2$. Under the sensing vector $\ba_1$, we assume that $H_1$ generates distribution $P_1$ for observation data; $H_2$ generates distribution $P_2$ for observation data. Under $\ba_2$, we assume that $H_1$ generates distribution $P_3$ for observation data; $H_2$ generates distribution $P_4$ for observation data. Please refer to Table \ref{tbl:measdist} for the observation distributions under different sensing vectors and different hypotheses.
\begin{table}[t]
\caption{ Hypothesis testing measurement distribution }
\centering
\setlength{\tabcolsep}{10pt}
\renewcommand{\arraystretch}{2}
\label{tbl:measdist}
\begin{tabular}{ccc}
                         & $H_1$                      & $H_2$                      \\ \cline{2-3}
\multicolumn{1}{c|}{$\ba_1$} & \multicolumn{1}{c|}{$P_1$} & \multicolumn{1}{c|}{$P_2$} \\ \cline{2-3}
\multicolumn{1}{c|}{$\ba_2$} & \multicolumn{1}{c|}{$P_3$} & \multicolumn{1}{c|}{$P_4$} \\ \cline{2-3}
\end{tabular}
\end{table}

From Neyman-Pearson lemma \cite[Theorem 11.7.1]{coverbook}, for a certain constant $T \geq 0$, the optimum test for two hypotheses is stated as the following likelihood ratio test:
\begin{align}\label{eq:likelihood_ratio}
\frac{P_1(X_1, X_2, ..., X_{m/2} | \ba_1) P_3(X_{m/2+1},X_{m/2+2},...,X_{m} | \ba_2)}{P_2(X_1, X_2, ..., X_{m/2} | \ba_1) P_4(X_{m/2+1},X_{m/2+2},...,X_{m} | \ba_2)} \geq T,
\end{align}
where $X_i$ is the $i$-th random variable. Then, suppose that $P$ is the empirical distribution of observation data under the sensing vector $\ba_1$, and that $P'$ is the empirical distribution of observation data under the sensing vector $\ba_2$. The likelihood ratio test in \eqref{eq:likelihood_ratio} is equivalent to
\par\noindent\small
\begin{align}
	D(P||P_2)-D(P||P_1)+ D(P'||P_4)-D(P'||P_3) \geq \frac{1}{n} \log(T),
\end{align}
\normalsize
where $D(P||Q) := \sum_{y \in \chi} P(y) \log \frac{P(y)}{Q(y)}$, which is the relative entropy or Kullback-Leibler distance between two probability mass functions $P$ and $Q$. By using the Sanov's theorem \cite[Theorem 11.4.1]{coverbook} to derive the probability of error, the error exponent of the second kind, i.e., wrongly deciding ``hypothesis $H_1$ is true'' when hypothesis $H_2$ is actually true, is stated by the following optimization problem:
\par\noindent\small
\begin{align}
    & \underset{P,P'}{\text{minimize}}\;\; D(P||P_2)+ D(P'||P_4) \nonumber\\
    & \text{subject to}\;\; D(P||P_2)-D(P||P_1)+ D(P'||P_4)-D(P'||P_3)\geq \frac{\log(T)}{n}, \nonumber\\
    & \quad\quad\quad\quad \sum_{y} P(y) = 1, \nonumber\\
    & \quad\quad\quad\quad \sum_{y} P'(y) = 1.
\end{align}
\normalsize

By using the Lagrange multiplier method, for the Lagrange function, we have
\par\noindent\small
\begin{align*}
 L(P,P',\lambda,v_1,v_2) & = D(P||P_2) + D(P'||P_4) \\
& \;\;+ \lambda \left(  D(P||P_2)-D(P||P_1) +  D(P'||P_4)-D(P'||P_3) - \frac{\log(T)}{n} \right ) \\
& \;\;+ v_1 \left( \sum_{y} P(y) - 1 \right) + v_2 \left( \sum_{y} P'(y) - 1 \right).
\end{align*}
\normalsize

From the first order condition for an optimal solution, by differentiating the Lagrange function, with respect to $P(y)$ and $P'(y)$, we have
\par\noindent\small
\begin{align*}
	\log\bigg(\frac{P(y)}{P_2(y)}\bigg)+1+\lambda \log \bigg(\frac{P_1(y)}{P_2(y)}\bigg) +v_1=0,\\
\log\bigg(\frac{P'(y)}{P_4(y)}\bigg)+1+\lambda \log \bigg(\frac{P_3(y)}{P_4(y)}\bigg) +v_2=0.
\end{align*}
\normalsize
From these equations, we can obtain the minimizing $P$ and $P'$,
\begin{align}\label{def:P_lambda}
    & P=\frac{P_1^\lambda(y)P_2^{1-\lambda}(y)}{\sum_{y \in \chi}{P_1^\lambda(y)P_2^{1-\lambda}(y)} } := P_{\lambda} (y|\ba_1), \nonumber \\
    & P'=\frac{P_3^\lambda(y)P_4^{1-\lambda}(y)}{\sum_{y \in \chi}{P_3^\lambda(y)P_4^{1-\lambda}(y)} } :=P_{\lambda} (y|\ba_2),
\end{align}
where $\lambda$ is chosen such that $D(P||P_2)-D(P||P_1)+ D(P'||P_4)-D(P'||P_3) = \frac{1}{n} \log(T)$.

By symmetry, the error exponent of the second kind and the error exponent of the first kind are stated as follows respectively:
\begin{align}
    & D(P_{\lambda} (y|\ba_1)~||~P_2) + D(P_{\lambda} (y|\ba_2)~||~P_4), \\
    & D(P_{\lambda} (y|\ba_1)~||~P_1) + D(P_{\lambda} (y|\ba_2)~||~P_3).
\end{align}
The first error exponent is a non-decreasing function in $\lambda$, and the second error exponent is a non-increasing function in $\lambda$. Therefore, the optimal error exponent, which is the minimum of these two exponents, is achieved when they are equal; namely
\begin{align} \label{eq:OC_min_error_exp_condition}
   D(P_{\lambda} (y|\ba_1)~||~P_2)+ D(P_{\lambda} (y|\ba_2)~||~P_4) = D(P_{\lambda} (y|\ba_1)~||~P_1)+ D(P_{\lambda} (y|\ba_2)~||~P_3).
\end{align}
And at this point, the probability of error in hypothesis testing is obtained as
\begin{align}\label{eq:prob_err_oc}
	\Pb_{err} = 2^{-\frac{m}{2}[D(P_{\lambda} (y|\ba_1)||P_2)+ D(P_{\lambda} (y|\ba_2)||P_4)]} = 2^{-\frac{m}{2}[D(P_{\lambda} (y|\ba_1)||P_1)+ D(P_{\lambda} (y|\ba_2)||P_3)]}.
\end{align}
Remark that $\frac{1}{2}[D(P_{\lambda} (y|\ba_1)||P_2)+ D(P_{\lambda} (y|\ba_2)||P_4)]$ represents $\E_{\bA}[ D(P_{\lambda} (y|\ba)||P(y | \ba,\; H_2) ]$, and in a general form, \eqref{eq:OC_min_error_exp_condition} is expressed as
\begin{align}\label{eq:general_min_error_exponent}
\E_{\bA} \bigg[ D \bigg(P_{\lambda} (y|\ba ) ~||~ P(y | \ba,\; H_v) \bigg) \bigg] = \E_{\bA}\bigg[ D \bigg(P_{\lambda} (y|\ba )~||~P(y | \ba,\; H_w) \bigg) \bigg],
\end{align}
which is the equation introduced in \eqref{eq:minimum_error_exponent}. We define the error exponent as the outer conditional Chernoff information. This finishes the characterization of the optimal error exponent in pairwise hypothesis testing under deterministic time-varying measurements introduced in Definition \ref{def:OC}.

We then prove that the error exponent \eqref{eq:general_min_error_exponent}, the outer conditional Chernoff information, is equivalent to
\par\noindent\small
\begin{align}
OC(P_{Y|\bA,H_v}, P_{Y|\bA,H_w}) &=-\min_{0\leq \lambda \leq 1} \int P_\bA(\ba) \log\left(\int{P_{Y|\bA,H_v}^{\lambda}(y|\ba,H_v)P_{Y|\bA,H_w}^{1-\lambda}(y|\ba,H_w)\,dy} \right)\,d\ba \nonumber\\
&=-\min_{0\leq \lambda \leq 1}  \E_{\bA} \left( \log \left(\int{P_{Y|\bA,H_v}^{\lambda}(y|\ba,H_v)P_{Y|\bA,H_w}^{1-\lambda}(y|\ba,H_w)\,dy}\right) \right).
\label{eq:conditionalchernoff inequality}
\end{align}
\normalsize
In the proof, we restrict our attention to $H_v=H_1$ and $H_w=H_2$. We will show that the $\lambda \in [0,1]$ minimizing $\E_{\bA} \left( \log \left(\int{P_{Y|\bA,H_v}^{\lambda}(y|\ba,H_v)P_{Y|\bA,H_w}^{1-\lambda}(y|\ba,H_w)\,dy}\right) \right)$ exactly leads to \eqref{eq:general_min_error_exponent}.
Especially, under that minimizer $\lambda$, we will obtain the following equalities:
\begin{align}
 -\E_{\bA} \left( \log \left(\int{P_{Y|\bA,H_1}^{\lambda}(y)P_{Y|\bA,H_2}^{1-\lambda}(y)\,dy}\right) \right) & = \frac{1}{2} D(P_{\lambda} (y|\ba_1)~||~P_1)+\frac{1}{2} D(P_{\lambda} (y|\ba_2)~||~P_3)\\
 & = \frac{1}{2} D(P_{\lambda} (y|\ba_1)~||~P_2)+\frac{1}{2} D(P_{\lambda} (y|\ba_2)~||~P_4).
\label{eq:optimallambda34}
\end{align}

On the one hand, we will show that from the definition of $OC(P_{Y|\bA,H_v}, P_{Y|\bA,H_w})$, the outer conditional Chernoff information is expressed as $-\E_{\bA} \left( \log \left(\int{P_{Y|\bA,H_v}^{\lambda}(y|\ba,H_v)P_{Y|\bA,H_w}^{1-\lambda}(y|\ba,H_w)\,dy}\right) \right)$ with an optimal $\lambda$. For that, from (\ref{eq:OC_min_error_exp_condition}), which is the definition of the outer conditional Chrenoff information  in this specific example, we obtain following equations:
\begin{align}\label{eq:intermediate}
0&\stackrel{\text{(\ref{eq:OC_min_error_exp_condition})}}{=}\;\frac{1}{2} [D(P_{\lambda} (y|\ba_1)~||~P_1)-D(P_{\lambda} (y|\ba_1)~||~P_2)]  +\frac{1}{2} [D(P_{\lambda} (y|\ba_2)~||~P_3)-D(P_{\lambda} (y|\ba_2)~||~P_4)]\nonumber\\
&\;\;\;=\;\frac{1}{2} \frac{\sum_{y}{P_1^{\lambda}(y)P_2^{1-\lambda}(y) \log(\frac{P_2(y)}{P_1(y)})} }{ \sum_{y} P_1^\lambda(y)P_2^{1-\lambda}(y)} +\frac{1}{2} \frac{\sum_{y}{P_3^{\lambda}(y)P_4^{1-\lambda}(y) \log(\frac{P_4(y)}{P_3(y)})} }{ \sum_{y} P_3^\lambda(y)P_4^{1-\lambda}(y)}.
\end{align}
Let a particular $\lambda$ satisfy (\ref{eq:intermediate}). Under this $\lambda$, the hypothesis testing error exponent with $m$ measurements (See \eqref{eq:prob_err_oc}) is equal to
\begin{eqnarray}\label{eq:OC_derivation}
	&& \hspace{-3em} \frac{1}{2} D(P_{\lambda} (y|\ba_1)~||~P_1)+\frac{1}{2} D(P_{\lambda} (y|\ba_2)~||~P_3) \nonumber \\
    &\overset{\eqref{def:P_lambda}}{=}& \frac{1}{2} \frac{1}{\sum_{y} P_1^\lambda(y)P_2^{1-\lambda}(y)}
        \sum_{y} \bigg( P_1^{\lambda}(y)P_2^{1-\lambda}(y) \bigg[(1-\lambda)\log(\frac{P_2(y)}{P_1(y)})-\log \big(\sum_{y} P_1^{\lambda}(y)P_2^{1-\lambda}(y) \big) \bigg] \bigg) \nonumber \\
    && \quad +\frac{1}{2} \frac{1}{\sum_{y} P_3^\lambda(y)P_4^{1-\lambda}(y)}\sum_{y} \bigg( P_3^{\lambda}(y)P_4^{1-\lambda}(y) \bigg[(1-\lambda)\log(\frac{P_4(y)}{P_3(y)})-\log \big(\sum_{y} P_3^{\lambda}(y)P_4^{1-\lambda}(y)\big) \bigg] \bigg),\nonumber \\
     & \overset{\eqref{eq:intermediate}}{=}& -\frac{1}{2} \log \big(\sum_{y} P_1^{\lambda}(y)P_2^{1-\lambda}(y) \big) - \frac{1}{2} \log \big(\sum_{y} P_3^{\lambda}(y)P_4^{1-\lambda}(y) \big) \nonumber \\
    &=&-\E_{\bA} \left( \log \left(\int{P_{Y|\bA,H_1}^{\lambda}(y|\ba,H_1)P_{Y|\bA,H_2}^{1-\lambda}(y|\ba,H_2)\,dy}\right) \right).
\end{eqnarray}

On the other hand, we will show that the minimizing $\E_{\bA} \left( \log \left(\int{P_{Y|\bA,H_1}^{\lambda}(y|\ba,H_1)P_{Y|\bA,H_2}^{1-\lambda}(y|\ba,H_2)\,dy}\right) \right)$ over $\lambda$  leads to the definition of the outer conditional Chrenoff information. In order to minimize
\begin{align}\label{eq:expectation_log}
	\E_{\bA} \left( \log \left(\int{P_{Y|\bA,H_1}^{\lambda}(y|\ba,H_1)P_{Y|\bA,H_2}^{1-\lambda}(y|\ba,H_2)\,dy}\right) \right),
\end{align}
which is $\frac{1}{2} \log \big(\sum_{y} P_1^{\lambda}(y)P_2^{1-\lambda}(y) \big) + \frac{1}{2} \log \big(\sum_{y} P_3^{\lambda}(y)P_4^{1-\lambda}(y) \big)$ in this specific example, over $\lambda$, by setting the derivative of \eqref{eq:expectation_log} with respect to $\lambda$ to $0$, we have
\begin{align}\label{eq:derv}
	0=\frac{1}{2} \frac{\sum_{y}{P_1^{\lambda}(y)P_2^{1-\lambda}(y) \log(\frac{P_2(y)}{P_1(y)})} }{ \sum_{y} P_1^\lambda(y)P_2^{1-\lambda}(y)} +\frac{1}{2} \frac{\sum_{y}{P_3^{\lambda}(y)P_4^{1-\lambda}(y) \log(\frac{P_4(y)}{P_3(y)})} }{ \sum_{y} P_3^\lambda(y)P_4^{1-\lambda}(y)}.
\end{align}
It is noteworthy that (\ref{eq:derv}) is the same as (\ref{eq:intermediate}). Let us denote a minimizer
\begin{align}
	\lambda_{min} = \underset{0\leq \lambda \leq 1}{\argmin} \; \E_{\bA} \left(\log \left(\int{P_{Y|\bA,H_1}^{\lambda}(y|\ba,H_1)P_{Y|\bA,H_2}^{1-\lambda}(y|\ba,H_2)\,dy}\right)\right).
\end{align}
Then, when $\lambda = \lambda_{min}$, \eqref{eq:derv} is satisfied. Furthermore, for $\lambda=\lambda_{min}$, we have \eqref{eq:OC_derivation} due to \eqref{eq:derv}. In the same reason, we can have the same relation between $-\E_{\bA} \left( \log \left(\int{P_{Y|\bA,H_1}^{\lambda}(y|\ba,H_1)P_{Y|\bA,H_2}^{1-\lambda}(y|\ba,H_2)\,dy}\right) \right)$ and $\frac{1}{2} D(P_{\lambda} (y|\ba_1)~||~P_2)+\frac{1}{2} D(P_{\lambda} (y|\ba_2)~||~P_4)$. Therefore, we can conclude the equivalence of the two different definitions of the outer conditional Chernoff information introduced in \eqref{eq:outer_CI}.



Overall, the smallest possible error exponent between any pair of hypotheses is
\begin{align*}
E=\min_{1\leq v,w \leq l, v\neq w} OC(P_{Y|\bA,H_v}, P_{Y|\bA,H_w}).
\end{align*}

Without loss of generality, we assume $H_1$ is the true hypothesis. Since the error probability $\Pb_{err}$ in the Neyman-Pearson testing is
\begin{align*}
	\Pb_{err} \leq 2^{-m OC(P_{Y|\bA,H_v}, P_{Y|\bA,H_w})} \leq 2^{-mE}.
\end{align*}
By the union bound over the $l-1$ possible pairs $(H_1, H_w)$, the probability that $H_1$ is not correctly identified as the true hypothesis is upper bounded by $l\times2^{-mE}$ in terms of scaling, where $l=\binom{n}{k}$. Therefore, $m=\Theta(k \log(n)E^{-1})$ samples are enough for identifying the $k$ anomalous samples with high probability.
\end{IEEEproof}

\end{document}